\documentclass[prodmode,acmtocl]{acmsmall}

\usepackage[dvips]{rotating}
\usepackage{algorithmic}
\usepackage{algorithm}
\usepackage{amscd}
\usepackage{amsfonts}
\usepackage{amsmath}
\usepackage{booktabs}
\usepackage{color}
\usepackage{epsfig}
\usepackage{graphicx}
\usepackage{latexsym}
\usepackage{multirow}
\usepackage{paralist}
\usepackage{proof} 
\usepackage{rotate} 
\usepackage{times} 
\usepackage{url}
\usepackage{verbatim} 
\usepackage{wrapfig}
\usepackage{xspace}
\usepackage{datetime}
\usepackage{todonotes}
\usepackage[FIGBOTCAP,TABBOTCAP,center,nooneline]{subfigure}
\usepackage{marginnote}

\newcommand{\blevel}{\ensuremath{\sf{blevel}}\xspace}

\newcommand{\satres}{\textsc{sat}\xspace}
\newcommand{\unsatres}{\textsc{unsat}\xspace}

\newcommand{\vi}{\ensuremath{\varphi}\xspace}
\newcommand{\vip}{\ensuremath{\varphi^p}\xspace}
\newcommand{\mup}{\ensuremath{\mu^p}\xspace}

\newcommand{\etap}{\ensuremath{\eta^p}\xspace}

\newcommand{\atoms}[1]{\ensuremath{Atoms(#1)}\xspace}

\newcommand{\T}{\ensuremath{\mathcal{T}}\xspace}
\newcommand{\Tone}{\ensuremath{\T_1}\xspace}
\newcommand{\Ttwo}{\ensuremath{\T_2}\xspace}
\newcommand{\Tonetwo}{\ensuremath{\Tone\cup \Ttwo}\xspace}
\newcommand{\smt}{SMT\xspace}
\newcommand{\smtt}{\ensuremath{\text{SMT}(\T)}\xspace}
\newcommand{\smttt}[1]{\ensuremath{\text{SMT}(#1)}\xspace}

\newcommand{\Ti}{\ensuremath{\T_i}\xspace}

\newcommand{\euf}{\ensuremath{\mathcal{EUF}}\xspace}

\newcommand{\la}{\ensuremath{\mathcal{LA}}\xspace}
\newcommand{\larat}{\ensuremath{\mathcal{LA}(\mathbb{Q})}\xspace}
\newcommand{\laint}{\ensuremath{\mathcal{LA}(\mathbb{Z})}\xspace}

\newcommand{\bv}{\ensuremath{\mathcal{BV}}\xspace}
\newcommand{\mem}{\ensuremath{\mathcal{AR}}\xspace}

\newcommand{\tonetwo}{\ensuremath{\T_1\cup\T_2}\xspace}

\newcommand{\smtlarat}{\smttt{\larat}}

\newcommand{\pmodels}{\models_p}

\newcommand{\Tsolver}{\TsolverGen{\T}}
\newcommand{\Tsolvers}{\TsolversGen{\T}}

\newcommand{\Tlemmas}{\T-lemmas\xspace}

\newcommand{\mathsat}{\textsc{MathSAT}\xspace}

\newcommand{\yices}{\textsc{Yices}\xspace}

\newcommand{\zthree}{\textsc{Z3}\xspace}

\newcommand{\mathsatfour}{\textsc{MathSAT4}\xspace}
\newcommand{\mathsatfive}{\textsc{MathSAT5}\xspace}

\newcommand{\omtt}{\ensuremath{\text{OMT}({\T})}\xspace}

\newcommand{\ub}{\ubgen{}{}}

\newcommand{\cost}{\ensuremath{{cost}}\xspace}
\newcommand{\mincost}{\ensuremath{{mincost}}\xspace}

\newcommand{\C}{\ensuremath{\mathcal{C}}\xspace}

\newcommand{\mut}{\ensuremath{\mu_{\T}}\xspace}
\newcommand{\mub}{\ensuremath{\mu_{\calb}}\xspace}

\newcommand{\I}{\ensuremath{\cali}\xspace}

\newcommand\mysout{\bgroup \markoverwith{{-}}\ULon}
\newcommand\nosout{\bgroup \markoverwith{{ }}\ULon}
\definecolor{mygray}{rgb}{0.90,0.90,0.90}
\definecolor{mywhite}{rgb}{1.00,1.00,1.00}

\definecolor {snow}                {rgb}{1.00,0.98,0.98}
\definecolor {ghostwhite}          {rgb}{0.97,0.97,1.00}
\definecolor {whitesmoke}          {rgb}{0.96,0.96,0.96}
\definecolor {gainsboro}           {rgb}{0.86,0.86,0.86}
\definecolor {floralwhite}         {rgb}{1.00,0.98,0.94}
\definecolor {oldlace}             {rgb}{0.99,0.96,0.90}
\definecolor {linen}               {rgb}{0.98,0.94,0.90}
\definecolor {antiquewhite}        {rgb}{0.98,0.92,0.84}
\definecolor {papayawhip}          {rgb}{1.00,0.94,0.84}
\definecolor {blanchedalmond}      {rgb}{1.00,0.92,0.80}
\definecolor {bisque}              {rgb}{1.00,0.89,0.77}
\definecolor {peachpuff}           {rgb}{1.00,0.85,0.73}
\definecolor {navajowhite}         {rgb}{1.00,0.87,0.68}
\definecolor {moccasin}            {rgb}{1.00,0.89,0.71}
\definecolor {cornsilk}            {rgb}{1.00,0.97,0.86}
\definecolor {ivory}               {rgb}{1.00,1.00,0.94}
\definecolor {lemonchiffon}        {rgb}{1.00,0.98,0.80}
\definecolor {seashell}            {rgb}{1.00,0.96,0.93}
\definecolor {honeydew}            {rgb}{0.94,1.00,0.94}
\definecolor {mintcream}           {rgb}{0.96,1.00,0.98}
\definecolor {azure}               {rgb}{0.94,1.00,1.00}
\definecolor {aliceblue}           {rgb}{0.94,0.97,1.00}
\definecolor {lavender}            {rgb}{0.90,0.90,0.98}
\definecolor {lavenderblush}       {rgb}{1.00,0.94,0.96}
\definecolor {mistyrose}           {rgb}{1.00,0.89,0.88}
\definecolor {white}               {rgb}{1.00,1.00,1.00}
\definecolor {black}               {rgb}{0.00,0.00,0.00}
\definecolor {darkslategray}       {rgb}{0.18,0.31,0.31}
\definecolor {dimgray}             {rgb}{0.41,0.41,0.41}
\definecolor {slategray}           {rgb}{0.44,0.50,0.56}
\definecolor {lightslategray}      {rgb}{0.47,0.53,0.60}
\definecolor {gray}                {rgb}{0.75,0.75,0.75}
\definecolor {lightgrey}           {rgb}{0.83,0.83,0.83}
\definecolor {midnightblue}        {rgb}{0.10,0.10,0.44}
\definecolor {navy}                {rgb}{0.00,0.00,0.50}
\definecolor {cornflowerblue}      {rgb}{0.39,0.58,0.93}
\definecolor {darkslateblue}       {rgb}{0.28,0.24,0.55}
\definecolor {slateblue}           {rgb}{0.42,0.35,0.80}
\definecolor {mediumslateblue}     {rgb}{0.48,0.41,0.93}
\definecolor {lightslateblue}      {rgb}{0.52,0.44,1.00}
\definecolor {mediumblue}          {rgb}{0.00,0.00,0.80}
\definecolor {royalblue}           {rgb}{0.25,0.41,0.88}
\definecolor {blue}                {rgb}{0.00,0.00,1.00}
\definecolor {dodgerblue}          {rgb}{0.12,0.56,1.00}
\definecolor {deepskyblue}         {rgb}{0.00,0.75,1.00}
\definecolor {skyblue}             {rgb}{0.53,0.81,0.92}
\definecolor {lightskyblue}        {rgb}{0.53,0.81,0.98}
\definecolor {steelblue}           {rgb}{0.27,0.51,0.71}
\definecolor {lightsteelblue}      {rgb}{0.69,0.77,0.87}
\definecolor {lightblue}           {rgb}{0.68,0.85,0.90}
\definecolor {powderblue}          {rgb}{0.69,0.88,0.90}
\definecolor {paleturquoise}       {rgb}{0.69,0.93,0.93}
\definecolor {darkturquoise}       {rgb}{0.00,0.81,0.82}
\definecolor {mediumturquoise}     {rgb}{0.28,0.82,0.80}
\definecolor {turquoise}           {rgb}{0.25,0.88,0.82}
\definecolor {cyan}                {rgb}{0.00,1.00,1.00}
\definecolor {lightcyan}           {rgb}{0.88,1.00,1.00}
\definecolor {cadetblue}           {rgb}{0.37,0.62,0.63}
\definecolor {mediumaquamarine}    {rgb}{0.40,0.80,0.67}
\definecolor {aquamarine}          {rgb}{0.50,1.00,0.83}
\definecolor {darkgreen}           {rgb}{0.00,0.39,0.00}
\definecolor {darkolivegreen}      {rgb}{0.33,0.42,0.18}
\definecolor {darkseagreen}        {rgb}{0.56,0.74,0.56}
\definecolor {seagreen}            {rgb}{0.18,0.55,0.34}
\definecolor {mediumseagreen}      {rgb}{0.24,0.70,0.44}
\definecolor {lightseagreen}       {rgb}{0.13,0.70,0.67}
\definecolor {palegreen}           {rgb}{0.60,0.98,0.60}
\definecolor {springgreen}         {rgb}{0.00,1.00,0.50}
\definecolor {lawngreen}           {rgb}{0.49,0.99,0.00}
\definecolor {green}               {rgb}{0.00,1.00,0.00}
\definecolor {chartreuse}          {rgb}{0.50,1.00,0.00}
\definecolor {mediumspringgreen}   {rgb}{0.00,0.98,0.60}
\definecolor {greenyellow}         {rgb}{0.68,1.00,0.18}
\definecolor {limegreen}           {rgb}{0.20,0.80,0.20}
\definecolor {yellowgreen}         {rgb}{0.60,0.80,0.20}
\definecolor {forestgreen}         {rgb}{0.13,0.55,0.13}
\definecolor {olivedrab}           {rgb}{0.42,0.56,0.14}
\definecolor {darkkhaki}           {rgb}{0.74,0.72,0.42}
\definecolor {khaki}               {rgb}{0.94,0.90,0.55}
\definecolor {palegoldenrod}       {rgb}{0.93,0.91,0.67}
\definecolor {lightgoldenrodyellow} {rgb}{0.98,0.98,0.82}
\definecolor {lightyellow}         {rgb}{1.00,1.00,0.88}
\definecolor {yellow}              {rgb}{1.00,1.00,0.00}
\definecolor {gold}                {rgb}{1.00,0.84,0.00}
\definecolor {lightgoldenrod}      {rgb}{0.93,0.87,0.51}
\definecolor {goldenrod}           {rgb}{0.85,0.65,0.13}
\definecolor {darkgoldenrod}       {rgb}{0.72,0.53,0.04}
\definecolor {rosybrown}           {rgb}{0.74,0.56,0.56}
\definecolor {indianred}           {rgb}{0.80,0.36,0.36}
\definecolor {saddlebrown}         {rgb}{0.55,0.27,0.07}
\definecolor {sienna}              {rgb}{0.63,0.32,0.18}
\definecolor {peru}                {rgb}{0.80,0.52,0.25}
\definecolor {burlywood}           {rgb}{0.87,0.72,0.53}
\definecolor {beige}               {rgb}{0.96,0.96,0.86}
\definecolor {wheat}               {rgb}{0.96,0.87,0.70}
\definecolor {sandybrown}          {rgb}{0.96,0.64,0.38}
\definecolor {tan}                 {rgb}{0.82,0.71,0.55}
\definecolor {chocolate}           {rgb}{0.82,0.41,0.12}
\definecolor {firebrick}           {rgb}{0.70,0.13,0.13}
\definecolor {brown}               {rgb}{0.65,0.16,0.16}
\definecolor {darksalmon}          {rgb}{0.91,0.59,0.48}
\definecolor {salmon}              {rgb}{0.98,0.50,0.45}
\definecolor {lightsalmon}         {rgb}{1.00,0.63,0.48}
\definecolor {orange}              {rgb}{1.00,0.65,0.00}
\definecolor {darkorange}          {rgb}{1.00,0.55,0.00}
\definecolor {coral}               {rgb}{1.00,0.50,0.31}
\definecolor {lightcoral}          {rgb}{0.94,0.50,0.50}
\definecolor {tomato}              {rgb}{1.00,0.39,0.28}
\definecolor {orangered}           {rgb}{1.00,0.27,0.00}
\definecolor {red}                 {rgb}{1.00,0.00,0.00}
\definecolor {hotpink}             {rgb}{1.00,0.41,0.71}
\definecolor {deeppink}            {rgb}{1.00,0.08,0.58}
\definecolor {pink}                {rgb}{1.00,0.75,0.80}
\definecolor {lightpink}           {rgb}{1.00,0.71,0.76}
\definecolor {palevioletred}       {rgb}{0.86,0.44,0.58}
\definecolor {maroon}              {rgb}{0.69,0.19,0.38}
\definecolor {mediumvioletred}     {rgb}{0.78,0.08,0.52}
\definecolor {violetred}           {rgb}{0.82,0.13,0.56}
\definecolor {magenta}             {rgb}{1.00,0.00,1.00}
\definecolor {violet}              {rgb}{0.93,0.51,0.93}
\definecolor {plum}                {rgb}{0.87,0.63,0.87}
\definecolor {orchid}              {rgb}{0.85,0.44,0.84}
\definecolor {mediumorchid}        {rgb}{0.73,0.33,0.83}
\definecolor {darkorchid}          {rgb}{0.60,0.20,0.80}
\definecolor {darkviolet}          {rgb}{0.58,0.00,0.83}
\definecolor {blueviolet}          {rgb}{0.54,0.17,0.89}
\definecolor {purple}              {rgb}{0.63,0.13,0.94}
\definecolor {mediumpurple}        {rgb}{0.58,0.44,0.86}
\definecolor {thistle}             {rgb}{0.85,0.75,0.85}
\definecolor {snow2}               {rgb}{0.93,0.91,0.91}
\definecolor {snow3}               {rgb}{0.80,0.79,0.79}
\definecolor {snow4}               {rgb}{0.55,0.54,0.54}
\definecolor {seashell2}           {rgb}{0.93,0.90,0.87}
\definecolor {seashell3}           {rgb}{0.80,0.77,0.75}
\definecolor {seashell4}           {rgb}{0.55,0.53,0.51}
\definecolor {antiquewhite1}       {rgb}{1.00,0.94,0.86}
\definecolor {antiquewhite2}       {rgb}{0.93,0.87,0.80}
\definecolor {antiquewhite3}       {rgb}{0.80,0.75,0.69}
\definecolor {antiquewhite4}       {rgb}{0.55,0.51,0.47}
\definecolor {bisque2}             {rgb}{0.93,0.84,0.72}
\definecolor {bisque3}             {rgb}{0.80,0.72,0.62}
\definecolor {bisque4}             {rgb}{0.55,0.49,0.42}
\definecolor {peachpuff2}          {rgb}{0.93,0.80,0.68}
\definecolor {peachpuff3}          {rgb}{0.80,0.69,0.58}
\definecolor {peachpuff4}          {rgb}{0.55,0.47,0.40}
\definecolor {navajowhite2}        {rgb}{0.93,0.81,0.63}
\definecolor {navajowhite3}        {rgb}{0.80,0.70,0.55}
\definecolor {navajowhite4}        {rgb}{0.55,0.47,0.37}
\definecolor {lemonchiffon2}       {rgb}{0.93,0.91,0.75}
\definecolor {lemonchiffon3}       {rgb}{0.80,0.79,0.65}
\definecolor {lemonchiffon4}       {rgb}{0.55,0.54,0.44}
\definecolor {cornsilk2}           {rgb}{0.93,0.91,0.80}
\definecolor {cornsilk3}           {rgb}{0.80,0.78,0.69}
\definecolor {cornsilk4}           {rgb}{0.55,0.53,0.47}
\definecolor {ivory2}              {rgb}{0.93,0.93,0.88}
\definecolor {ivory3}              {rgb}{0.80,0.80,0.76}
\definecolor {ivory4}              {rgb}{0.55,0.55,0.51}
\definecolor {honeydew2}           {rgb}{0.88,0.93,0.88}
\definecolor {honeydew3}           {rgb}{0.76,0.80,0.76}
\definecolor {honeydew4}           {rgb}{0.51,0.55,0.51}
\definecolor {lavenderblush2}      {rgb}{0.93,0.88,0.90}
\definecolor {lavenderblush3}      {rgb}{0.80,0.76,0.77}
\definecolor {lavenderblush4}      {rgb}{0.55,0.51,0.53}
\definecolor {mistyrose2}          {rgb}{0.93,0.84,0.82}
\definecolor {mistyrose3}          {rgb}{0.80,0.72,0.71}
\definecolor {mistyrose4}          {rgb}{0.55,0.49,0.48}
\definecolor {azure2}              {rgb}{0.88,0.93,0.93}
\definecolor {azure3}              {rgb}{0.76,0.80,0.80}
\definecolor {azure4}              {rgb}{0.51,0.55,0.55}
\definecolor {slateblue1}          {rgb}{0.51,0.44,1.00}
\definecolor {slateblue2}          {rgb}{0.48,0.40,0.93}
\definecolor {slateblue3}          {rgb}{0.41,0.35,0.80}
\definecolor {slateblue4}          {rgb}{0.28,0.24,0.55}
\definecolor {royalblue1}          {rgb}{0.28,0.46,1.00}
\definecolor {royalblue2}          {rgb}{0.26,0.43,0.93}
\definecolor {royalblue3}          {rgb}{0.23,0.37,0.80}
\definecolor {royalblue4}          {rgb}{0.15,0.25,0.55}
\definecolor {blue2}               {rgb}{0.00,0.00,0.93}
\definecolor {blue4}               {rgb}{0.00,0.00,0.55}
\definecolor {dodgerblue2}         {rgb}{0.11,0.53,0.93}
\definecolor {dodgerblue3}         {rgb}{0.09,0.45,0.80}
\definecolor {dodgerblue4}         {rgb}{0.06,0.31,0.55}
\definecolor {steelblue1}          {rgb}{0.39,0.72,1.00}
\definecolor {steelblue2}          {rgb}{0.36,0.67,0.93}
\definecolor {steelblue3}          {rgb}{0.31,0.58,0.80}
\definecolor {steelblue4}          {rgb}{0.21,0.39,0.55}
\definecolor {deepskyblue2}        {rgb}{0.00,0.70,0.93}
\definecolor {deepskyblue3}        {rgb}{0.00,0.60,0.80}
\definecolor {deepskyblue4}        {rgb}{0.00,0.41,0.55}
\definecolor {skyblue1}            {rgb}{0.53,0.81,1.00}
\definecolor {skyblue2}            {rgb}{0.49,0.75,0.93}
\definecolor {skyblue3}            {rgb}{0.42,0.65,0.80}
\definecolor {skyblue4}            {rgb}{0.29,0.44,0.55}
\definecolor {lightskyblue1}       {rgb}{0.69,0.89,1.00}
\definecolor {lightskyblue2}       {rgb}{0.64,0.83,0.93}
\definecolor {lightskyblue3}       {rgb}{0.55,0.71,0.80}
\definecolor {lightskyblue4}       {rgb}{0.38,0.48,0.55}
\definecolor {slategray1}          {rgb}{0.78,0.89,1.00}
\definecolor {slategray2}          {rgb}{0.73,0.83,0.93}
\definecolor {slategray3}          {rgb}{0.62,0.71,0.80}
\definecolor {slategray4}          {rgb}{0.42,0.48,0.55}
\definecolor {lightsteelblue1}     {rgb}{0.79,0.88,1.00}
\definecolor {lightsteelblue2}     {rgb}{0.74,0.82,0.93}
\definecolor {lightsteelblue3}     {rgb}{0.64,0.71,0.80}
\definecolor {lightsteelblue4}     {rgb}{0.43,0.48,0.55}
\definecolor {lightblue1}          {rgb}{0.75,0.94,1.00}
\definecolor {lightblue2}          {rgb}{0.70,0.87,0.93}
\definecolor {lightblue3}          {rgb}{0.60,0.75,0.80}
\definecolor {lightblue4}          {rgb}{0.41,0.51,0.55}
\definecolor {lightcyan2}          {rgb}{0.82,0.93,0.93}
\definecolor {lightcyan3}          {rgb}{0.71,0.80,0.80}
\definecolor {lightcyan4}          {rgb}{0.48,0.55,0.55}
\definecolor {paleturquoise1}      {rgb}{0.73,1.00,1.00}
\definecolor {paleturquoise2}      {rgb}{0.68,0.93,0.93}
\definecolor {paleturquoise3}      {rgb}{0.59,0.80,0.80}
\definecolor {paleturquoise4}      {rgb}{0.40,0.55,0.55}
\definecolor {cadetblue1}          {rgb}{0.60,0.96,1.00}
\definecolor {cadetblue2}          {rgb}{0.56,0.90,0.93}
\definecolor {cadetblue3}          {rgb}{0.48,0.77,0.80}
\definecolor {cadetblue4}          {rgb}{0.33,0.53,0.55}
\definecolor {turquoise1}          {rgb}{0.00,0.96,1.00}
\definecolor {turquoise2}          {rgb}{0.00,0.90,0.93}
\definecolor {turquoise3}          {rgb}{0.00,0.77,0.80}
\definecolor {turquoise4}          {rgb}{0.00,0.53,0.55}
\definecolor {cyan2}               {rgb}{0.00,0.93,0.93}
\definecolor {cyan3}               {rgb}{0.00,0.80,0.80}
\definecolor {cyan4}               {rgb}{0.00,0.55,0.55}
\definecolor {darkslategray1}      {rgb}{0.59,1.00,1.00}
\definecolor {darkslategray2}      {rgb}{0.55,0.93,0.93}
\definecolor {darkslategray3}      {rgb}{0.47,0.80,0.80}
\definecolor {darkslategray4}      {rgb}{0.32,0.55,0.55}
\definecolor {aquamarine2}         {rgb}{0.46,0.93,0.78}
\definecolor {aquamarine4}         {rgb}{0.27,0.55,0.45}
\definecolor {darkseagreen1}       {rgb}{0.76,1.00,0.76}
\definecolor {darkseagreen2}       {rgb}{0.71,0.93,0.71}
\definecolor {darkseagreen3}       {rgb}{0.61,0.80,0.61}
\definecolor {darkseagreen4}       {rgb}{0.41,0.55,0.41}
\definecolor {seagreen1}           {rgb}{0.33,1.00,0.62}
\definecolor {seagreen2}           {rgb}{0.31,0.93,0.58}
\definecolor {seagreen3}           {rgb}{0.26,0.80,0.50}
\definecolor {palegreen1}          {rgb}{0.60,1.00,0.60}
\definecolor {palegreen2}          {rgb}{0.56,0.93,0.56}
\definecolor {palegreen3}          {rgb}{0.49,0.80,0.49}
\definecolor {palegreen4}          {rgb}{0.33,0.55,0.33}
\definecolor {springgreen2}        {rgb}{0.00,0.93,0.46}
\definecolor {springgreen3}        {rgb}{0.00,0.80,0.40}
\definecolor {springgreen4}        {rgb}{0.00,0.55,0.27}
\definecolor {green2}              {rgb}{0.00,0.93,0.00}
\definecolor {green3}              {rgb}{0.00,0.80,0.00}
\definecolor {green4}              {rgb}{0.00,0.55,0.00}
\definecolor {chartreuse2}         {rgb}{0.46,0.93,0.00}
\definecolor {chartreuse3}         {rgb}{0.40,0.80,0.00}
\definecolor {chartreuse4}         {rgb}{0.27,0.55,0.00}
\definecolor {olivedrab1}          {rgb}{0.75,1.00,0.24}
\definecolor {olivedrab2}          {rgb}{0.70,0.93,0.23}
\definecolor {olivedrab4}          {rgb}{0.41,0.55,0.13}
\definecolor {darkolivegreen1}     {rgb}{0.79,1.00,0.44}
\definecolor {darkolivegreen2}     {rgb}{0.74,0.93,0.41}
\definecolor {darkolivegreen3}     {rgb}{0.64,0.80,0.35}
\definecolor {darkolivegreen4}     {rgb}{0.43,0.55,0.24}
\definecolor {khaki1}              {rgb}{1.00,0.96,0.56}
\definecolor {khaki2}              {rgb}{0.93,0.90,0.52}
\definecolor {khaki3}              {rgb}{0.80,0.78,0.45}
\definecolor {khaki4}              {rgb}{0.55,0.53,0.31}
\definecolor {lightgoldenrod1}     {rgb}{1.00,0.93,0.55}
\definecolor {lightgoldenrod2}     {rgb}{0.93,0.86,0.51}
\definecolor {lightgoldenrod3}     {rgb}{0.80,0.75,0.44}
\definecolor {lightgoldenrod4}     {rgb}{0.55,0.51,0.30}
\definecolor {lightyellow2}        {rgb}{0.93,0.93,0.82}
\definecolor {lightyellow3}        {rgb}{0.80,0.80,0.71}
\definecolor {lightyellow4}        {rgb}{0.55,0.55,0.48}
\definecolor {yellow2}             {rgb}{0.93,0.93,0.00}
\definecolor {yellow3}             {rgb}{0.80,0.80,0.00}
\definecolor {yellow4}             {rgb}{0.55,0.55,0.00}
\definecolor {gold2}               {rgb}{0.93,0.79,0.00}
\definecolor {gold3}               {rgb}{0.80,0.68,0.00}
\definecolor {gold4}               {rgb}{0.55,0.46,0.00}
\definecolor {goldenrod1}          {rgb}{1.00,0.76,0.15}
\definecolor {goldenrod2}          {rgb}{0.93,0.71,0.13}
\definecolor {goldenrod3}          {rgb}{0.80,0.61,0.11}
\definecolor {goldenrod4}          {rgb}{0.55,0.41,0.08}
\definecolor {darkgoldenrod1}      {rgb}{1.00,0.73,0.06}
\definecolor {darkgoldenrod2}      {rgb}{0.93,0.68,0.05}
\definecolor {darkgoldenrod3}      {rgb}{0.80,0.58,0.05}
\definecolor {darkgoldenrod4}      {rgb}{0.55,0.40,0.03}
\definecolor {rosybrown1}          {rgb}{1.00,0.76,0.76}
\definecolor {rosybrown2}          {rgb}{0.93,0.71,0.71}
\definecolor {rosybrown3}          {rgb}{0.80,0.61,0.61}
\definecolor {rosybrown4}          {rgb}{0.55,0.41,0.41}
\definecolor {indianred1}          {rgb}{1.00,0.42,0.42}
\definecolor {indianred2}          {rgb}{0.93,0.39,0.39}
\definecolor {indianred3}          {rgb}{0.80,0.33,0.33}
\definecolor {indianred4}          {rgb}{0.55,0.23,0.23}
\definecolor {sienna1}             {rgb}{1.00,0.51,0.28}
\definecolor {sienna2}             {rgb}{0.93,0.47,0.26}
\definecolor {sienna3}             {rgb}{0.80,0.41,0.22}
\definecolor {sienna4}             {rgb}{0.55,0.28,0.15}
\definecolor {burlywood1}          {rgb}{1.00,0.83,0.61}
\definecolor {burlywood2}          {rgb}{0.93,0.77,0.57}
\definecolor {burlywood3}          {rgb}{0.80,0.67,0.49}
\definecolor {burlywood4}          {rgb}{0.55,0.45,0.33}
\definecolor {wheat1}              {rgb}{1.00,0.91,0.73}
\definecolor {wheat2}              {rgb}{0.93,0.85,0.68}
\definecolor {wheat3}              {rgb}{0.80,0.73,0.59}
\definecolor {wheat4}              {rgb}{0.55,0.49,0.40}
\definecolor {tan1}                {rgb}{1.00,0.65,0.31}
\definecolor {tan2}                {rgb}{0.93,0.60,0.29}
\definecolor {tan4}                {rgb}{0.55,0.35,0.17}
\definecolor {chocolate1}          {rgb}{1.00,0.50,0.14}
\definecolor {chocolate2}          {rgb}{0.93,0.46,0.13}
\definecolor {chocolate3}          {rgb}{0.80,0.40,0.11}
\definecolor {firebrick1}          {rgb}{1.00,0.19,0.19}
\definecolor {firebrick2}          {rgb}{0.93,0.17,0.17}
\definecolor {firebrick3}          {rgb}{0.80,0.15,0.15}
\definecolor {firebrick4}          {rgb}{0.55,0.10,0.10}
\definecolor {brown1}              {rgb}{1.00,0.25,0.25}
\definecolor {brown2}              {rgb}{0.93,0.23,0.23}
\definecolor {brown3}              {rgb}{0.80,0.20,0.20}
\definecolor {brown4}              {rgb}{0.55,0.14,0.14}
\definecolor {salmon1}             {rgb}{1.00,0.55,0.41}
\definecolor {salmon2}             {rgb}{0.93,0.51,0.38}
\definecolor {salmon3}             {rgb}{0.80,0.44,0.33}
\definecolor {salmon4}             {rgb}{0.55,0.30,0.22}
\definecolor {lightsalmon2}        {rgb}{0.93,0.58,0.45}
\definecolor {lightsalmon3}        {rgb}{0.80,0.51,0.38}
\definecolor {lightsalmon4}        {rgb}{0.55,0.34,0.26}
\definecolor {orange2}             {rgb}{0.93,0.60,0.00}
\definecolor {orange3}             {rgb}{0.80,0.52,0.00}
\definecolor {orange4}             {rgb}{0.55,0.35,0.00}
\definecolor {darkorange1}         {rgb}{1.00,0.50,0.00}
\definecolor {darkorange2}         {rgb}{0.93,0.46,0.00}
\definecolor {darkorange3}         {rgb}{0.80,0.40,0.00}
\definecolor {darkorange4}         {rgb}{0.55,0.27,0.00}
\definecolor {coral1}              {rgb}{1.00,0.45,0.34}
\definecolor {coral2}              {rgb}{0.93,0.42,0.31}
\definecolor {coral3}              {rgb}{0.80,0.36,0.27}
\definecolor {coral4}              {rgb}{0.55,0.24,0.18}
\definecolor {tomato2}             {rgb}{0.93,0.36,0.26}
\definecolor {tomato3}             {rgb}{0.80,0.31,0.22}
\definecolor {tomato4}             {rgb}{0.55,0.21,0.15}
\definecolor {orangered2}          {rgb}{0.93,0.25,0.00}
\definecolor {orangered3}          {rgb}{0.80,0.22,0.00}
\definecolor {orangered4}          {rgb}{0.55,0.15,0.00}
\definecolor {red2}                {rgb}{0.93,0.00,0.00}
\definecolor {red3}                {rgb}{0.80,0.00,0.00}
\definecolor {red4}                {rgb}{0.55,0.00,0.00}
\definecolor {deeppink2}           {rgb}{0.93,0.07,0.54}
\definecolor {deeppink3}           {rgb}{0.80,0.06,0.46}
\definecolor {deeppink4}           {rgb}{0.55,0.04,0.31}
\definecolor {hotpink1}            {rgb}{1.00,0.43,0.71}
\definecolor {hotpink2}            {rgb}{0.93,0.42,0.65}
\definecolor {hotpink3}            {rgb}{0.80,0.38,0.56}
\definecolor {hotpink4}            {rgb}{0.55,0.23,0.38}
\definecolor {pink1}               {rgb}{1.00,0.71,0.77}
\definecolor {pink2}               {rgb}{0.93,0.66,0.72}
\definecolor {pink3}               {rgb}{0.80,0.57,0.62}
\definecolor {pink4}               {rgb}{0.55,0.39,0.42}
\definecolor {lightpink1}          {rgb}{1.00,0.68,0.73}
\definecolor {lightpink2}          {rgb}{0.93,0.64,0.68}
\definecolor {lightpink3}          {rgb}{0.80,0.55,0.58}
\definecolor {lightpink4}          {rgb}{0.55,0.37,0.40}
\definecolor {palevioletred1}      {rgb}{1.00,0.51,0.67}
\definecolor {palevioletred2}      {rgb}{0.93,0.47,0.62}
\definecolor {palevioletred3}      {rgb}{0.80,0.41,0.54}
\definecolor {palevioletred4}      {rgb}{0.55,0.28,0.36}
\definecolor {maroon1}             {rgb}{1.00,0.20,0.70}
\definecolor {maroon2}             {rgb}{0.93,0.19,0.65}
\definecolor {maroon3}             {rgb}{0.80,0.16,0.56}
\definecolor {maroon4}             {rgb}{0.55,0.11,0.38}
\definecolor {violetred1}          {rgb}{1.00,0.24,0.59}
\definecolor {violetred2}          {rgb}{0.93,0.23,0.55}
\definecolor {violetred3}          {rgb}{0.80,0.20,0.47}
\definecolor {violetred4}          {rgb}{0.55,0.13,0.32}
\definecolor {magenta2}            {rgb}{0.93,0.00,0.93}
\definecolor {magenta3}            {rgb}{0.80,0.00,0.80}
\definecolor {magenta4}            {rgb}{0.55,0.00,0.55}
\definecolor {orchid1}             {rgb}{1.00,0.51,0.98}
\definecolor {orchid2}             {rgb}{0.93,0.48,0.91}
\definecolor {orchid3}             {rgb}{0.80,0.41,0.79}
\definecolor {orchid4}             {rgb}{0.55,0.28,0.54}
\definecolor {plum1}               {rgb}{1.00,0.73,1.00}
\definecolor {plum2}               {rgb}{0.93,0.68,0.93}
\definecolor {plum3}               {rgb}{0.80,0.59,0.80}
\definecolor {plum4}               {rgb}{0.55,0.40,0.55}
\definecolor {mediumorchid1}       {rgb}{0.88,0.40,1.00}
\definecolor {mediumorchid2}       {rgb}{0.82,0.37,0.93}
\definecolor {mediumorchid3}       {rgb}{0.71,0.32,0.80}
\definecolor {mediumorchid4}       {rgb}{0.48,0.22,0.55}
\definecolor {darkorchid1}         {rgb}{0.75,0.24,1.00}
\definecolor {darkorchid2}         {rgb}{0.70,0.23,0.93}
\definecolor {darkorchid3}         {rgb}{0.60,0.20,0.80}
\definecolor {darkorchid4}         {rgb}{0.41,0.13,0.55}
\definecolor {purple1}             {rgb}{0.61,0.19,1.00}
\definecolor {purple2}             {rgb}{0.57,0.17,0.93}
\definecolor {purple3}             {rgb}{0.49,0.15,0.80}
\definecolor {purple4}             {rgb}{0.33,0.10,0.55}
\definecolor {mediumpurple1}       {rgb}{0.67,0.51,1.00}
\definecolor {mediumpurple2}       {rgb}{0.62,0.47,0.93}
\definecolor {mediumpurple3}       {rgb}{0.54,0.41,0.80}
\definecolor {mediumpurple4}       {rgb}{0.36,0.28,0.55}
\definecolor {thistle1}            {rgb}{1.00,0.88,1.00}
\definecolor {thistle2}            {rgb}{0.93,0.82,0.93}
\definecolor {thistle3}            {rgb}{0.80,0.71,0.80}
\definecolor {thistle4}            {rgb}{0.55,0.48,0.55}
\definecolor {gray1}               {rgb}{0.01,0.01,0.01}
\definecolor {gray2}               {rgb}{0.02,0.02,0.02}
\definecolor {gray3}               {rgb}{0.03,0.03,0.03}
\definecolor {gray4}               {rgb}{0.04,0.04,0.04}
\definecolor {gray5}               {rgb}{0.05,0.05,0.05}
\definecolor {gray6}               {rgb}{0.06,0.06,0.06}
\definecolor {gray7}               {rgb}{0.07,0.07,0.07}
\definecolor {gray8}               {rgb}{0.08,0.08,0.08}
\definecolor {gray9}               {rgb}{0.09,0.09,0.09}
\definecolor {gray10}              {rgb}{0.10,0.10,0.10}
\definecolor {gray11}              {rgb}{0.11,0.11,0.11}
\definecolor {gray12}              {rgb}{0.12,0.12,0.12}
\definecolor {gray13}              {rgb}{0.13,0.13,0.13}
\definecolor {gray14}              {rgb}{0.14,0.14,0.14}
\definecolor {gray15}              {rgb}{0.15,0.15,0.15}
\definecolor {gray16}              {rgb}{0.16,0.16,0.16}
\definecolor {gray17}              {rgb}{0.17,0.17,0.17}
\definecolor {gray18}              {rgb}{0.18,0.18,0.18}
\definecolor {gray19}              {rgb}{0.19,0.19,0.19}
\definecolor {gray20}              {rgb}{0.20,0.20,0.20}
\definecolor {gray21}              {rgb}{0.21,0.21,0.21}
\definecolor {gray22}              {rgb}{0.22,0.22,0.22}
\definecolor {gray23}              {rgb}{0.23,0.23,0.23}
\definecolor {gray24}              {rgb}{0.24,0.24,0.24}
\definecolor {gray25}              {rgb}{0.25,0.25,0.25}
\definecolor {gray26}              {rgb}{0.26,0.26,0.26}
\definecolor {gray27}              {rgb}{0.27,0.27,0.27}
\definecolor {gray28}              {rgb}{0.28,0.28,0.28}
\definecolor {gray29}              {rgb}{0.29,0.29,0.29}
\definecolor {gray30}              {rgb}{0.30,0.30,0.30}
\definecolor {gray31}              {rgb}{0.31,0.31,0.31}
\definecolor {gray32}              {rgb}{0.32,0.32,0.32}
\definecolor {gray33}              {rgb}{0.33,0.33,0.33}
\definecolor {gray34}              {rgb}{0.34,0.34,0.34}
\definecolor {gray35}              {rgb}{0.35,0.35,0.35}
\definecolor {gray36}              {rgb}{0.36,0.36,0.36}
\definecolor {gray37}              {rgb}{0.37,0.37,0.37}
\definecolor {gray38}              {rgb}{0.38,0.38,0.38}
\definecolor {gray39}              {rgb}{0.39,0.39,0.39}
\definecolor {gray40}              {rgb}{0.40,0.40,0.40}
\definecolor {gray42}              {rgb}{0.42,0.42,0.42}
\definecolor {gray43}              {rgb}{0.43,0.43,0.43}
\definecolor {gray44}              {rgb}{0.44,0.44,0.44}
\definecolor {gray45}              {rgb}{0.45,0.45,0.45}
\definecolor {gray46}              {rgb}{0.46,0.46,0.46}
\definecolor {gray47}              {rgb}{0.47,0.47,0.47}
\definecolor {gray48}              {rgb}{0.48,0.48,0.48}
\definecolor {gray49}              {rgb}{0.49,0.49,0.49}
\definecolor {gray50}              {rgb}{0.50,0.50,0.50}
\definecolor {gray51}              {rgb}{0.51,0.51,0.51}
\definecolor {gray52}              {rgb}{0.52,0.52,0.52}
\definecolor {gray53}              {rgb}{0.53,0.53,0.53}
\definecolor {gray54}              {rgb}{0.54,0.54,0.54}
\definecolor {gray55}              {rgb}{0.55,0.55,0.55}
\definecolor {gray56}              {rgb}{0.56,0.56,0.56}
\definecolor {gray57}              {rgb}{0.57,0.57,0.57}
\definecolor {gray58}              {rgb}{0.58,0.58,0.58}
\definecolor {gray59}              {rgb}{0.59,0.59,0.59}
\definecolor {gray60}              {rgb}{0.60,0.60,0.60}
\definecolor {gray61}              {rgb}{0.61,0.61,0.61}
\definecolor {gray62}              {rgb}{0.62,0.62,0.62}
\definecolor {gray63}              {rgb}{0.63,0.63,0.63}
\definecolor {gray64}              {rgb}{0.64,0.64,0.64}
\definecolor {gray65}              {rgb}{0.65,0.65,0.65}
\definecolor {gray66}              {rgb}{0.66,0.66,0.66}
\definecolor {gray67}              {rgb}{0.67,0.67,0.67}
\definecolor {gray68}              {rgb}{0.68,0.68,0.68}
\definecolor {gray69}              {rgb}{0.69,0.69,0.69}
\definecolor {gray70}              {rgb}{0.70,0.70,0.70}
\definecolor {gray71}              {rgb}{0.71,0.71,0.71}
\definecolor {gray72}              {rgb}{0.72,0.72,0.72}
\definecolor {gray73}              {rgb}{0.73,0.73,0.73}
\definecolor {gray74}              {rgb}{0.74,0.74,0.74}
\definecolor {gray75}              {rgb}{0.75,0.75,0.75}
\definecolor {gray76}              {rgb}{0.76,0.76,0.76}
\definecolor {gray77}              {rgb}{0.77,0.77,0.77}
\definecolor {gray78}              {rgb}{0.78,0.78,0.78}
\definecolor {gray79}              {rgb}{0.79,0.79,0.79}
\definecolor {gray80}              {rgb}{0.80,0.80,0.80}
\definecolor {gray81}              {rgb}{0.81,0.81,0.81}
\definecolor {gray82}              {rgb}{0.82,0.82,0.82}
\definecolor {gray83}              {rgb}{0.83,0.83,0.83}
\definecolor {gray84}              {rgb}{0.84,0.84,0.84}
\definecolor {gray85}              {rgb}{0.85,0.85,0.85}
\definecolor {gray86}              {rgb}{0.86,0.86,0.86}
\definecolor {gray87}              {rgb}{0.87,0.87,0.87}
\definecolor {gray88}              {rgb}{0.88,0.88,0.88}
\definecolor {gray89}              {rgb}{0.89,0.89,0.89}
\definecolor {gray90}              {rgb}{0.90,0.90,0.90}
\definecolor {gray91}              {rgb}{0.91,0.91,0.91}
\definecolor {gray92}              {rgb}{0.92,0.92,0.92}
\definecolor {gray93}              {rgb}{0.93,0.93,0.93}
\definecolor {gray94}              {rgb}{0.94,0.94,0.94}
\definecolor {gray95}              {rgb}{0.95,0.95,0.95}
\definecolor {gray97}              {rgb}{0.97,0.97,0.97}
\definecolor {gray98}              {rgb}{0.98,0.98,0.98}
\definecolor {gray99}              {rgb}{0.99,0.99,0.99}
\definecolor {darkgrey}            {rgb}{0.66,0.66,0.66}

\newcommand{\TODO}[1]{{}}

\newcommand{\ignore}[1]{}
\newcommand{\RSCHANGE}[1]{\textcolor{darkgreen}{{#1}}}
\newcommand{\RSTODO}[1]{{\bf \textcolor{darkgreen}{{\fbox{RS TODO:} #1}}}}

\newcommand{\marg}[1]{\marginpar{\ \\{\em #1\/}}}
\renewcommand{\marg}[1]{\marginpar{\ \\\textcolor{blue}{\ {\sf #1\/}}}}

\newenvironment{rschange}{\color{blue}}{\normalcolor}

 \newcommand{\ignoreinshort}[1]{}
 \newcommand{\ignoreinlong}[1]{{#1}}

\def\makenewenumerate#1#2{%
\newcounter{cnt#1}
\newenvironment{#1}%
{\begin{list}{\makebox[0pt][r]{#2}}%
{\setlength{\itemsep}{0pt}%
 \setlength{\parsep}{.2em}%
 \setlength{\leftmargin}{1.5em}%
 \setlength{\labelwidth}{.4em}%
 \usecounter{cnt#1}}}
{\end{list}}}

\makenewenumerate{myenumerate}{\arabic{cntmyenumerate}.}

\makenewenumerate{renumerate}{\rm(\roman{cntrenumerate})}
\makenewenumerate{renumerateprime}{\rm(\roman{cntrenumerateprime}$'$)}
\makenewenumerate{renumeratesecond}{\rm(\roman{cntrenumeratesecond}$''$)}

\makenewenumerate{aenumerate}{({\it\alph{cntaenumerate}})}
\makenewenumerate{aenumerateprime}{({\it\alph{cntaenumerateprime}$'$})}
\makenewenumerate{aenumeratesecond}{({\it\alph{cntaenumeratesecond}$''$})}

\def\newplaintheorem#1#2{%
\newtheorem{#1plain}{#2}[section]%
\newenvironment{#1}{\begin{#1plain}\rm }{\end{#1plain}}}

\newplaintheorem{notation}{Notation}
\newplaintheorem{cexample}{Counter-Example}

\newcommand{\sref}[1]{\S{}\ref{#1}}

\newcommand{\pair}[2]{\ensuremath{\langle{#1},{#2}\rangle}\xspace}

\newcommand{\tuple}[1]{\ensuremath{\langle{#1}\rangle}\xspace}
\newcommand{\set}[1]{\ensuremath{\{{#1}\}}\xspace}
\newcommand{\imp}{\ensuremath{\rightarrow}\xspace}

\renewcommand{\iff}{\ensuremath{\leftrightarrow}\xspace}
\newcommand{\defas}{\ensuremath{\stackrel{\text{\tiny def}}{=}}\xspace}
\newcommand{\thus}{\ensuremath{\Longrightarrow}\xspace}

\newcommand\calb{\ensuremath{\mathcal{B}}\xspace}
\newcommand\calc{\ensuremath{\mathcal{C}}\xspace}

\newcommand\cale{\ensuremath{\mathcal{E}}\xspace}

\newcommand\cali{\ensuremath{\mathcal{I}}\xspace}

\newcommand\calm{\ensuremath{\mathcal{M}}\xspace}

\newcommand\calp{\ensuremath{\mathcal{P}}\xspace}

\newcommand\calx{\ensuremath{\mathcal{X}}\xspace}

\renewcommand{\todo}[1] {\noindent \textcolor{red}{\fbox{{\bf TO DO:}} #1 }}

\renewenvironment{rschange}{\color{blue}}{\normalcolor}
\renewcommand{\RSCHANGE}[1]{\textcolor{blue}{{#1}}}

\newcommand{\onlyinsilviathesis}[1]{}

\renewcommand{\ignoreinshort}[1]{%
#1%
}  
\renewcommand{\ignoreinlong}[1]{}

\renewcommand{\marg}[1]{}

\newcommand{\pr}[1]{\ensuremath{p_{#1}}\xspace}

\newcommand{\omtlcft}{\ensuremath{\text{OMT-LCF}(\T)}\xspace}

\newcommand{\lb}{\ensuremath{\mathsf{lb}}\xspace}
\renewcommand{\ub}{\ensuremath{\mathsf{ub}}\xspace}
\newcommand{\omtplus}[1]{\ensuremath{\text{OMT}({#1})}\xspace}
\newcommand{\omlarat}{\ensuremath{\text{OMT}(\larat)}\xspace}
\newcommand{\omlaratplus}{\ensuremath{\text{OMT}(\larat\cup\T)}\xspace}
\newcommand{\varphicost}{\ensuremath{\varphi_C}\xspace}
\newcommand{\pivot}{\ensuremath{\mathsf{pivot}}\xspace}
\newcommand{\status}{\ensuremath{\mathsf{status}}\xspace}
\renewcommand{\status}{\ensuremath{\mathsf{res}}\xspace}

\newcommand{\smtlaratplus}{\smttt{\larat\cup\T}\xspace}
\newcommand{\laratplus}{\ensuremath{\larat\cup\T}\xspace}

\renewcommand{\cost}{\ensuremath{{\sf cost}}\xspace}

\newcommand{\currlb}{\ensuremath{\mathsf{l}}\xspace}
\newcommand{\currub}{\ensuremath{\mathsf{u}}\xspace}
\newcommand{\range}{\ensuremath{[\lb,\ub[}\xspace}
\newcommand{\currrange}{\ensuremath{[\currlb,\currub[}\xspace}
\newcommand{\lpivotrange}{\ensuremath{[\currlb,\pivot[}\xspace}
\newcommand{\rpivotrange}{\ensuremath{[\pivot,\currub[}\xspace}
\newcommand{\computepivot}{\ensuremath{{\sf ComputePivot}}\xspace}
\newcommand{\dopivoting}{\ensuremath{{\sf BinSearchMode()}}\xspace}

\newcommand{\ubliti}[1]{\ensuremath{(\cost < #1)}\xspace}
\newcommand{\lbliti}[1]{\ensuremath{\neg(\cost < #1)}\xspace}
\newcommand{\pivotatom}{\ensuremath{\mathsf{PIV}}\xspace}
\newcommand{\currlblit}{\ensuremath{\neg(\cost < \currlb)}\xspace}
\newcommand{\currublit}{\ensuremath{(\cost < \currub)}\xspace}
\newcommand{\currubliti}[1]{\ensuremath{(\cost < \currub_{#1})}\xspace}

\newcommand{\minvalue}{\ensuremath{\mathsf{min}}\xspace}
\newcommand{\maxvalue}{\ensuremath{\mathsf{max}}\xspace}
\newcommand{\mvalue}{\ensuremath{\mathsf{m}}\xspace}
\renewcommand{\mincost}{\ensuremath{\mathsf{min}_\cost}\xspace}

\newcommand{\minimize}{\ensuremath{\mathsf{Minimize}}\xspace}

\newcommand{\incrementalsmt}{\ensuremath{\mathsf{SMT.IncrementalSolve}}\xspace}
\newcommand{\smtcoreextract}{\ensuremath{\mathsf{SMT.ExtractUnsatCore}}\xspace}

\renewcommand{\Tsolver}{\T-\ensuremath{\mathsf{Solver}}\xspace}
\renewcommand{\Tsolvers}{\T-\ensuremath{\mathsf{Solvers}}\xspace}
\newcommand{\laratsolver}{\larat-\ensuremath{\mathsf{Solver}}\xspace}

\newcommand{\vplus}[1]{\ensuremath{\pair{#1}{#1_{\delta}}}}

\newcommand{\optimathsat}{\optmathsat}

\newcommand{\ar}[2]{\ensuremath{\underline{{\bf #1}}^{#2}}}
\newcommand{\mat}[2]{\ensuremath{\underline{{\bf #1}}^{#2}}}
\renewcommand{\ar}[2]{\ensuremath{{{\bf #1}}^{#2}}}
\renewcommand{\mat}[2]{\ensuremath{{{\bf #1}}^{#2}}}

\newcommand{\ari}[3]{\ensuremath{\underline{{\bf #1}}^{#2}_{#3}}}
\newcommand{\mati}[3]{\ensuremath{\underline{{\bf #1}}^{#2}_{#3}}}
\renewcommand{\ari}[3]{\ensuremath{{{\bf #1}}^{#2}_{#3}}}
\renewcommand{\mati}[3]{\ensuremath{{{\bf #1}}^{#2}_{#3}}}

\newcommand{\enc}[1]{\ensuremath{\left[\left[{#1}\right]\right]}}
\renewcommand{\enc}[1]{\ensuremath{[[{#1}]]}}

\newcommand{\cplex}{\textsc{Cplex}\xspace}

\newcommand{\trueval}{{\ensuremath{\mathsf{true}}}}
\newcommand{\falseval}{{\ensuremath{\mathsf{false}}}}

\newcommand{\eqij}{\ensuremath{(x_i=x_j)}\xspace}
\newcommand{\neqij}{\ensuremath{\neg(x_i=x_j)}\xspace}
\newcommand{\dij}{\ensuremath{(x_i<x_j)}\xspace}
\newcommand{\dji}{\ensuremath{(x_i>x_j)}\xspace}
\newcommand{\mularat}{\ensuremath{\mu_{\larat}}\xspace}
\newcommand{\mubool}{\ensuremath{\mu_{\mathbb{B}}}\xspace}

\newcommand{\mue}{\ensuremath{\mu_{e}}\xspace}
\newcommand{\mud}{\ensuremath{\mu_{d}}\xspace}
\newcommand{\mui}{\ensuremath{\mu_{i}}\xspace}
\newcommand{\mued}{\ensuremath{\mu_{ed}}\xspace}
\newcommand{\muei}{\ensuremath{\mu_{ei}}\xspace}
\newcommand{\mueid}{\ensuremath{\mu_{eid}}\xspace}

\newcommand{\murelaxed}{\ensuremath{\mu_{rel}}\xspace}

\newcommand{\IE}[1]{\ensuremath{\cali\cale{(#1)}}\xspace}

\newcommand{\Tk}{\ensuremath{\T_k}\xspace}

\newcommand{\edmuext}{{\ensuremath{\cale\calx_{ed}(\mu)}\xspace}}
\newcommand{\edimuext}{{\ensuremath{\cale\calx_{edi}(\mu)}\xspace}}

\renewcommand{\algorithmiccomment}[1]{\ \ \ \ \ // #1}

\renewcommand{\RSCHANGE}[1]{#1}
\newcommand{\RSCHANGEONE}[1]{\rone{}~\textcolor{blue}{{#1}}}
\newcommand{\RSCHANGETWO}[1]{\rtwo{}~\textcolor{blue}{{#1}}}
\newcommand{\RSCHANGETHREE}[1]{\rthree{}~\textcolor{blue}{{#1}}}
\newcommand{\RSCHANGEONETHREE}[1]{\ronethree{}~\textcolor{blue}{{#1}}}
\newcommand{\RSCHANGEONETWO}[1]{\ronetwo{}~\textcolor{blue}{{#1}}}

\newcommand{\RSCHANGEFOUR}[1]{\rfour{}~\textcolor{blue}{{#1}}}

\renewcommand{\RSCHANGEONE}[1]{#1}
\renewcommand{\RSCHANGETWO}[1]{#1}
\renewcommand{\RSCHANGETHREE}[1]{#1}
\renewcommand{\RSCHANGEONETHREE}[1]{#1}
\renewcommand{\RSCHANGEONETWO}[1]{#1}

\renewcommand{\RSCHANGEFOUR}[1]{#1}

\renewenvironment{rschange}{}{\normalcolor}

\newenvironment{stchange}{\color{darkviolet}}{\normalcolor}

\newcommand{\best}[1]{\textcolor{red}{\bf {#1}}}

\newcommand{\optmathsat}{\textsc{OptiMathSAT}\xspace}
\newcommand{\optmathsatv}[2]{\textsc{OptiMathSAT}{#1}{#2}\xspace}
  
\newcommand{\pboptmathsat}{\textsc{PB-}\mathsat}
\newcommand{\mathsatv}[2]{\textsc{MathSAT}{#1}{#2}\xspace}
\newcommand{\pboptmathsatv}[2]{\textsc{PB-}\mathsatv{#1}{#2}\xspace}

\newcommand{\renc}{\textsc{lgdp2smt}\xspace}

\newcommand{\encone}{\textsc{smt2lgdp$_1$}\xspace}
\newcommand{\enctwo}{\textsc{smt2lgdp$_2$}\xspace}

\newcommand{\dirgenerated}{``directly generated''\xspace}
\newcommand{\encoded}{``encoded''\xspace}
\newcommand{\oneencoded}{``\encone encoded''\xspace}
\newcommand{\twoencoded}{``\enctwo encoded''\xspace}

\newcommand{\bin}{\textsc{-BIN}\xspace}
\newcommand{\lin}{\textsc{-LIN}\xspace}
\newcommand{\ada}{\textsc{-ADA}\xspace}
\newcommand{\of}{\textsc{-OF}\xspace}
\newcommand{\inl}{\textsc{-IN}\xspace}

\newcommand{\binof}{\textsc{-BIN-OF}\xspace}
\newcommand{\linof}{\textsc{-LIN-OF}\xspace}

\newcommand{\binin}{\textsc{-BIN-IN}\xspace}
\newcommand{\linin}{\textsc{-LIN-IN}\xspace}
\newcommand{\adain}{\textsc{-ADA-IN}\xspace}

\newcommand{\bininm}{\textsc{-BIN-IN-NMIN}\xspace}
\newcommand{\lininm}{\textsc{-LIN-IN-NMIN}\xspace}
\newcommand{\adainm}{\textsc{-ADA-IN-NMIN}\xspace}

\newcommand{\optlinin}[1]{\optmathsatv{#1}{\linin}}
\newcommand{\optbinin}[1]{\optmathsatv{#1}{\binin}}
\newcommand{\optadain}[1]{\optmathsatv{#1}{\adain}}

\newcommand{\optlininm}[1]{\optmathsatv{#1}{\lininm}}
\newcommand{\optbininm}[1]{\optmathsatv{#1}{\bininm}}
\newcommand{\optadainm}[1]{\optmathsatv{#1}{\adainm}}

\newcommand{\optpblin}[1]{\pboptmathsatv{#1}{\lin}}
\newcommand{\optpbbin}[1]{\pboptmathsatv{#1}{\bin}}

\newcommand{\shortoptlinof}[1]{\textsc{OM}#1\linof\xspace}
\newcommand{\shortoptbinof}[1]{\textsc{OM}#1\binof\xspace}

\newcommand{\shortoptlinin}[1]{\textsc{OM}#1\linin\xspace}
\newcommand{\shortoptbinin}[1]{\textsc{OM}#1\binin\xspace}
\newcommand{\shortoptadain}[1]{\textsc{OM}#1\adain\xspace}

\newcommand{\gams}{\textsc{GAMS}\xspace}
\newcommand{\logmip}{\textsc{LogMIP}\xspace}

\newcommand{\jams}{\textsc{JAMS}\xspace}
\newcommand{\emp}{\textsc{EMP}\xspace}

\newcommand{\logmipG}{\textsc{LogMIP+CPLEX}\xspace}

\newcommand{\logmipCH}{\textsc{LogMIP(CH)+CPLEX}\xspace}
\newcommand{\logmipBM}{\textsc{LogMIP(BM)+CPLEX}\xspace}

\newcommand{\jamsCH}{\textsc{JAMS(CH)+CPLEX}\xspace}
\newcommand{\jamsBM}{\textsc{JAMS(BM)+CPLEX}\xspace}
\newcommand{\jamsBMmult}{\textsc{JAMS(BM)+CPLEX-mc}\xspace}

\newcommand{\shortlogmipCH}{\textsc{LogMIP(CH)}\xspace}
\newcommand{\shortlogmipBM}{\textsc{LogMIP(BM)}\xspace}

\newcommand{\shortjamsCH}{\textsc{JAMS(CH)}\xspace}
\newcommand{\shortjamsBM}{\textsc{JAMS(BM)}\xspace}
\newcommand{\shortjamsBMmult}{\textsc{JAMS(BM)-mc}\xspace}

\renewcommand{\jamsBMmult}{\textsc{JAMS(BM)+CPLEX-4cores}\xspace}

\renewcommand{\shortjamsBMmult}{\textsc{JAMS(BM)-4cores}\xspace}

\newcommand{\bmc}{\textsc{BMC}\xspace}
\newcommand{\kind}{\textsc{K-ind}\xspace}

\newcommand{\Deltaublin}{\ensuremath{\Delta \ub_{lin}}\xspace}
\newcommand{\Deltaubbin}{\ensuremath{\Delta \ub_{bin}}\xspace}

\newcommand{\Deltanconflin}{\ensuremath{\Delta \# {\sf conf}_{lin}}\xspace}
\newcommand{\Deltanconfbin}{\ensuremath{\Delta \# {\sf conf}_{bin}}\xspace}

\newcommand{\mytitle}{Optimization Modulo Theories with Linear Rational Costs}

\title{%
\mytitle%
}

\author{
ROBERTO SEBASTIANI and SILVIA TOMASI
\affil{DISI, University of Trento, Italy}
}

\begin{abstract}

In the contexts of automated reasoning (AR) and formal verification (FV), 
important {\em decision} problems are effectively encoded into 
Satisfiability Modulo Theories (SMT).
In the last decade efficient SMT
solvers have been developed
for several 
theories of
practical interest (e.g., linear arithmetic, arrays, bit-vectors). 
Surprisingly,  little work has been done to extend SMT
to deal with {\em optimization} problems; in particular, we are 
not aware of any previous work on SMT solvers able to produce solutions 
which minimize cost functions over {\em arithmetical} variables. 
This is unfortunate, since some problems of interest require this
functionality. 

In the work described in this paper we start filling this gap. We
present and discuss two general 
procedures for leveraging SMT to handle the minimization of linear
rational 
cost functions, combining SMT  with standard minimization 
techniques. 
We have implemented the procedures within the MathSAT SMT solver.
Due to the absence of competitors in the AR, FV and SMT domains, 
we have experimentally
evaluated our implementation  against state-of-the-art  tools for the
domain of {\em linear
generalized disjunctive programming (LGDP)}, which is closest in
spirit to our domain, on sets of problems 
 which have been previously proposed 
as benchmarks for the latter tools. 
The results show that our tool is very competitive with, and 
 often outperforms, these tools on these problems,  
clearly demonstrating the potential of the
approach.

\ignore{

}

\end{abstract}


\begin{document}

\ignore{
\pagestyle{plain}
\pagenumbering{roman}

\noindent
{\large {\bf \mytitle}}

 \setcounter{tocdepth}{4}
 \tableofcontents 
\ \\
\begin{center}
 Last update: \today, \currenttime
\end{center}
\begin{rschange}
\newpage
\section*{TODO \& PLAN:}


\begin{itemize}
\item Cambiare titolo?
\item citare SAT13 paper
\item citare paper mcmillan
\item citatre paper ILP modulo theories
\item citare paper \mathsatfive TACAS13
\item \mathsat \thus \mathsatfive
\item NOTA: occhio al solito ``variable'' vs. ``constant''
\item \sref{sec:optsmt}: Add a formal part
  \begin{itemize}
  \item nozione di polarita' della variabile cost
  \item spostare qui parte dei concetti in \sref{sec:algorithms_omlaratplus}
  \item beware: total vs partial assignments
  \end{itemize}
\item \sref{sec:algorithms}: Add examples!
  \begin{itemize}
  \item 
  \end{itemize}
\item \sref{sec:algorithms}: failed attempts of optimizations
  \begin{itemize}
  \item 
  \end{itemize}
\item \sref{sec:expeval}: Add analysis section
  \begin{itemize}
  \item solving, minimization \& certify times
  \item 
  \end{itemize}
\item  \sref{sec:expeval}: add comparison on PB versus PB-MATHSAT
  \begin{itemize}
  \item 
  \end{itemize}
\item  \sref{sec:expeval}: add comparison wrt naive binary search?
  \begin{itemize}
  \item 
  \end{itemize}
\item 
\end{itemize}

\end{rschange}

\newpage

\pagestyle{plain}
\pagenumbering{arabic}
}


\makeatletter
\newtheorem{rep@theorem}{\rep@title}
\newcommand{\newreptheorem}[2]{%
\newenvironment{rep#1}[1]{%
 \def\rep@title{#2 \ref{##1}}%
 \begin{rep@theorem}}%
 {\end{rep@theorem}}}
\makeatother

\newreptheorem{theorem}{Theorem}

\begin{bottomstuff}
Authors' addresses: 
Roberto Sebastiani ({\tt roberto.sebastiani@unitn.it}),
Silvia Tomasi ({\tt silvia.tomasi@disi.unitn.it}),
\textsc{DISI}, Universit\`a di Trento, via Sommarive 9, I-38123 Povo, Trento, Italy.
Roberto Sebastiani is supported by Semiconductor Research
Corporation (SRC) under
GRC Research Project 2012-TJ-2266 WOLF.
\end{bottomstuff}

\maketitle


\section{Introduction}
\label{sec:intro}

\marg{SMT generalities}
In the contexts of automated reasoning (AR) and formal verification (FV), 
important {\em decision} problems are effectively encoded into and solved as
Satisfiability Modulo Theories (SMT) problems.
In the last decade efficient SMT
solvers have been developed, that combine the power of modern 
conflict-driven clause-learning (CDCL)  SAT solvers
with dedicated decision procedures (\Tsolvers)
for several first-order 
theories of practical interest like, e.g., those of 
equality with uninterpreted functions (\euf),
of linear arithmetic over the
rationals (\larat{}) or the integers (\laint{}), of arrays (\mem),
of bit-vectors (\bv), and their combinations. 
\ignoreinshort{We refer the reader to \cite{sebastiani07,BSST09HBSAT} for an overview.}
\ignoreinlong{(See \cite{BSST09HBSAT} for an overview.)}

 \marg{need for \omtt}%
 Many SMT-encodable problems of interest, however, may require also
 the capability of finding models that are {\em optimal} wrt. some
 cost function over continuous arithmetical variables.~%
%
%
\marg{examples: \\resource planning\\MC with parameters\\others from LGDP}%
For example, in (SMT-based) {\em planning with resources} \cite{wolfman1} a
plan for achieving a certain goal must be found which not only
fulfills some resource constraints (e.g. on time, gasoline consumption,
among others) but that also minimizes the usage of some of such resources; in
SMT-based {\em model checking with timed or hybrid systems}
\ignoreinshort{(e.g. \cite{acks_forte02,AudemardBCS05})}
\ignoreinlong{(e.g. \cite{acks_forte02})}
you may want to find
executions which minimize some parameter (e.g. elapsed time), or which
minimize/maximize the value of some constant parameter (e.g., a clock
timeout value) while fulfilling/violating some property (e.g.,
minimize the closure time interval of a rail-crossing while preserving
safety).
This also involves, as particular subcases, problems which are
traditionally addressed as {\em linear
disjunctive programming (LDP)} \cite{Egon19983} or
{\em linear
generalized disjunctive programming (LGDP)}
\cite{RamanGross94,SawayaGrossmann2012}, or as SAT/SMT with
Pseudo-Boolean (PB) constraints and (weighted partial) MaxSAT/SMT
 problems
 \cite{RM09HBSAT,LM09HBSAT,nieuwenhuis_sat06,cimattifgss10,cgss_sat13_maxsmt}.~%
Notice that the two latter problems can be  encoded into each other.

\marg{no previous work}
Surprisingly, little work has been done so far to extend SMT 
to deal with {\em optimization} problems 
\cite{nieuwenhuis_sat06,cimattifgss10,st-ijcar12,dilligdma12,cgss_sat13_maxsmt,manoliosp13}
(see \sref{sec:related}). In particular, 
to the best of our knowledge, most such works aim at minimizing 
cost functions over {\em Boolean} variables (i.e., SMT with PB cost
functions or MaxSMT), whilst 
we are not aware of any previous work on SMT solvers able to produce solutions 
which minimize cost functions over {\em arithmetical} variables.
Notice that the former can be encoded into the latter, but not vice
versa.

\smallskip 
In this this work  we start filling this gap.  
We present two general procedures for adding to
\smt{} the functionality of finding
models which minimize some \larat cost variable 
---\T being some possibly-empty stably-infinite theory
s.t. \T and \larat are signature disjoint.
These two procedures combine standard SMT
and minimization techniques:
the first, called {\em offline}, is much simpler to implement, since
it uses an incremental SMT solver as a black-box, whilst the second,
called {\em inline}, is more sophisticate and efficient, but it
requires modifying the code of the SMT solver.  This distinction is
important, since the source code of many SMT solvers is not publicly
available.

We have implemented these procedures within the \mathsatfive SMT
solver \cite{mathsat5_tacas13}.  Due to the absence of competitors
from AR, FV and SMT domains (\sref{sec:related}), we have
experimentally evaluated our implementation against state-of-the-art
tools for the domain of LGDP, which is closest in spirit to our
domain, on sets of problems which have been previously proposed as
benchmarks for the latter tools, and on other problem sets.  
(Notice that LGDP is limited to
plain \larat{}, so that, e.g., it cannot handle combinations of
theories like \laratplus.)
The results show that our tool is very competitive with, and 
 often outperforms, these tools on these problems,  
clearly demonstrating the potential of the approach.

\marg{content}
\smallskip\noindent
{\bf Content.} 
The rest of the paper is organized as follows:
\ignoreinshort{%
in \sref{sec:background} we provide some background knowledge about 
\RSCHANGETWO{ SAT, SMT, and LGDP};
}
in \sref{sec:optsmt} we formally define the problem addressed, 
{provide the necessary formal results for its solution}, 
and show how the problem generalizes many known optimization problems;
in \sref{sec:algorithms} we present our novel procedures;
in \sref{sec:expeval} we present an extensive experimental evaluation;
{in \sref{sec:related} we survey the related work;}
in \sref{sec:concl} we briefly conclude and highlight directions
for future work. 
In Appendix~\ref{sec:appendix} we provide the proofs of all the
theorems presented in the paper. 

\smallskip\noindent
\RSCHANGETWO{%
{\bf Disclaimer.}  
This work was presented in a preliminary form in a much shorter paper
at IJCAR~2012~conference~\cite{st-ijcar12}. Here the content
is extended in many ways:
first, we provide the theoretical foundations of the procedures,
including formal definitions, theorems and relative proofs;
second, we provide a much more detailed description and analysis of
the procedures, describing in details issues which were only hinted in
the conference paper;
third, we introduce novel improvements to the procedures;
fourth, we provide a much more extended empirical evaluation;
finally, we provide a detailed description of the background and of the
related work. 
}

\ignoreinshort{%

\section{Background}
\label{sec:background}
In this section we provide the necessary background about 
\RSCHANGETWO{
SAT (\sref{sec:background_sat}), 
SMT (\sref{sec:background_smt}),
and 
LGDP (\sref{sec:background_lgdp}).
}
We assume a basic background knowledge about 
logic and operational research. 
%
We
provide 
a uniform notation for
SAT and SMT:
we use 
boldface lowcase letters $\ar{a}{}, \ar{y}{}$ for arrays and 
boldface upcase letters $\ar{A}{}, \ar{Y}{}$  for matrices
\RSCHANGEONE{(i.e., two-dimensional arrays)},
 standard lowcase letters a, y for single rational 
variables/constants or indices
and 
standard upcase letters ${A}$, ${Y}$ for Boolean atoms and index sets;
we use the first five letters in the various forms 
$\ar{a}{},...\ar{e}{}$,
...
${A},...{E}$, 
to denote {\em constant} values, 
the last five 
$\ar{v}{},...\ar{z}{}$,
$...$
${V},...{Z}$ 
to denote {\em variables}, and the letters
${i}, {j}, {k}, {I}, {J}, {K}$ for indexes and index sets
respectively,
subscripts $._{j}$ denote the $j$-th element of an array or matrix, 
whilst superscripts $.^{ij}$ are just indexes, being  part of the name
of the element. 
We use lowcase Greek letters $\vi,\phi,\psi,\mu,\eta$ for denoting formulas
and upcase ones $\Phi,\Psi$ for denoting sets of formulas.

\label{remark:constantsvsvariables}
\begin{remark}
\RSCHANGEFOUR{Although we refer to quantifier-free formulas, as it is standard 
 practice in SAT, SMT,  CSP and OR communities,
 with a little abuse of terminology we 
 call ``Boolean variables'' the propositional atoms  and we call
 ``variables'' the
free constants $x_i$ in quantifier-free 
\larat-atoms like ``$(3x_1-2x_2+x_3\le 3)$''.
}
\end{remark}

\ignoreinlong{
We assume the standard syntactic and semantic notions of propositional
and first-order
logic (including the standard notions of formula, atom, literal, CNF formula,
truth assignment, clause, unit clause).
}
\ignoreinshort{
We assume the standard syntactic and semantic notions of propositional logic.
Given a non-empty set of primitive propositions  
$\calp = \{\pr{1},\pr{2},\ldots\}$, the
language of propositional logic is the least set of formulas containing 
\calp and the primitive constants $\top$ and $\bot$ (``true'' and ``false'') and closed under the set of standard
propositional connectives $\{\neg,\wedge,\vee,\imp,\iff\}$.
We call a {\em propositional atom} every primitive proposition in \calp,
and  a {\em propositional literal} every propositional atom ({\em
  positive literal}) or its negation ({\em
  negative literal}). 
We implicitly remove double negations: e.g., 
if $l$ is the negative literal $\neg \pr{i}$, then by $\neg l$ we mean $\pr{i}$ 
rather than $\neg\neg \pr{i}$. 
} 
With a little abuse of notation, we represent a truth assignment $\mu$ 
indifferently either as 
a {\em set} of literals $\{l_i\}_i$, with the intended meaning
that a positive [resp. negative] literal $\pr{i}$ means that $\pr{i}$ is
assigned to true [resp. false],
or as a {\em conjunction} of literals $\bigwedge_i l_i$; 
thus, e.g., we may say ``$l_i\in\mu$'' or ``$\mu_1\subseteq\mu_2$'',
but also 
``$\neg\mu$'' meaning the clause ``$\bigvee_i \neg l_i$''.
%

\ignoreinshort{
A propositional formula is in {\em conjunctive
  normal form, CNF,} if it is written as a conjunction of disjunctions
of literals: $\bigwedge_i \bigvee_j l_{ij}$. 
Each disjunction of
literals $\bigvee_j l_{ij}$ is called a {\em clause}.
A {\em unit clause} is a clause with only one literal.
}
%

\RSCHANGEFOUR{The above notation and terminology about
  [positive/negative] literals, truth assignments, CNF and [unit]
  clauses extend straightforwardly to quantifier-free first-order
  formulas.} 

\subsection{SAT and CDCL SAT solvers}
\label{sec:background_sat}

We present here a brief description on how a Conflict-Driven
Clause-Learning (CDCL) SAT solver works. We refer the reader, e.g., to
\RSCHANGETWO{\cite{sil96a,mzzm01,MSLM09HBSAT}} for a detailed description.

We assume the input propositional formula $\vi$ is in CNF. (If not, it is first
CNF-ized as in \cite{plaisted6}.)  
The assignment $\mu$ is initially empty, and it is updated in a stack-based 
manner. 
The SAT solver performs an external loop, alternating
three main phases: {\em Decision}, {\em Boolean Constraint
  Propagation (BCP)} and {\em Backjumping and Learning}.  

During {\em Decision}  an unassigned 
literal $l$ from $\varphi$ is selected according to some 
heuristic criterion, and it is pushed  into $\mu$.
$l$ is called \emph{decision literal}
and the number of decision literals 
\RSCHANGETHREE{which are contained 
in $\mu$ immediately after deciding $l$} is called
the \emph{decision level} of $l$.

Then {\em BCP} iteratively deduces the literals $l_1,l_2,...$ deriving from the current
assignment and pushes them into $\mu$. 
{\em BCP} is based on the iterative application of
{\em unit propagation}: if all but one literals in a clause 
are false, then the only unassigned literal $l$ 
is added to $\mu$, all negative occurrences of $l$ in other clauses 
are declared false and 
all clauses with positive occurrences of $l$ are declared satisfied. 
Current SAT solvers include rocket-fast implementations of {\em BCP} based 
on the {\em two-watched-literal scheme}, see \RSCHANGETWO{\cite{mzzm01,MSLM09HBSAT}}.
{\em BCP} is repeated until either 
no more literals can be deduced, 
so that the loop goes back to another 
  decision  step, or 
no more Boolean variable can be assigned,
so that the SAT solver ends returning \satres,
or 
$\mu$ falsifies some clause $\psi$ of $\vi$ ({\em conflicting
  clause}). 

In the latter case, {\em Backjumping and Learning} are performed. 
A process of {\em conflict analysis}~%
 \ignoreinshort{
\footnote{%
When a clause $\psi$ is
falsified by the current assignment 
a
{\em conflict clause} $\psi'$ is computed from $\psi$ s.t. $\psi'$
contains only one literal $l_u$ which has been assigned at the last
decision level.
$\psi'$ is computed starting from $\psi'=\psi$ by iteratively
resolving $\psi'$ with the clause $\psi_l$ causing the
unit-propagation of some literal $l$ in $\psi'$ 
until some stop criterion is met.
}  
} 
detects a subset $\eta$ of $\mu$ which actually caused the
falsification of $\psi$ (\emph{conflict set})~%
\footnote{%
  That is, $\eta$ is enough to force the unit-propagation of the
  literals causing the failure of $\psi$.  } and the decision level
\blevel{} where to backtrack.
%
Additionally,
the {\em conflict clause} $\psi'\defas\neg\eta$ is added to $\varphi$
(\emph{Learning}) and the procedure 
backtracks up to \blevel (\emph{Backjumping}),
popping out of $\mu$ all literals whose decision level is greater than
\blevel.
When two contradictory literals $l,\neg l$ are assigned at level 0, 
the loop terminates, returning \unsatres.

Notice that CDCL SAT solvers implement ``safe'' strategies for
deleting clauses when no more necessary, which guarantee the use of
polynomial space without affecting the termination, correctness and
completeness of the procedure. (See
e.g. \cite{MSLM09HBSAT,nieot-jacm-06}.)

Many modern CDCL SAT solvers provide a {\em stack-based 
  incremental interface} (see e.g. 
\RSCHANGETWO{\cite{eensorensson:sat2003}}), by
which it is possible to push/pop 
sub-formulas $\phi_i$ into a stack of formulas
$\Phi\defas\set{\phi_1,...,\phi_k}$, and check incrementally the
satisfiability of $\bigwedge_{i=1}^k \phi_i$. 
The interface maintains \RSCHANGETWO{most of the information about the
  {\em status}} of the search from one call 
to the other, in particular it records the learned clauses (plus other information).
Consequently, when invoked on
$\Phi$ 
the solver can reuse a clause $\psi$ which was learned during a previous
call on some $\Phi'$ 
if $\psi$ was derived only from clauses 
which are still in $\Phi$
  ---provided $\psi$ was not discharged in the meantime;
 in particular, if $\Phi'\subseteq\Phi$, then the solver can reuse
all clauses learned while solving $\Phi'$.  

Another important feature of many incremental CDCL SAT
solvers is their capability, when $\Phi$ is found unsatisfiable, to
return a subset of formulas in $\Phi$ which caused the
unsatisfiability of $\Phi$. 
This is related to the problem of finding an {\em unsatisfiable core} of a
  formula, see e.g. \cite{lynce_unsatcore_sat04}.
Notice that such subset is not unique, and it is not
necessarily minimal. 

\subsection{SMT and Lazy SMT solvers} 
\label{sec:background_smt}
\noindent
{We assume a basic background knowledge on first-order logic.
to {\em ground} formulas/literals/atoms in the language of \T
(\T-formulas/literals/atoms hereafter).}  
\RSCHANGEFOUR{Notice that, for better readability, with a little abuse
of notation we often refer to a theory \T instead of its corresponding
signature; also by ``empty theory'' we mean the empty theory over the
empty signature; finally, by adopting the terminology in
Remark~2.1, we say that a {\em variable}
belongs to the signature of a theory \T (or simply that it belongs to
a theory \T).}

A {\em theory solver for \T}, \Tsolver, is a procedure able to
decide the \T-satisfiability of a conjunction/set $\mu$ of
\T-literals.
If $\mu$ is \T-unsatisfiable, then \Tsolver returns
\unsatres and a set/conjunction $\eta$ of \T-literals in $\mu$ which was
found \T-unsatisfiable;  $\eta$ is called a
\emph{\T-conflict set}, and $\neg\eta$ a \emph{\T-conflict clause}.~%
If $\mu$ is \T-satisfiable, then \Tsolver returns \satres; it may also
be able to return some unassigned \T-literal $l \not\in \mu$
 from a set of all available \T-literals,
s.t. $\{l_1,...,l_n\}\models_{\T} l$, where
$\{l_1,...,l_n\}\subseteq\mu$.  We call this process
{\em \T-deduction} and $(\bigvee_{i=1}^n \neg l_i \vee l)$ a {\em
  \T-deduction clause}.
Notice that \T-conflict and \T-deduction clauses are valid in
\T. We call them {\em \Tlemmas}.

Given a \T-formula $\vi$, 
the formula \vip obtained by rewriting each \T-atom in \vi into a fresh atomic
proposition is the {\em Boolean abstraction} of \vi, and \vi is the
{\em refinement} of \vip.
Notationally, we indicate by \vip and \mup the Boolean abstraction 
of \vi and $\mu$, and by \vi and $\mu$ the refinements of \vip and
\mup respectively. 
\ignoreinshort{%
With a little abuse of notation, we say that \mup 
is \T-(un)satisfiable iff $\mu$ 
is \T-(un)satisfiable.
}
%
\ignoreinshort{%
We say that  the truth assignment $\mu$ 
\emph{propositionally satisfies} the formula $\varphi$, 
written $\mu \pmodels \varphi$, if $\mu^p \models \varphi^p$.
}

In a lazy \smtt{} solver, 
%
the {Boolean abstraction}
\vip of the input formula \vi 
is given as input to a CDCL SAT solver, and 
whenever a satisfying assignment $\mup$ is found s.t. $\mup\models  \vip$,
the corresponding set of \T-literals $\mu$ is fed to the \Tsolver;
if $\mu$ is found \T{}-consistent, then $\vi$ is
\T{}-consistent; otherwise, 
\Tsolver{} returns a \T-conflict set $\eta$ causing the
inconsistency, so that the clause $\neg\etap$ 
is used to drive 
the backjumping and learning mechanism of the SAT solver.
The process proceeds until either a \T{}-consistent assignment $\mu$
is found, 
or no more assignments are available
(\vi is \T-inconsistent).

%
Important optimizations are \emph{early pruning} and
\emph{\T-propagation}. The \Tsolver is invoked also when an assignment
$\mu$ is still under construction: if it is \T-unsatisfiable, then the
procedure  backtracks, without exploring the (possibly 
many) extensions of $\mu$;
if not, and if the \Tsolver is able to perform  a \T-deduction
$\{l_1,...,l_n\}\models_{\T} l$, 
then $l$ can be unit-propagated, and the \T-deduction 
clause  $(\bigvee_{i=1}^n
\neg l_i \vee l)$ can be  used in backjumping and learning. 
\RSCHANGEFOUR{
To this extent, in order to maximize the efficiency, most \T-solvers are
{\em incremental} 
and {\em backtrackable}, that is, they are called via a push\&pop interface,
maintaining and reusing the status of the search from one call and the
other. 
}

%
Another optimization is {\em pure-literal filtering}: if some
\larat-atoms occur only positively [resp. negatively] in the original
formula (learned clauses are ignored), then we can safely drop
every negative [resp. positive] occurrence of them from the assignment
$\mu$ to be checked by the \Tsolver \cite{sebastiani07}.
Intuitively, since such occurrences play no role in satisfying the
formula, {the resulting partial assignment ${\mup}'$ still satisfies
  $\vip$.}
The benefits of this action are twofold:
(i) it reduces the workload for the \Tsolver by feeding to it smaller sets;
(ii) it 
increases the chance of finding a \T-consistent satisfying
  assignment by removing   ``useless'' \T-literals
  which may cause the \T-inconsistency of $\mu$.

The above schema is a coarse abstraction of the
procedures underlying all the
state-of-the-art lazy \smt{} tools.
The interested reader is pointed  to, e.g., 
\cite{nieot-jacm-06,sebastiani07,BSST09HBSAT}
for details and further references.
 %
Importantly, some SMT solvers, including \mathsat,
 inherit from their embedded SAT solver the capabilities
of working incrementally and of returning the subset of input formulas
causing the inconsistency, as described in \sref{sec:background_sat}.

\smallskip
The {\it Theory of Linear Arithmetic 
 on the rationals} (\larat) and on the
integers (\laint) is one of the theories of main interest in SMT.
It is a first-order theory 
whose atoms
are of the form 
$(a_1x_1 + \ldots + a_nx_n \diamond b)$, 
i.e. $(\ar{a}{}\ar{x}{}\diamond b)$, 
 s.t $\diamond\in\set{=,\neq,<,>,\le,\ge}$.
%

%
Efficient incremental and backtrackable procedures have been conceived
in order to decide \larat \cite{demoura_cav06} and \laint
\cite{griggio-jsat11}. 
%
%
In particular, for \larat 
most SMT solvers implement variants of the 
simplex-based algorithm by Dutertre and de Moura \cite{demoura_cav06} which is
 specifically designed for integration in a lazy SMT
solver, since it is fully incremental and backtrackable and
allows for aggressive \T-deduction.
%
%
Another benefit of such algorithm 
is that it handles {\em strict inequalities} directly. 
\ignoreinshort{
Its method is based on the fact that 
  a set of \larat atoms $\Gamma$ containing strict
  inequalities $S = \{ 0 < t_1, \ldots, 0 < t_n\}$ is satisfiable iff there
  exists a rational number $\epsilon > 0$ such that $\Gamma_\epsilon \defas
  (\Gamma \cup S_\epsilon) \setminus S$ is satisfiable, s.t.
  $S_\epsilon \defas \{\epsilon \leq t_1, \ldots, \epsilon \leq t_n \}$.
%
The idea of \cite{demoura_cav06} is that of treating the
\emph{infinitesimal parameter} $\epsilon$
symbolically instead of explicitly computing its value. Strict bounds $(x <
b)$ are replaced with weak ones $(x \leq b - \epsilon)$, and the operations
on bounds are adjusted to take  
$\epsilon$ into account.
} 
 We refer the reader to \cite{demoura_cav06} for details.

%
\ignore{%
Remarkably, in \cite{demoura_cav06} strict inequalities $(t>0)$ are
implicitly treated as non-strict ones $(t\ge\epsilon)$, $\epsilon$ being
an infinitesimal parameter which is not computed explicitly but
treated symbolically. %
To do that, the values of bounds and variable assignments are
represented over the pairs $\mathbb{Q}_{\epsilon}$ of rationals
$\vplus{v}$, whose intended value is given by $v + \epsilon
v_{\epsilon}$.~%
\footnote{%
  Standard 
  operations over
  $\mathbb{Q}_{\epsilon}$ are defined as:
  $\vplus{v}+\vplus{u}\defas\pair{v+u}{v_\epsilon+u_\epsilon}$,
  $a\vplus{v}\defas\vplus{av}$,
  $\vplus{v}\le\vplus{u}\Longleftrightarrow (v<u)\ or\ ((v=u)\ and\
  v_{\epsilon}\le u_{\epsilon})$.  }
}

 \subsection{Linear Generalized Disjunctive Programming}
 \label{sec:background_lgdp}
\emph{Mixed Integer Linear Programming} (MILP) is an extension of Linear
Programming (LP) involving both discrete and continuous variables \cite{lodi}. 
MILP problems have the following form:
\begin{equation} \label{eq:milp}
\min \{ \ar{c}{}\ar{x}{} : \mat{A}{}\ar{x}{} \ge \ar{b}{}, \ar{x}{} \ge 0, \ari{x}{}{j} \in \mathbb{Z} \mbox{ }\forall j \in I \}
\end{equation}
where \mat{A}{} is a matrix, \ar{c}{} and \ar{b}{} are constant vectors and
\ar{x}{} the variable vector.
A large variety of techniques and
tools for MILP are available, mostly based on 
efficient combinations of LP, \emph{branch-and-bound}
search mechanism and \emph{cutting-plane} methods, resulting in a
\emph{branch-and-cut} approach (see e.g. \cite{lodi}). 
%
SAT techniques have also been incorporated into these procedures for MILP 
(see \cite{Achterberg08}).

The branch-and-bound search iteratively partitions the solution space 
of the original MILP problem into subproblems and solves their LP relaxation  
(i.e. a MILP problem where the integrality constraint on the variables 
$\ari{x}{}{j}$, for all $j \in I$, is dropped) until all variables are
integral in \RSCHANGEONE{the optimal solution of}
the LP relaxation. 
\RSCHANGETWO{}
\ignore{
 The solutions that are infeasible in the original 
problem guide the search in the following way.
When the optimal solution of a LP relaxation is greater than or equal to the 
optimal solution found so far, the search backtracks since there cannot exist 
a better solution. 
Otherwise, if a variable $\ari{x}{}{j}$ is required to be integral in the 
original problem, the rounding of its value $a$ in the LP 
relaxation suggests how to branch  by requiring 
$\ari{x}{}{j} \leq \lfloor a \rfloor $ in one branch 
and $\ari{x}{}{j} \geq \lfloor a \rfloor +1$ in the other.
}
Cutting planes (e.g. Gomory mixed-integer and lift-and-project cuts
\cite{lodi})  
are linear inequalities that can be inferred and added to the original MILP 
problem and its subproblems in order to cut away non-integer solutions of the 
LP relaxation and obtain tighter relaxations.
%
%

\smallskip
\emph{Linear Disjunctive Programming} (LDP) problems are LP problems where linear constraints 
are connected by the logical operations of conjunction and disjunction 
(see, e.g., \cite{Egon19983}).
\RSCHANGEONETHREE{The constraint set can be expressed by a disjunction of linear
systems ({\em Disjunctive Normal Form}):
%
\ignoreinshort{
\begin{equation} 
\bigvee_{i \in I} (\mat{A}{i} \ar{x}{} \ge \ar{b}{i})
\label{disjset}
\end{equation}
or, alternatively, as a conjunction ({\em Conjunctive Normal Form}):
\begin{equation} 
(\mat{A}{}\ar{x}{} \ge \ar{b}{}) \wedge \bigwedge_{j=1}^t \bigvee_{k \in I_j} (\ar{c}{k} \ar{x}{} \ge d^k) 
\label{disjset2}
\end{equation}
}
%
%
%
\noindent
or in an intermediate form called {\em Regular Form} (see, e.g., \cite{balas83}).
Notice that \eqref{disjset2} can be obtained from
  \eqref{disjset} by factoring out the common
inequalities $(\mat{A}{}\ar{x}{} \ge \ar{b}{})$
 and then by applying the
distributivity of $\wedge$ and $\vee$, although the latter step can
cause a blowup in size.
}
LDP problems are effectively solved by the lift-and-project approach which
combines a family of cutting planes, called lift-and-project cuts,
and the branch-and-bound schema (see, e.g., \cite{BalasB07}).

\smallskip
\emph{Linear Generalized Disjunctive Programming} (LGDP), is a
generalization of LDP which has been proposed in 
\cite{RamanGross94} as an alternative model to the MILP problem.
Unlike MILP, which
is based entirely on algebraic equations and inequalities,
the LGDP model allows for combining algebraic and
logical equations with Boolean propositions 
through Boolean operations, 
providing a much more natural representation of discrete decisions.
%
Current approaches successfully address LGDP 
by reformulating and solving it as a MILP
problem \cite{RamanGross94,VecchGross04,SawayaG05,SawayaGrossmann2012}; 
these reformulations focus on efficiently 
encoding disjunctions and logic propositions 
into MILP, so as to be fed to an efficient MILP solver like \cplex.


The general formulation of a LGDP problem is the following \cite{RamanGross94}:
%
\ignoreinshort{
\RSCHANGEONETHREE{
\begin{eqnarray} \label{eq:lgdp}
\mbox{min } & \sum_{ k \in K} \ari{z}{}{k} + \ar{d}{} \ar{x}{} & \nonumber  \\ 
\mbox{s.t. } & \mat{B}{} \ar{x}{} \le \ar{b}{} & \nonumber \\
& \bigvee_{j \in J_k} \Biggr[
\begin{array}{c}
Y^{jk} \\
\mat{A}{jk} \ar{x}{} \le \ar{a}{jk}\\
\ari{z}{}{k} = c^{jk} \\
\end{array} \Biggr]  & \forall k \in K \\
& \phi  &  \nonumber \\
& \ar{0}{} \le \ar{x}{} \le \ar{e}{} & \nonumber \\
& \ari{z}{}{k},c^{jk} \in \mathbb{R}_+^1, Y^{jk} \in \{True,False\} & \forall j \in J_k, \forall k \in K \nonumber
\end{eqnarray}
}
}
\noindent
\RSCHANGEONE{%
where $\ar{x}{}$ is a vector of positive rational variables, 
$\ar{d}{}$ is a vector of positive rational values representing the
cost-per-unit of each variable in $\ar{x}{}$, 
$\ar{z}{}$ is a vector of positive rational variables representing the cost assigned to 
each disjunction, $c^{jk}$ are positive constant values,
$\ar{e}{}$ is a vector of upper bounds for \ar{x}{} and
$Y^{jk}$ are Boolean variables.
}

\RSCHANGEONE{%
The disequalities
$\mat{B}{} \ar{x}{} \le \ar{b}{}$, where $(\mat{B}{},\ar{b}{})$ 
is a $m\times(n+1)$ matrix,
are the ``common'' constraints that must always hold.
}

\RSCHANGEONE{%
Each disjunction $k \in K$ consists in at least two 
disjuncts $j \in J_k$, s.t. the ${jk}$-th disjunct contains:
\begin{renumerate}
\item the Boolean variable $Y^{jk}$, representing discrete decisions,
\item a set of linear
constraints $\mat{A}{jk} \ar{x}{} \le \ar{a}{jk}$,
where $(\mat{A}{jk},\ar{a}{jk})$ is a $m_{jk} \times (n+1)$ matrix,
\item the equality $\ari{z}{}{k} = c^{jk}$, assigning 
the value $c^{jk}$ to the cost variable $\ari{z}{}{k}$.
\end{renumerate}
\noindent
Each disjunct is true if and only if all the three 
elements (i)-(iii) above are true. 
$\phi$ is a propositional formula, expressed in Conjunctive Normal Form,
which must contain
the ``xor'' constraints $\bigoplus_{j\in J_k} Y^{jk}$ for
each $k\in K$, plus possibly other constraints. 
}
\RSCHANGEONE{%
Intuitively, for each $k \in K$, 
the only variable $Y^{jk}$ which is set to true selects the 
set of disequalities $\mat{A}{jk} \ar{x}{} \le \ar{a}{jk}$ 
which are enforced and hence it selects the relative
cost $c^{jk}$ of this choice
to be assigned to the cost variable $\ari{z}{}{k}$.
}

LGDP problems can be solved using MILP solvers by reformulating
the original problem in different ways; big-M (BM) and convex hull 
(CH) are the two most common reformulations.
In BM, the Boolean variables $Y^{jk}$ and the logic constraints 
$\phi$ are replaced by binary variables $\mati{Y}{}{jk}$
and linear inequalities as follows \cite{RamanGross94}:
%
%

\RSCHANGEONETHREE{ 
\ignoreinshort{
\begin{eqnarray}\label{eq:bm}
\mbox{min } & \sum_{ k \in K}  \sum_{ j \in J_k} c^{jk} \mati{Y}{}{jk} + \ar{d}{} \ar{x}{} & \nonumber \\ 
\mbox{s.t. } & \mat{B}{} \ar{x}{} \le \ar{b}{} & \nonumber \\
& \mat{A}{jk} \ar{x}{} - \ar{a}{jk} \leq \mat{M}{jk} (1 - \mati{Y}{}{jk}) & \forall j \in J_k, \forall k \in K \nonumber \\
&  \sum_{ j \in J_k} \mati{Y}{}{jk} = 1 & \forall k \in K \\
& \mat{D}{} \mat{Y}{} \le \ar{D'}{} & \nonumber \\
& \ar{x}{} \in \mathbb{R}^n\ s.t.\ \ar{0}{} \le \ar{x}{} \le \ar{e}{}, \mati{Y}{}{jk} \in \{0,1\} & \forall j \in J_k, \forall k \in K \nonumber 
\end{eqnarray}
}
}
where $\mat{M}{jk}$ are the `''big-M`` parameters that makes redundant 
the system of constraint $j\in J_k$ in the disjunction $k \in K$ 
when $\ari{Y}{}{jk} = 0$ and the constraints 
$\mat{D}{} \mat{Y}{} \le \ar{D'}{}$ are derived from $\phi$.

In CH, the Boolean variables $Y^{jk}$ are replaced by
binary variables \mati{Y}{}{jk} and the variables $\ar{x}{} \in \mathbb{R}^n$
are disaggregated into new variables $\ar{v}{} \in \mathbb{R}^n$ 
in the following way:
%
\ignoreinshort{
\RSCHANGEONETHREE{
\begin{eqnarray}\label{eq:cm}
\mbox{min } & \sum_{ k \in K} \sum_{ j \in J_k} c^{jk} \mati{Y}{}{jk} + \ar{d}{} \ar{x}{} & \nonumber \\ 
\mbox{s.t. } & \mat{B}{} \ar{x}{} \le \ar{b}{} & \nonumber \\
& \mat{A}{kj} \ar{v}{jk} \leq \ar{a}{jk} \mati{Y}{}{jk} & \forall j \in J_k, \forall k \in K \nonumber \\
& \ar{x}{} = \sum_{ j \in J_k} \ari{v}{}{jk} & \forall k \in K\\
& \ari{v}{}{jk} \le \mati{Y}{}{jk} \ar{e}{jk} & \forall j \in J_k, \forall k \in K \nonumber \\
& \sum_{ j \in J_k} \mati{Y}{}{jk} = 1 & \forall k \in K \nonumber \\
& \mat{D}{} \mat{Y}{} \le \ar{D'}{} & \nonumber \\  
& \ar{x}{},\ar{v}{} \in \mathbb{R}^n\ s.t.\ \ar{0}{} \le \ar{x}{},\ar{v}{},\ \mati{Y}{}{jk} \in \{0,1\} &   \forall j \in J_k, \forall k \in K \nonumber
\end{eqnarray}
}
}
where constant \ar{e}{jk} are upper bounds for variables \ar{v}{} chosen 
to match the upper bounds on the variables \ar{x}{}.

Sawaya and Grossman \cite{SawayaG05} observed two facts. 
First, the relaxation of BM is often weak causing a higher number 
of nodes examined in the branch-and-bound search. 
Second, the disaggregated variables and new constraints 
increase the size of the reformulation leading to 
a high computational effort.
In order to overcome these issues, 
they proposed a cutting plane method that consists in 
solving a sequence of BM relaxations with cutting
planes that are obtained from CH relaxations.
They provided an evaluation of the presented algorithm
on three different problems:
strip-packing, retrofit planning
and zero-wait job-shop scheduling problems.

}%
\section{Optimization in \smtlaratplus }
\label{sec:optsmt}
  In this section we define the problem addressed
  (\sref{sec:optsmt_formal}), we introduce the formal foundations for its
  solution (\sref{sec:optsmt_results}), and we show how it generalizes
  many known optimization problems from the literature 
(\sref{sec:optsmt_comparison}).

\ignoreinlong{We assume the reader is familiar with the main concepts of
Boolean and first-order logic.}

\subsection{Basic Definitions and Notation}
\label{sec:optsmt_formal}

\ignore{
\noindent
  In very-general terms, we define {\em Optimization Modulo
  Theory} ({\em OMT}) as
  follows. 

  \begin{definition}[\omtplus{\T_{\preceq} \cup\bigcup_i \T_i}]
\label{def:omtplus}
Let \vi be a ground formula in some background theory $\T_{\preceq}
\cup \bigcup_i \T_i$, \RSCHANGEFOUR{where the signature of
$\T_{\preceq}$ contains a predicate 
$\preceq$ which is axiomatized  in
$\T_{\preceq}$ as a {\em total order},}
 and let \cost be a $\T_{\preceq}$-variable occurring in
$\varphi$.  The other theories $\T_i$'s may possibly be empty.
%
We call {\em Optimization Modulo $\T_{\preceq} \cup\bigcup_i \T_i$},
written {\em \omtplus{\T_{\preceq} \cup\bigcup_i \T_i}}, the problem
of finding a model for $\varphi$ whose value \minvalue of \cost is minimum
according to $\preceq$, \RSCHANGEFOUR{that is, s.t. the
  quantifier-free formula 
$
(x\preceq \minvalue)\wedge\neg(x=\minvalue)
$ is inconsistent in $\T_{\preceq}$.}
   
  \end{definition}

In this paper we consider signature-disjoint stably-infinite
theories  with equality (``Nelson-Oppen theories'' \cite{nelson3}) 
and we restrict our interest to \larat as $\T_{\preceq}$.
Thus we assume $\T$ to be some stably-infinite theory with equality,
s.t.  \larat and \T are signature-disjoint.  \T
can also be  a combination of Nelson-Oppen theories.
}

\RSCHANGEFOUR{In this paper we 
consider only signature-disjoint stably-infinite
theories  with equality $\T_i$ (``Nelson-Oppen theories'' \cite{nelson3}) 
and we focus our interest on  \larat.
In particular, in 
what follows we assume $\T$ to be some stably-infinite theory with equality,
s.t.  \larat and \T are signature-disjoint.  \T
can also be  a combination of Nelson-Oppen theories.}

\RSCHANGEFOUR{We assume the standard model of \larat,
  whose domain is the set of rational numbers $\mathbb{Q}$.}

\begin{definition}[\omlaratplus, \omlarat, and \mincost{}.]
\label{def:omlaratplus}
\marg{definizione \omlaratplus}
Let $\varphi$ be a  ground \smtlaratplus formula, 
and $\cost$ be a \larat
variable occurring in \vi.
We call an {\em Optimization Modulo $\larat\cup\T$ problem}, written
{\em \omlaratplus}, the problem of finding a model for \vi (if
any) whose value of \cost is minimum.  We denote such value as
$\mincost(\vi)$.
If \vi is \laratplus-unsatisfiable, then $\mincost(\vi)$ is $+\infty$;
if there is no minimum value for \cost, then $\mincost(\vi)$ is $-\infty$.
We call an {\em Optimization Modulo $\larat$} problem, written
  {\em \omlarat}, an \omlaratplus problem where \T is the empty
  theory.%
\end{definition}

\noindent
A dual definition where we look for a {\em maximum} value is easy to formulate.

In order to make the discussion simpler, we assume w.l.o.g. that all
\laratplus formulas are {\em pure} \cite{nelson3}.~%
\RSCHANGEFOUR{With a little abuse of notation, we say} that an atom in a ground \Tonetwo formula 
 is $\Ti$-pure if it contains only variables
and symbols from the signature of \Ti, for every $i\in\{1,2\}$; 
a \Tonetwo ground formula is pure iff all its atoms are either 
\Tone-pure or \Ttwo-pure. 
Although the purity assumption is not necessary (see \cite{bds02}), it 
simplifies the explanation, since it allows us to speak about 
``$\larat$-atoms'' or ``\T-atoms'' without further specifying. Moreover,
every non-pure formula can be easily purified \cite{nelson3}.

We also assume w.l.o.g. that 
all \larat-atoms containing the variable 
\cost are in the form $(t\bowtie\cost)$,
s.t. $\bowtie\ \in\set{=,
\le,\ge,<,>}$ and
\cost does not occur in  $t$. 

\ignore{
\marg{polarity of \cost}
\begin{definition}[polarity of \cost]
\label{def:cost_polarity}
We say that \cost occurs {\em positively} in \larat-atoms like 
$(t \le \cost)$, $(t < \cost)$, 
{\em negatively} \larat-atoms like 
$(t \ge \cost)$, $(t > \cost)$, and {\em both positively and
  negatively} in \larat-atoms like $(t = \cost)$, $(t \neq \cost)$;
we say that \cost occurs {\em only positively} [resp. negatively] in \vi 
iff it occurs positively [resp. negatively] in all atoms occurring
positively  in \vi and negatively [resp. positively] in all these
occurring negatively in \vi.   
\end{definition}
}

\begin{definition}[Bounds and range for \cost]
\label{def:bounds}
If $\vi$ is in the form $\vi'\wedge\ubliti{c}$ [resp. $\vi'\wedge
\lbliti{c}$] for some value $c\in\mathbb{Q}$, then we call $c$
an {\em upper bound} [resp. {\em lower bound}] for \cost.
If \ub [resp \lb] is the minimum upper bound [resp. the maximum lower
bound] for \vi,
we also call the interval \range the  {\em range} of \cost.~%
  \end{definition}
\ignoreinshort{
   We adopt the convention of defining upper bounds to be
  strict and lower bounds to be non-strict for a practical reason:
  typically an upper bound \ubliti{c}  derives from the fact that a
  model $\cali$ of cost $c$ has been previously found,
  whilst a lower bound \lbliti{c} derives either from the user's
  knowledge (e.g. ``the cost cannot be lower than zero'') of from
  the fact that the formula $\vi\wedge (\cost<c)$ has been previously
  found \T-unsatisfiable whilst \vi has not.  
}

\subsection{Theoretical Results}
\label{sec:optsmt_results}

We present here the theoretical foundations of our procedures. 
The proofs of the novel results are reported in Appendix~\ref{sec:appendix}.

The following facts follow straightforwardly from Definition~\ref{def:omlaratplus}.
\begin{proposition}
\label{prop:cost_monotonicity}
Let $\vi,\vi_1,\vi_2$ be \laratplus-formulas and $\mu_1,\mu_2$ 
be truth assignments.
  \begin{aenumerate}
    \item 
If $\vi_1\models\vi_2$, then $\mincost(\vi_1)\ge\mincost(\vi_2) $.
    \item 
If $\mu_1\supseteq\mu_2$, then $\mincost(\mu_1)\ge\mincost(\mu_2) $.
    \item 
\vi is \laratplus-satisfiable if and only if $\mincost(\vi)<+\infty$.
  \end{aenumerate}

\end{proposition}

We recall first some definitions and results from the literature. 
\begin{definition}
\marg{complete collection of assignments}
\label{def:complete_assignment_set}
  We say that a collection
${\cal M\/} := \{\mu_1,\ldots,\mu_n\}$ of
(possibly partial) assignments propositionally satisfying
 $\vi$ is  {\em complete\/} iff, 
for every {total} assignment  $\eta$ s.t. $\eta\pmodels\vi$,
there exists $\mu_j\in {\cal M\/}$ s.t. $\mu_j\subseteq \eta$.
\end{definition}

\begin{theorem}[\cite{sebastiani07}]
\label{teo:smt_foundations}
Let $\vi$ be a  \T{}-formula and let ${\cal M\/} := \{\mu_1,\ldots,\mu_n\}$
be a complete collection of (possibly partial)
truth assignments propositionally satisfying $\vi$.
Then, $\vi$ is \T{}-satisfiable if and only if $\mu_j$ is
\T{}-satisfiable for some $\mu_j\in{\cal M}$.
\end{theorem}

Theorem~\ref{teo:smt_foundations} is the theoretical foundation of 
the lazy SMT approach described in \sref{sec:background_smt}, where a
CDCL SAT solver enumerates a complete collection ${\cal M\/}$ of
truth assignments as above, whose \T-satisfiability is checked by a
\Tsolver. Notice that in Theorem~\ref{teo:smt_foundations} the theory
\T can be any combination of theories \Ti, including \larat. 

Here we extend Theorem~\ref{teo:smt_foundations} to 
\omlaratplus as follows. 


\begin{theorem}
\label{teo:main}
\marg{main\\ theorem}
Let $\vi$ be a  \laratplus-formula and let 
$\calm  \defas \{\mu_1,\ldots,\mu_n\}$
be a complete collection of (possibly-partial)
 truth assignments which propositionally satisfy $\vi$.
Then $\mincost(\vi)=min_{\mu\in\calm} \mincost(\mu)$.
\end{theorem}

\noindent
Notice that we implicitly define
  $min_{\mu\in\calm}\mincost(\mu)\defas+\infty$ if $\cal M$ is empty.
Since  $\mincost(\mu)$ is
$+\infty$ if $\mu$ is \laratplus-unsatisfiable, we can safely
restrict the search for minima to the \laratplus-satisfiable  assignments
in \calm.

If \T is the empty theory, then the notion of $\mincost(\mu)$ is
straightforward, since each $\mu$ is a conjunction
 of Boolean literals and of
\larat constraints, so that Theorem~\ref{teo:main} provides 
the theoretical foundation for \omlarat. 

If  instead \T is not the empty theory, then each $\mu$ is a set of Boolean
literals and of pure \T-literals and \larat constraints sharing
variables, so that the notion of
$\mincost(\mu)$ is not straightforward. 
To cope with this fact, we first recall  from the literature 
some definitions and an important result.

\begin{definition}[Interface variables, interface equalities.]
\label{def:interfaceequalities}
Let \Tone and \Ttwo be two stably-infinite theories with equality and
disjoint signatures, and  
let \vi be a \tonetwo-formula.
%
%
We call {\em interface variables} of \vi the variables occurring in both
\Tone-pure and \Ttwo-pure atoms of \vi, 
 and {\em interface equalities} of \vi
the equalities $\eqij$ on the interface variables of \vi.
\end{definition}

\noindent
As it is common practice in SMT (see e.g. \cite{TinHar-FROCOS-96}) 
hereafter we consider only interface equalities modulo reflexivity and
symmetry, that is, 
we implicitly assume some total order $\preceq$ on the interface
variables $x_i$ of $\vi$, and we restrict w.l.o.g. the set of interface
equalities on $\vi$ to
\mbox{$\IE{\vi}\defas\{(x_i=x_j)\ |\ x_i\prec x_j \}$}, 
dropping thus uninformative equalities like $(x_i=x_i)$ and
considering only the first equality in each pair
$\{(x_i=x_j),(x_j=x_i)\}$. 

Notation-wise, in what follows
 we use the subscripts $e,d,i$ in ``$\mu_{...}$'', like in
``\mued'', to denote conjunctions of 
  equalities, 
disequalities and inequalities between {interface}
variables respectively.

\begin{theorem}[\cite{TinHar-FROCOS-96}]
\label{teo:NO}
Let \Tone and \Ttwo be two stably-infinite theories with equality and
disjoint signatures; 
let $\mu \defas\mu_{\Tone}\wedge\mu_{\Ttwo}$ be a
conjunction of \Tonetwo-literals
s.t. each $\mu_{\Ti}$ is pure for $\T_i$.
%
\ignore{
(We call {\em interface variables} the variables occurring in both
$\mu_{\Tone}$ and $\mu_{\Ttwo}$, and {\em interface equalities} 
the equalities $\eqij$ on interface variables.)
} 
Then
$\mu$ is \tonetwo-satisfiable  if and only if
  there exists an equivalence class 
$Eq\subseteq\IE{\mu}$ over the interface variables of $\mu$ 
and the corresponding 
total truth assignment \mued to the interface equalities over $\mu$:~%
%
  \begin{equation}
\label{eq:no}
   \mued \defas \mue \wedge \mud,\ \ s.t. \
\mue\defas\bigwedge_{(x_i,x_j)\ \in\  Eq} \eqij,\ \ 
\mud\defas\bigwedge_{(x_i,x_j)\ \not\in\  Eq} \neqij 
  \end{equation}
  s.t.  $\mu_{\Tk}\wedge \mued$ is $\T_k$-satisfiable for every $k \in
  \{1,2\}$.
\end{theorem}

\ignore{
It is trivial to notice that
Theorem~\ref{teo:NO} still holds under this restriction, because
assigning $(x_i=x_i)$ to false or the elements of
$\{(x_i=x_j),(x_j=x_i)\}$ to different truth values would obviously
make the assignment $\mu_{\Tk}\wedge \mued$ \Tk-unsatisfiable for each $k \in
\{1,2\}$.
}
%

Theorem~\ref{teo:NO} is the theoretical foundation of, among others,
the {\em Delayed Theory Combination} SMT technique for combined
theories \cite{bozzanobcjrrs06}, where a CDCL SAT solver enumerates a
complete collection of extended assignments $\mu\wedge\mued$,
  which propositionally satisfy the input formula, and
dedicated \Tk-solvers check independently the \Tk-satisfiability of
$\mu_{\Tk}\wedge \mued$, for each $k \in \{1,2\}$.

\smallskip
We consider now a \laratplus formula \vi and a (possibly-partial)
truth assignment $\mu$ 
which propositionally satisfies it. 
$\mu$ can be written as $\mu \defas
\mubool\wedge
\mularat\wedge\mut$, s.t. 
\mubool is a consistent conjunction of Boolean literals, 
\mularat and \mut are \larat-pure
and \T-pure conjunctions of literals respectively. 
(Notice that the \mub component does not affect the
\laratplus-satisfiability of $\mu$.)
Then the following definitions and theorems show
how $\mincost(\mu)$ can be defined and computed.

\begin{definition}
\label{def:mued-extension}
    Let $\mu \defas \mubool\wedge \mularat\wedge\mut$ be a 
    truth assignment satisfying some \laratplus ground formula, s.t.  \mubool is
  a consistent conjunction of Boolean literals, \mularat and \mut are
  \larat-pure and \T-pure conjunctions of literals respectively.
We call 
the {\em complete set of $ed$-extensions of $\mu$} 
the set $\edmuext\defas\{\eta_1,...,\eta_n\}$ 
of all possible assignments in the form $\mu\wedge\mued$, where $\mued$ 
is in the form \eqref{eq:no}, for every  equivalence class
$Eq$ in \IE{\mu}. 
\end{definition}

\begin{theorem}
\label{teo:mued-extension}
  Let $\mu$ be as in
  Definition~\ref{def:mued-extension}.   
Then 
\begin{aenumerate}
\item
$\mincost(\mu)=min_{\eta\in\edmuext} \mincost(\eta)$
\item forall $\eta \in\edmuext$, \\
$
\mincost(\eta) = 
\left \{ 
  \begin{array}{lll}
+\infty && \mbox{if $\mut\wedge\mued$ is
  \T-unsatisfiable or} \\  
 &&    \mbox{if $\mularat\wedge\mued$ is
  \larat-unsatisfiable} \\
\mincost(\mularat\wedge\mued) & \ \ \ \ &  \mbox{otherwise.}
  \end{array}
\right .
$
\end{aenumerate}
\end{theorem}

We notice that, at least in principle,
 computing $\mincost(\mularat\wedge\mued)$ is an
operation which can 
be performed by standard linear-programming techniques (see
\sref{sec:algorithms}). 
Thus, by combining Theorems~\ref{teo:main} and \ref{teo:mued-extension}
we have a general method for computing $\mincost(\vi)$
also in the general case of non-empty theory \T.

\smallskip 
In practice, however, it is often the case that \larat-solvers/optimizers
cannot handle efficiently negated equalities like, e.g., \neqij (see
\cite{demoura_cav06}).  Thus, a technique which is adopted by most SMT
solver is to expand them into the corresponding disjunction of strict
inequalities $\dij\vee\dji$.  This ``case split'' is typically
efficiently handled directly by the embedded SAT solver.

We notice, however, that such case-split may be applied also to 
interface equalities \eqij, and that the resulting ``interface
inequalities'' \dij and \dji cannot be handled by the other theory \T, 
because ``$<$'' and ``$>$'' are \larat-specific symbols. 
In order to cope with this fact, 
some more theoretical discussion is needed.

\begin{definition}
\label{def:muedi-extension}
Let $\mu$ be as in  Definition~\ref{def:mued-extension}.  
 We call 
the {\em complete set of ${edi}$-extensions of $\mu$} 
the set $\edimuext\defas\{\rho_1,...,\rho_n\}$ 
of all possible truth assignments in the form $\mu\wedge\mued\wedge\mu_i$,
where $\mued$ is as in Definition~\ref{def:mued-extension} and 
\mui 
is a total truth assignment to the atoms 
in $\{\dij, \dji\ | \eqij\in\IE{\mu}\}$
s.t. 
$\mued\wedge\mui$ is \larat-consistent.
\end{definition}
 
\mui assigns both \dij and \dji to false if \eqij
is true in \mued, one of them to true and the other to false if \eqij
is false in \mued. 
Intuitively, the presence of each negated interface equalities \neqij
in \mued forces 
the choice of one of the two parts
\tuple{\dij,\dji} of the solution space.

\begin{theorem}
\label{teo:muedi-extension}
  Let $\mu$ be as in
  Definition~\ref{def:mued-extension}.  
Then 
\begin{aenumerate}
\item \label{item:a} $\mu$ is \laratplus-satisfiable iff some $\rho\in\edimuext$ 
is \laratplus-satisfiable. 
\item \label{item:b}
$\mincost(\mu)=min_{\rho\in\edimuext} \mincost(\rho)$.
\item \label{item:c} forall $\rho\in\edimuext$,
 $\rho$ is \laratplus-satisfiable iff $\mut\wedge\mued$ is
  \T-satisfiable and $\mularat\wedge\mue\wedge\mui$ is
  \larat-satisfiable.  
\item \label{item:d} forall $\rho \in\edimuext$, \\
$
\mincost(\rho) = 
\left \{ 
  \begin{array}{lll}
+\infty && \mbox{if $\mut\wedge\mued$ is
  \T-unsatisfiable or} \\  
 &&    \mbox{if $\mularat\wedge\mue\wedge\mui$ is
  \larat-unsatisfiable} \\
\mincost(\mularat\wedge\mue\wedge\mui) & \  \ &  \mbox{otherwise.}
  \end{array}
\right .
$
\end{aenumerate}
\end{theorem}

Thus, by combining Theorems~\ref{teo:main} and \ref{teo:muedi-extension}
we have a general method for computing $\mincost(\vi)$ 
in the case of non-empty theory \T, which is compliant with an efficient 
usage of standard \larat-solvers/optimizers.

\subsection{\omlaratplus wrt. other Optimization Problems}
\label{sec:optsmt_comparison}

In this section we show that \omlaratplus captures many interesting
optimizations problems.

%
%
\ignoreinshort{
LP
is a particular subcase of \omlarat with no Boolean component, such that
 $\vi\defas\vi'\wedge(\cost = \sum_{i} \ari{a}{}{i} \ari{x}{}{i})$ 
and $\vi'=\bigwedge_j (\sum_i \ari{A}{}{ij} \ari{x}{}{i} \le \ari{b}{}{j})$.
%

LDP can also be encoded into 
\omlarat, 
since
(\ref{disjset}) 
\ignoreinshort{and (\ref{disjset2})}
can be written respectively as
\begin{equation}
%
\bigvee_i \bigwedge_j (\mati{A}{i}{j} \ar{x}{}\ge \ari{b}{i}{j})
\label{eq:ldp_encoding}
\end{equation}
\begin{equation}
\bigwedge_{j} (\mati{A}{}{j} \ar{x}{}\ge \ari{b}{}{j}) 
\wedge  \bigwedge_{j=1}^t \bigvee_{k \in I_j} (\ari{c}{k}{}  \ar{x}{} \ge d^k),
\label{eq:ldp_encoding2} 
\end{equation}
where 
\mati{A}{i}{j} and \mati{A}{}{j} are respectively
 the $j$th row of the matrices \mati{A}{i}{} and \mati{A}{}{}, 
\ari{b}{i}{j} and \ari{b}{}{j} are respectively
 the $j$th row of the vectors \ari{b}{i}{} and \ari{b}{}{}. 
 Since  \eqref{eq:ldp_encoding} is not in CNF, the
 CNF-ization process of \cite{plaisted6} is then applied.


LGDP \eqref{eq:lgdp}
is straightforwardly encoded into a \omlarat problem \pair{\vi}{\cost}:
\RSCHANGEONETHREE{ 
\begin{equation}
\textstyle
\label{eq:ldgp_enc}
\begin{array}{lll}
\vi&\defas&  (\cost = \sum_{ k \in K} \ari{z}{}{k} + \ar{d}{}
\ar{x}{}) 
\wedge
\enc{\mat{B}{} \ar{x}{} \le \ar{b}{}}
\wedge
\phi
\wedge
\enc{ \ar{0}{} \le \ar{x}{}}
\wedge
\enc{\ar{x}{} \le \ar{e}{}}
\\
%
\textstyle
&\wedge & 
\bigwedge_{k \in K} 
\bigvee_{j \in J_k} 
(
Y^{jk}
\wedge 
\enc{\mat{A}{jk} \ar{x}{} \le \ar{a}{jk}}
\wedge
(
\ari{z}{}{k} = c^{jk}
)
) 
\\ &\wedge& 
 \bigwedge_{k \in K} 
(
(\ari{z}{}{k} \ge min_{j\in J_k}c^{jk})
\wedge
(\ari{z}{}{k} \le max_{j\in J_k}c^{jk})
)
\end{array}
\end{equation}
}
\noindent
s.t. $\enc{\ar{x}{}\bowtie \ar{a}{}}$ 
and $\enc{\mat{A}{}\ar{x}{}\bowtie \ar{a}{}}$ 
are abbreviations respectively for 
$\bigwedge_i (\ari{x}{}{i}\bowtie \ari{a}{}{i})$ and
$\bigwedge_i (\mati{A}{}{i\cdot}\ari{x}{}{}\bowtie \ari{a}{}{i})$,
$\bowtie\ \in\set{=,\neq\le,\ge,<,>}$. 
\RSCHANGEONE{The last conjunction ``$\bigwedge_{k \in K}
  ((\ari{z}{}{k} \ge\ ...\ )) $''
in \eqref{eq:ldgp_enc} is not necessary, but it improves
the performances of the \smtlarat solver, because it allows for exploiting
the early-pruning SMT technique (see \sref{sec:background_smt}) by providing a range for the values of the
\ari{z}{}{k}'s before the respective $Y^{jk}$'s are assigned.
}
 Since  \eqref{eq:ldgp_enc} is not in CNF, the
 CNF-ization process of \cite{plaisted6} is then applied.
} 


Pseudo-Boolean (PB) constraints (see \cite{RM09HBSAT}) 
in the form $(\sum_i\ari{a}{}{i} X^i \le b)$ s.t. $X^i$ are Boolean
atoms and $\ari{a}{}{i}$ constant values in $\mathbb{Q}$, 
 and cost functions $\cost = \sum_i \ari{a}{}{i}X^i $,
 are encoded into \omlarat by rewriting each PB-term
$\sum_i \ari{a}{}{i}X^i $ into the \larat-term 
$\sum_i \ari{x}{}{i} $,  $\ari{x}{}{}$ being an array of fresh
\larat variables, and by conjoining to \vi the formula:~%
%
\begin{equation}
\label{eq:pb2smt-encoding}
\textstyle
\bigwedge_i((\neg X^i \vee (\ari{x}{}{i}=\ari{a}{}{i}))\wedge (X^i \vee
(\ari{x}{}{i}=0))
\ignoreinshort{
\wedge (\ari{x}{}{i}\ge 0) \wedge (\ari{x}{}{i}\le \ari{a}{}{i})~%
}
).  
\end{equation}

\noindent
\RSCHANGEONE{
The term 
``$(\ari{x}{}{i}\ge 0) \wedge (\ari{x}{}{i}\le \ari{a}{}{i})$'' 
in \eqref{eq:pb2smt-encoding} is not necessary, but it improves
the performances of the \smtlarat solver, because it allows for exploiting
the early-pruning technique by providing a range for the values of the
\ari{x}{}{i}'s before the respective $X^i$'s are assigned.
}

A (partial weighted) MaxSMT problem (see \cite{nieuwenhuis_sat06,cimattifgss10,cgss_sat13_maxsmt})
is a pair \tuple{\vi_{h},\vi_{s}} where $\vi_{h}$ is a set of
``hard'' \T-clauses and $\vi_{s}$ is a set of weighted ``soft''
\T-clauses, s.t. a positive weight \ari{a}{}{i} is associated to each
soft \T-clause $C_i\in\vi_{s}$; the problem consists in finding a
maximum-weight set of soft \T-clauses $\psi_{s}$ s.t.
$\psi_{s}\subseteq\vi_{s}$ and $\vi_{h}\cup\psi_{s}$ is
\T-satisfiable. Notice that one can see \ari{a}{}{i} as a penalty to
pay when the
corresponding soft clause is not satisfied.
A MaxSMT problem \tuple{\vi_{h},\vi_{s}}
can be encoded straightforwardly into an SMT problem with PB cost
function \tuple{\vi',\cost} by augmenting each soft \T-clause $C_j$ with 
a fresh Boolean variables $X^j$ as follows:
\begin{eqnarray}
\label{eq:maxsmt2pb-encoding}
\textstyle
\vi' \defas \vi_{h} \cup\bigcup_{\C_j\in\vi_{s}} \{(X^j\vee\C_j)\};
\ \
\cost \defas \sum_{\C_j\in\vi_{s}} \ari{a}{}{j}X^j. 
%
\end{eqnarray}
Vice versa,
\tuple{\vi',\cost\defas \sum_j \ari{a}{}{j}X^j} can be encoded into MaxSMT:
\begin{eqnarray}
\label{eq:pb2maxsmt-encoding}
\textstyle
\vi_{h} \defas \vi';
\ \
\vi_{s} \defas \bigcup_{j} \{\underbrace{(\neg X^j)}_{\ari{a}{}{j}}\}.
\end{eqnarray}

\noindent
Thus, combining \eqref{eq:pb2smt-encoding} and
\eqref{eq:maxsmt2pb-encoding}, optimization problems for SAT with PB
constraints and MaxSAT can be encoded into \omlarat, whilst those for
\smtt with PB constraints and MaxSMT can be encoded into \omlaratplus{},
under the assumption that \T matches the definition in \sref{sec:optsmt_formal}.

\begin{remark}
 We notice the deep difference between \omlarat/\omlaratplus and the
 problem of SAT/SMT with PB constraints and cost functions (or
 MaxSAT/ MaxSMT) addressed in \cite{nieuwenhuis_sat06,cimattifgss10,cgss_sat13_maxsmt}.
 With the latter problems, the value of \cost is a deterministic consequence of a
 truth assignment to the atoms of the formula, so that the search has
 only a Boolean component, consisting in finding the cheapest truth
 assignment. With \omlarat/ \omlaratplus, instead, for every satisfying
 assignment $\mu$ it is also necessary to find the minimum-cost
 \larat-model for $\mu$, so that the search has both a Boolean and a
 \larat-component.
\end{remark}

\ignore{
With respect to existing works related to optimization in SAT 
\cite{bebmaxsat,Xing2005MaxSolver,HerasLO08,RM09HBSAT,Barth95adavis-putnam,bebmethods} 
and SMT \cite{NieuwenhuisOliveras2006SAT,CimattiFGSS10},
we express cost functions as linear functions over domain variables  
rather than as purely pseudo-Boolean functions.
To this extent, notice that \smt with pseudo-boolean functions 
\cite{NieuwenhuisOliveras2006SAT,CimattiFGSS10}
can be easily encoded into \omtlcft
by introducing novel cost variables $x_i$, adding
clauses of the form $(B_i \rightarrow (x_i=1))$ and $(\neg B_i \rightarrow (x_i=0))$
to $\varphi$ (where $B_i$ are fresh Boolean atoms), 
and writing the cost function as in (\ref{cfsum}).
Obviously, SAT optimization problems are specializations of \smtt.
}

\section{Procedures for \omlarat and \omlaratplus} 
\label{sec:algorithms}
It might be noticed that very naive \omlarat or \omlaratplus procedures
could be straightforwardly  implemented by performing a sequence of calls 
to an SMT solver on formulas like
$\vi\wedge(\cost\ge \currlb_i)\wedge(\cost<\currub_i)$,
each time restricting the range $[\currlb_i,\currub_i[$ according to a
linear-search or binary-search schema.
With the linear-search schema, 
\ignoreinshort{every time the SMT solver returns a model
of cost $c_i$, a new constraint $(\cost<c_i)$ would be added 
to \vi, and the solver would be invoked again;
}
however, the SMT solver would repeatedly generate
the same \larat-satisfiable truth assignment, each time 
finding a cheaper model for it. 
%
With the binary-search schema the efficiency should improve;
however, an initial lower-bound should be necessarily required as input
(which is not the case, e.g., of the problems in \sref{sec:expeval_smtlib}.)

In this section we present more sophisticate procedures,
based on the combination of SMT and minimization techniques.  
We first 
present and discuss 
an {\em offline} schema  (\sref{sec:algorithms_offline}) and an
 {\em inline} (\sref{sec:algorithms_inline}) schema
 for an \omlarat procedure;
then we show how to extend them to the \omlaratplus{} case
(\sref{sec:algorithms_omlaratplus}). 

\ignoreinlong{
In what follows we assume the reader is familiar with the basics about
 CDCL SAT solvers and lazy SMT solvers. 
A detailed background section on that is
available on the extended version of this paper
\cite{ST-ijcar12-extended}; for a much more detailed description, we
refer the reader, e.g., to \cite{MSLM09HBSAT,BSST09HBSAT}  respectively.
}

\subsection{An offline schema for \omlarat }
\label{sec:algorithms_offline}

%
\marg{I/O description}
The general schema for the offline \omlarat procedure 
is displayed in Algorithm~\ref{algo:mixsearch}.
%
It takes as input an instance
of the \omlarat problem plus optionally values for \lb and \ub,
which are implicitly considered to be $-\infty$ and $+\infty$ if not
present, and returns the model \calm of minimum cost and its 
cost \currub (the value \ub if \vi
is \larat-inconsistent).~%
\ignoreinshort{%
Notice that, by providing a lower bound \lb [resp. an upper bound \ub],
the user implicitly assumes the responsibility of asserting there is no model 
whose cost is lower than \lb [resp. there is a model whose cost is \ub].
}

We represent \vi as a set of clauses, which may be pushed or
popped from the input formula-stack of an incremental SMT solver. 
\RSCHANGEONETWO{To this extent, every operation like 
``$\varphi \leftarrow \varphi \cup \set{...}$''
[resp. ``$\varphi \leftarrow \varphi \setminus \set{...}$''] in
Algorithm~\ref{algo:mixsearch}, ``$\set{...}$'' being a clause set, 
should be interpreted as 
``push $\set{...}$ into $\varphi$'' 
[resp. ``pop $\set{...}$ from $\varphi$''].
}

\begin{algorithm}[t]
\caption{Offline \omlarat Procedure based on Mixed Linear/Binary Search.
\label{algo:mixsearch}
} 
\algsetup{indent=2em}
\renewcommand{\varphicost}{\vi}
\begin{algorithmic}[1]
\REQUIRE
 $\langle \varphi, \cost, \lb, \ub \rangle$ 
$\{$\ub can be $+\infty$, \lb can be $-\infty$$\}$
\STATE $
\currlb \leftarrow\lb;
\currub \leftarrow\ub;
\pivotatom \leftarrow \top;
\calm \leftarrow \emptyset
$ \label{algomix:init}
\STATE $ \varphicost \leftarrow \varphi \cup \set{\currlblit,\currublit}$
\label{algomix:constraints}
\RSCHANGEONETWO{\COMMENT Push bound constraints into \vi}
\WHILE{($\currlb < \currub $ )} \label{algomix:while}
   \IF[Binary-search Mode]{(\dopivoting{})} \label{algomix:pivotingcond}
     \STATE $\pivot \leftarrow \computepivot{}(\currlb,\currub)$ 
\label{algomix:initpivoting}
     \STATE $\pivotatom \leftarrow (\cost < \pivot)$ 
     \STATE $\vi \leftarrow \vi \cup\set{
       \pivotatom}$ \label{algomix:addpivot}
\RSCHANGEONETWO{\COMMENT Push \pivotatom into \vi}
   \STATE $\tuple{\status,\mu} \leftarrow
   \incrementalsmt(\varphicost)$
\label{algomix:smtsolverone}
\RSCHANGEONE{
\IF{$(\status = \unsatres)$}
   \STATE $\eta \leftarrow
   \smtcoreextract(\varphicost)$ \label{algomix:smtsolvercore}
\ELSE
\STATE $\eta\leftarrow \emptyset$
\ENDIF\label{algomix:endpivoting}
}
   \ELSE[Linear-search Mode] 
   \STATE $\tuple{\status,\mu} \leftarrow \incrementalsmt(\varphicost)$ \label{algomix:smtsolver}
\STATE $\eta\leftarrow \emptyset$
   \ENDIF 
   \IF{$(\status = \satres)$}\label{algomix:issat} 
      \STATE $\pair{\calm}{\currub} \leftarrow \minimize(\cost, \mu)$ \label{algomix:min}
     \STATE $\varphicost \leftarrow \varphicost \cup \set{\currublit}
$ 
\label{algomix:add}
\RSCHANGEONETWO{\COMMENT Push new upper-bound constraint into \vi}
   \ELSE[\status = \unsatres]
   \IF{$({\pivotatom \not\in \eta}) $} \label{algomix:unsatcondition}
      \STATE $\currlb \leftarrow \currub$ \label{algomix:end}
   \ELSE 
     \STATE $\currlb \leftarrow \pivot$ \label{algomix:lbound}
     \STATE $\varphicost \leftarrow \varphicost \ \setminus\
     \set{\ \pivotatom}$\label{algomix:poppivot}   
\RSCHANGEONETWO{\COMMENT Pop $\pivotatom$ from  \vi}
     \STATE $\varphicost \leftarrow \varphicost \cup\set{\neg\pivotatom}$ \label{algomix:pushnegpivot}    
\RSCHANGEONETWO{\COMMENT Push $\neg\pivotatom$ into  \vi}
   \ENDIF
   \ENDIF
\ENDWHILE
\label{algomix:endwhile}
\RETURN $\pair{\calm}{\currub}$ \label{algomix:return}
\end{algorithmic}
\end{algorithm}


\marg{initialization} 
First, the variables \currlb, \currub defining the current range
 are initialized to \lb and \ub respectively,
the atom \pivotatom to $\top$, 
and \calm is initialized to be an empty model.
Then the procedure adds to \vi the bound constraints, if
present, which restrict the search within
the range \currrange (row \ref{algomix:constraints}).~%
(Obviously literals like \lbliti{-\infty} 
and \ubliti{+\infty} 
are not added.)
%
The solution space is then explored iteratively (rows
\ref{algomix:while}-\ref{algomix:endwhile}), reducing 
the current range \currrange to explore at each loop, until the range is empty.
Then \pair{\calm}{\currub} 
is returned ---\pair{\emptyset}{\ub}
if there is no solution in \range---
\calm being the model of minimum cost \currub.
Each loop may work in either {\em linear-search} or {\em
binary-search} mode, driven by the heuristic \dopivoting. 
Notice that if $\currub = +\infty$ or $\currlb = -\infty$, 
then \dopivoting returns \falseval. 

\marg{linear-search mode} 
In {\bf linear-search mode}, 
steps \ref{algomix:initpivoting}-\ref{algomix:endpivoting} and
\ref{algomix:lbound}-\ref{algomix:pushnegpivot}    
are not executed. 
First,  an incremental
\smtlarat solver is invoked on \vi (row
\ref{algomix:smtsolver}). 
(Notice that, given the incrementality of the solver,
 every operation in the form 
``$\vi \leftarrow \vi \cup \set{\phi_i}$" 
[resp. $\vi \leftarrow \vi \setminus\set{\phi_i}$] 
is implemented as a ``push'' [resp. ``pop'']
operation  on the stack representation of
$\vi$\ignoreinshort{,\ see \sref{sec:background_sat}};
it is also very important to recall that  
during the SMT call $\vi$ is updated with 
the clauses which are learned during the 
SMT search.)
Then $\eta$ is set to be empty, which  forces condition
\ref{algomix:unsatcondition}  to hold. 

\marg{new minimum} 
If \vi is
\larat-satisfiable, then it is returned \status=\satres and a
\larat-satisfiable truth assignment $\mu$ for $\vi$.
Thus \minimize is invoked on (the subset of \larat-literals of) 
$\mu$,\ignoreinshort{~\footnote{%
Possibly after applying pure-literal filtering to $\mu$
(see \sref{sec:background_smt}).}} returning 
the model \calm for $\mu$ of minimum cost
$\currub$ ($-\infty$ iff the problem is unbounded).
%
The current solution 
$\currub$ becomes the new upper bound, 
thus the \larat-atom $(\cost < \currub)$
is added to $\vi$ (row \ref{algomix:add}). 
Notice that, if the problem is unbounded, then for some $\mu$ 
\minimize{} will return $-\infty$, forcing
condition~\ref{algomix:while} to be \falseval{} and 
the whole process to stop.
If \vi is
\larat-unsatisfiable, then no model in the current cost range
\currrange can be found;
hence the flag \currlb is set to \currub, forcing the end of the loop. 

\marg{binary-search mode} 
In {\bf binary-search mode}
at the beginning of the loop 
(steps \ref{algomix:initpivoting}-\ref{algomix:endpivoting}), 
the value $\pivot\in\ ]\currlb,\currub[$ is computed by the heuristic
function \computepivot{}
(in the simplest form, $\pivot$ is $(\currlb + \currub)/2$), 
the possibly-new atom $\pivotatom\defas(\cost<\pivot)$ is pushed
into the formula stack,
so that to
temporarily restrict the cost range to \lpivotrange.
Then the incremental SMT solver is invoked on \vi;
\RSCHANGEONE{if the result is \unsatres, the  feature \smtcoreextract{} is
activated,} which returns also a
subset $\eta$ of formulas in (the formula stack of) $\vi$ which caused
the unsatisfiability of $\vi$\ignoreinshort{\ (see
\sref{sec:background_sat})}.
This exploits techniques similar to unsat-core extraction
\cite{lynce_unsatcore_sat04}.
\RSCHANGEONE{(In practical implementations, it is not strictly
  necessary to explicitly produce the unsat core $\eta$; rather, it
  suffices to check if $\pivotatom\in\eta$.)}

If \vi is \larat-satisfiable, then the procedure behaves as in
linear-search mode. 
If instead \vi is \larat-unsatisfiable, we look at $\eta$ and
distinguish two subcases.  
If $\pivotatom$ does not occur in $\eta$,
this means that $\vi\setminus\set{\pivotatom}$ is \larat-inconsistent,
i.e. there is no model in the whole cost range \currrange. Then the
procedure behaves as in linear-search mode, forcing the end
of the loop.
Otherwise, 
we can only conclude that there is no model in the cost range \lpivotrange,
so that we still need exploring the cost range \rpivotrange.
Thus \currlb is set to \pivot, \pivotatom is popped from \vi and its negation is pushed into
\vi. Then the search proceeds, investigating the cost range \rpivotrange.

\medskip
\marg{no pure binary search}
We notice an important fact:  if \dopivoting always returned \trueval{}, then
Algorithm~\ref{algo:mixsearch} would not necessarily terminate. 
In fact, 
an SMT solver invoked on $\vi$ may 
return a set $\eta$ containing $\pivotatom$ even 
if $\vi\setminus\pivotatom$ is \larat-inconsistent.~%
\ignoreinshort{%
\footnote{A CDCL-based SMT solver implicitly builds a
resolution refutation whose leaves are either clauses in $\vi$ 
or \larat-lemmas, and the set $\eta$ represents the subset of clauses 
in $\vi$ which occur as leaves of such proof
(see e.g. \cite{cimattigs11_unsatcore} for details).
If  the SMT solver
is invoked on $\vi$ even $\vi\setminus\pivotatom$ is \larat-inconsistent,
 then it can ``use'' \pivotatom and return 
a proof involving it even though 
another \pivotatom-less proof exists.%
} 
}
Thus, e.g., the procedure might got stuck into a 
\ignoreinshort{``Zeno''~\footnote{%
In the famous Zeno's paradox, Achilles never reaches the tortoise for a
similar reason.}}
  infinite loop,
 each time halving the cost range right-bound
(e.g., $[-1,0[$, $[-1/2,0[$, $[-1/4,0[$,...).

To cope with this fact, however, 
it suffices to guarantee that \dopivoting 
returns \falseval{} after a finite number of such steps,
guaranteeing thus that eventually a linear-search loop will be forced,
which detects the inconsistency.  In our implementations,
we have empirically chosen to force one linear-search
loop after {\em every} binary-search loop returning \unsatres, because
satisfiable calls are typically much cheaper than
unsatisfiable ones. 
\RSCHANGETWO{%
      We have empirically verified in previous tests that this 
      was in general the best option, since introducing this test
      caused no significant overhead and prevented the chains of 
      (very expensive) unsatisfiable calls where they used to occur.
}

\smallskip
\marg{correctness, completeness and termination}
Under such hypothesis, as a consequence of
Theorem~\ref{teo:main} of
\sref{sec:optsmt_results}, 
we have that:

\begin{renumerate}
\item 
{%
Algorithm~\ref{algo:mixsearch} terminates.
In linear-search mode it terminates 
because there are only a finite number
of candidate truth assignments $\mu$ to be enumerated, and steps
\ref{algomix:min}-\ref{algomix:add} guarantee that the same assignment
$\mu$ will never be returned twice by the SMT solver. In 
mixed linear/binary-search mode, as above, it terminates since there
can be at most finitely-many binary-search loops between two
consequent linear-search loops;
}
\item Algorithm~\ref{algo:mixsearch}
 returns a model of minimum cost, because it explores the whole
search space of candidate truth assignments, and for every suitable
assignment $\mu$
\minimize finds the minimum-cost model for $\mu$;
\item Algorithm~\ref{algo:mixsearch}
requires polynomial space, under the assumption that the underlying
CDCL SAT solver adopts a polynomial-size clause-deleting
  strategy (which is typically the
case of SMT solvers, including \mathsat). 
\end{renumerate}

%
\subsubsection{Handling strict inequalities}

\marg{\minimize \&\\ strict inequalities} 
%
\minimize is a simple extension of the simplex-based
\laratsolver of \cite{demoura_cav06},
which is invoked after one solution is found, 
  minimizing it by standard simplex techniques.
We recall that the algorithm in \cite{demoura_cav06} can handle
strict inequalities.
\RSCHANGEONE{Thus, if $\mu$ contains strict inequalities,
 then \minimize temporarily relaxes them into non-strict ones and then
 it finds a  
 solution \ar{x}{} of minimum cost $\minvalue$ of the relaxed problem, namely 
\murelaxed. 
(Notice that this could be done also by any standard LP package.)}
 \begin{rschange}
Then:
\begin{enumerate}
\item 
if such minimum-cost solution \ar{x}{} 
lays only on
 non-strict inequalities, then \ar{x}{} is a also solution of the non-relaxed
 problem $\mu$, hence \minvalue can be returned;

\item
otherwise, we temporarily add the constraint $(\cost\le 
\minvalue)$ to the non-relaxed version of $\mu$  and then we invoke on
it the \larat-solving procedure of 
\cite{demoura_cav06}  (without minimization),
since such algorithm  can handle strict
inequalities. Then:

\begin{renumerate}
\item if the procedure returns \satres, then $\mu$ has a model   of cost $\minvalue$.~%
If so, then the value \minvalue can be returned, and $(\cost<\minvalue)$
 can be pushed into $\vi$;

\item 
 otherwise, $\mu$ has no model of cost $\minvalue$.
 If so, since 
 $\mu$ has a convex set of solutions whose cost is strictly
 greater than  $\minvalue$  and 
 there is a solution of cost \minvalue for the relaxed
 problem \murelaxed, then for some $\delta>0$ and for every cost $c\in\
 ]\minvalue,\minvalue+\delta]$ there exists a solution for $\mu$ of cost $c$.
 (If  needed explicitly, such solution can be computed using the
 techniques for handling strict inequalities described in
 \cite{demoura_cav06}.)  Thus the value \minvalue can be tagged
 as a non-strict minimum and returned, so that the constraint
 $(\cost\le\minvalue)$, rather than  $(\cost<\minvalue)$, is pushed into \vi.

\end{renumerate}
\end{enumerate}
%

Notice that situation (2).(i) is very rare in practice but it is
possible in principle, as illustrated in the following example.

\begin{example}
  Suppose we have that $\mu=\{(\cost \ge 1),(\cost > y), (\cost
> -y)\}$. If we temporarily relax strict inequalities into non-strict
ones, then $\{\cost=1,y=1\}$ is a minimum-cost solution which lays on
the strict inequality $(\cost > y)$. Nevertheless, there is a solution 
of \cost 1 for the un-relaxed problem, e.g., $\{\cost=1,y=0.9999\}$. 
\end{example}

\RSCHANGEFOUR{Notice also that $(\cost\le\minvalue)$ is pushed into
  \vi only if the minimum cost of the 
current assignment $\mu$ is strictly greater than $\minvalue$, as in 
situation (2).(ii).
This prevents the SMT solver from returning $\mu$ again. Therefore
 the termination, the correctness and the completeness of the
 algorithm are guaranteed  also in the case 
 some truth assignments have strict minimum costs.}

\end{rschange}

\ignore{
Then:
\begin{enumerate}
\item 
if such minimum-cost solution \ar{x}{} of cost $\minvalue$ lays only on
 non-strict inequalities, then \ar{x}{} is a solution of the original
 problem, hence \minvalue can be returned;

\item
otherwise, we may have two alternative subcases:
\begin{renumerate}
\item there is some other solution  of cost $\minvalue$.~%
\footnote{%
This subcase is rare in practice but it is possible in principle. 
For instance, suppose we have that $\mu=\{(\cost \ge 1),(\cost > y), (\cost
> -y)\}$. If we temporarily relax strict inequalities into non-strict
ones, then $\{\cost=1,y=1\}$ is a minimum-cost solution which lays on
the strict inequality $(\cost > y)$. Nevertheless, there is a solution 
of \cost 1, e.g., $\{\cost=1,y=0.9999\}$. 
}
If so, the value \minvalue can be returned; 

\item 
 there is no solution of cost $\minvalue$.
 If so,  then for some $\delta>0$ and for every cost $c\in\
 ]\minvalue,\minvalue+\delta]$ there exists a solution of cost $c$.
 (If  needed explicitly, such solution can be computed using the
 techniques for handling strict inequalities described in
 \cite{demoura_cav06}.)  Thus the value \minvalue can be tagged
 as a non-strict minimum and returned, so that the constraint
 $(\cost\le\minvalue)$, rather than  $(\cost<\minvalue)$, is added to \vi.

\end{renumerate}
\end{enumerate}

\noindent
Condition 2.(i) can be checked easily, e.g., by
temporarily adding the constraint $(\cost\le 
\minvalue)$ to $\mu$  and then by invoking again the \larat-solving procedure of
\cite{demoura_cav06}  on $\mu$  (without minimization),
since such algorithm  can handle strict
inequalities.~
(%
In our implementation, we have directly modified the algorithm of
\cite{demoura_cav06}
so that to perform this check internally.)

}

\ignore{
An alternative way is to omit this check  and behave as in
(ii) in both cases, so that $(\cost\le\minvalue)$ 
is learned; if so, in case (i), it may happen that $\mu\cup\{(\cost\le\minvalue)\}$ is generated 
passed to \minimize
}


\ignore{
\ignoreinshort{%
Notice that fact 2.  never happens  
if 
the SMT solver (like \mathsat) implements 
pure-literal filtering (\sref{sec:background_smt}) 
and 
if no strict inequalities occur  in $\vi$ with positive polarity 
[resp. no non-strict inequality occur  in $\vi$ with negative polarity], 
   which is the case of many problems in this paper.
In fact, due to pure-literal filtering, the only strict inequalities
which may be fed to \minimize are upper-bound constraints
\currubliti{i}, which by construction do not touch the minimum-cost
solution \ar{x}{} above.    
 }
}

\subsubsection{Discussion}
%

We remark a few facts about this procedure.
\begin{enumerate}

\item \label{item:approximatesolution}
If Algorithm~\ref{algo:mixsearch} is interrupted (e.g., by
a timeout device), then \currub can be returned, representing the best
approximation of the minimum
cost found so far. 

\item 
\marg{role of \\incrementality}
The incrementality of the SMT solver 
\ignoreinshort{(see \sref{sec:background_sat} and \sref{sec:background_smt})}
plays an essential role here, since at every call \incrementalsmt
resumes the status of the search at the end of the previous call, only with 
tighter cost range constraints. (Notice that at each
call here the solver can reuse all previously-learned clauses.)
To this extent, one can see the whole process mostly as only one SMT process, 
which is interrupted and resumed each time a new model is found,
in which cost range constraints are progressively tightened.

\item 
\marg{early-pruning as ``bound''}
In Algorithm~\ref{algo:mixsearch} all the
literals constraining the cost range 
(i.e., $\neg (\cost < \currlb)$, $(\cost <\currub)$) 
are added to \vi as unit clauses;
thus inside \incrementalsmt they are immediately
unit-propagated, becoming part of each truth 
assignment $\mu$ from the very beginning of its construction.
\ignoreinlong{(We recall that the SMT solver invokes incrementally  \laratsolver also
while building an assignment $\mu$ ({\em early pruning calls} 
\cite{BSST09HBSAT}.))} 
As soon as novel \larat-literals
are added to $\mu$ which prevent it from 
having a \larat-model of cost in  \currrange, 
the \larat-solver invoked on $\mu$ by early-pruning  calls 
\ignoreinshort{(see \sref{sec:background_smt})} returns
\unsatres and the \larat-lemma $\neg\eta$ describing the conflict $\eta\subseteq\mu$,
triggering theory-backjumping and -learning. 
%
%
To this extent, \incrementalsmt implicitly plays a
form of {\em branch \& bound}: (i) decide a new literal $l$ and
\ignoreinshort{\ unit- or theory-}propagate the literals which derive
from $l$ (``branch'') and (ii)
backtrack as soon as the current branch can no more be expanded into 
models in the current cost range (``bound'').

\item \label{item:lowerboundpruning}
\ignoreinshort{The unit clause \currlblit 
plays a role even in linear-search mode, since it helps pruning the
search inside \incrementalsmt.}

%
\item 
\marg{better than pure binary}
%
In binary-search mode,  the range-partition strategy  may be even
more aggressive than that of standard binary search, because the
minimum cost \currub returned in row~\ref{algomix:min}
 can be  smaller than
\pivot, so that the cost range is more then halved. 

\item \label{item:binvslin} 
\marg{binary vs linear}
Unlike with other domains (e.g., search in sorted arrays) 
here binary search  
is not ``obviously faster'' than linear search, because the
unsatisfiable calls to \incrementalsmt are typically much more
expensive than the satisfiable ones, since they must explore the whole
Boolean search space rather than only a portion of it ---although with a
higher pruning power, due to the stronger constraint induced by the
presence of \pivot.
Thus, we have a tradeoff between a typically much-smaller number of calls plus
a stronger pruning power in  binary search versus an average much
smaller cost of the calls in linear search. 
To this extent, it is possible \RSCHANGETWO{in principle} to use 
dynamic/adaptive strategies for \computepivot{} 
(see \cite{sellmannk08}).

\end{enumerate}

\subsection{An inline schema for \omlarat}
\label{sec:algorithms_inline}

\marg{description of \\inline schema}
With the inline schema, the whole optimization procedure is pushed
inside the SMT solver by embedding
the range-minimization loop inside the CDCL Boolean-search loop of the
standard lazy SMT schema\ignoreinshort{\ of
  \sref{sec:background_smt}}. 
 The SMT solver, which is thus called only once, is modified as follows.

\ignoreinshort{
\begin{figure}[th]
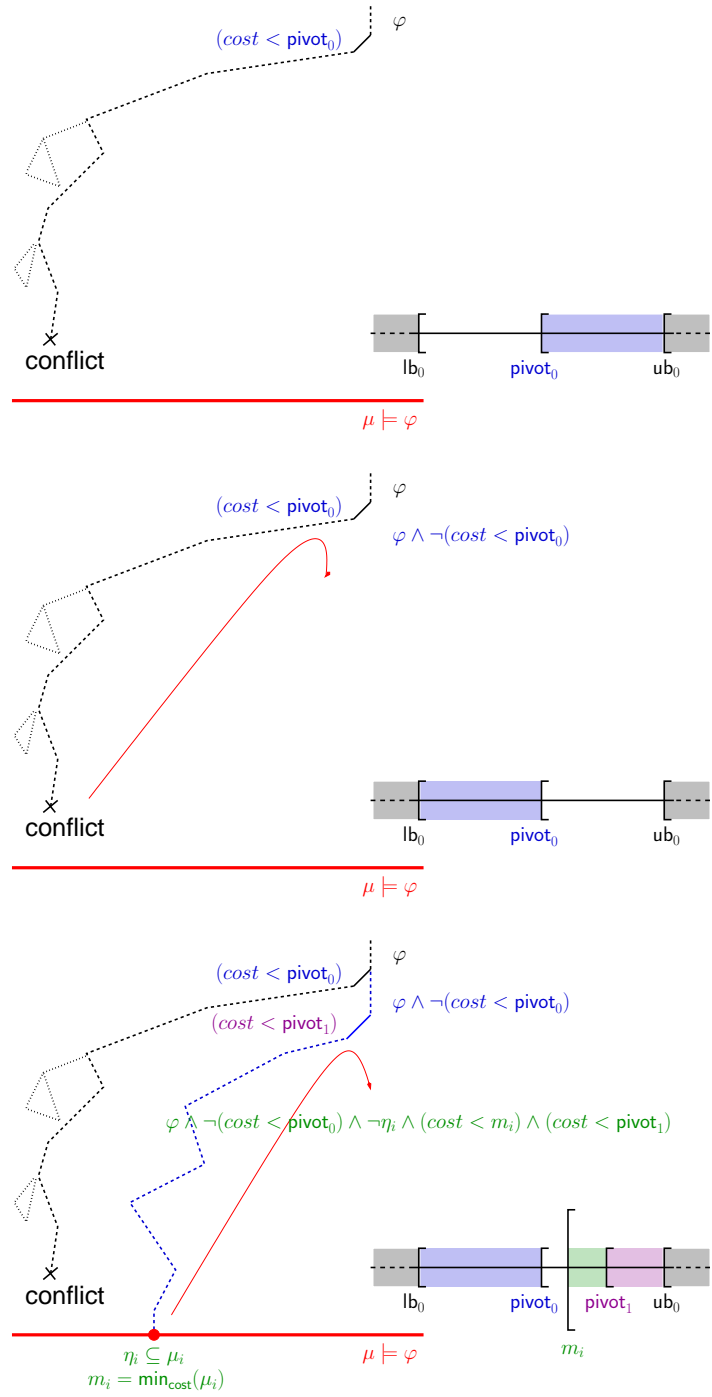

  \centering
\scalebox{.32}{\input{bin_0.pstex_t}} 
\scalebox{.32}{\input{bin_1.pstex_t}} 
 \scalebox{.32}{\input{bin_4.pstex_t}}
  \caption{One piece of possible execution of an inline procedure. 
  (i) Pivoting on $(\cost < \pivot_0)$. 
  (ii) Increasing the lower bound to $\pivot_0$.
  (iii) Decreasing the upper bound to $\mincost(\mu_i)$.
  \label{fig:algo_inline}
}
\end{figure}
}

%
%

\noindent {\bf Initialization.}
The variables $\lb,\ub,\currlb,\currub,\pivotatom,\pivot,\calm$ 
are brought inside the SMT solver, and are 
initialized as in 
Algorithm~\ref{algo:mixsearch},  steps
\ref{algomix:init}-\ref{algomix:constraints}. 

\noindent{\bf Range Updating \& Pivoting.}
Every time the search of the CDCL SAT solver gets back to decision level 0,
the range \currrange is updated s.t. 
\currub [resp. {\currlb] is assigned the lowest [resp. highest] value
  $\currub_i$ [resp. $\currlb_i$] such that the atom $(\cost <
  \currub_i)$ [resp. $\neg (\cost < \currlb_i)$] is currently assigned
  at level 0.
(If $\currub \le \currlb$, or two literals 
 $l,\neg l$ are both assigned at level 0,
then the procedure
  terminates, returning the current value of \currub.) 
%
Then \dopivoting{} is invoked: if it returns \trueval, then
\computepivot{} computes $\pivot\in\
]\currlb,\currub[$, and the (possibly new) atom $\pivotatom\defas
(\cost < \pivot)$ is decided to be true (level 1) by the SAT solver.  This
mimics steps \ref{algomix:initpivoting}-\ref{algomix:addpivot}
in Algorithm~\ref{algo:mixsearch}, temporarily restricting the cost range to
  $[\currlb,\pivot[$.  

%

\noindent{\bf Decreasing the Upper Bound.} 
When an assignment $\mu$ propositionally satisfying \vi is generated
which is found \larat-consistent by \laratsolver, $\mu$ is also fed to
\minimize, returning the minimum cost \minvalue of $\mu$; then the unit
clause $(\cost < \minvalue)$ is learned and fed to the
backjumping mechanism, which forces the SAT solver to backjump to
level 0 and then to unit-propagate $(\cost < \minvalue)$. This case
mirrors steps \ref{algomix:issat}-\ref{algomix:add} in
Algorithm~\ref{algo:mixsearch}, permanently restricting the cost range
to $[\currlb,\minvalue[$.
\minimize is embedded within the \laratsolver, so that it is
 called incrementally after it, without restarting its search from scratch.  

As a result of these modifications, we also have the following typical
scenario\ignoreinshort{\ (see Figure~\ref{fig:algo_inline})}.

\marg{right part}
\noindent{\bf Increasing the Lower Bound.} 
In binary-search mode, when a conflict
occurs s.t. the conflict analysis of the SAT solver produces a conflict clause in the
form $\neg\pivotatom\vee\neg\eta'$ s.t. all literals in $\eta'$ are
assigned \trueval{} at level 0 (i.e., $\vi\wedge\pivotatom$ is
\larat-inconsistent), then the SAT solver backtracks to level 0,
unit-propagating $\neg\pivotatom$.  This case mirrors steps
\ref{algomix:lbound}-\ref{algomix:pushnegpivot} in
Algorithm~\ref{algo:mixsearch}, permanently restricting the cost range
to $[\pivot,\currub[$.

\smallskip
Although the modified SMT solver mimics to some extent the behaviour of
Algorithm~\ref{algo:mixsearch}, the ``control'' of the
range-restriction process is handled by the standard SMT search. 
\marg{not only binary search}
To this extent, notice that also other situations may allow for restricting the
cost range: e.g., if 
$\vi\wedge \neg (\cost < \currlb)\wedge(\cost < \currub)\models 
(\cost \bowtie \mvalue)$ for some atom $(\cost \bowtie \mvalue)$ occurring in \vi s.t. 
$\mvalue \in \currrange$ and $\bowtie\ \in
\{\le,<,\ge,>\}$, then the SMT solver may backjump to decision level 0 
and propagate $(\cost \bowtie \mvalue)$, further restricting the cost
range.

The same facts (\ref{item:approximatesolution})-(\ref{item:binvslin}) 
 about the offline procedure in
 \sref{sec:algorithms_offline} hold for the inline version. 
The efficiency of the inline procedure can be further improved as follows.

\ignore{%
First, an early-pruning call to \laratsolver is performed
  also at level 0, before the first decision is performed, so that to
  cut the search if the current set of unit clauses is
  \larat-inconsistent.%
}

\smallskip 
\noindent\RSCHANGEONE{{\bf Activating previously-learned
    clauses.}  In binary-search mode, when an assignment $\mu$ with a
  novel minimum \minvalue is found, not only $(\cost < \minvalue)$ but
  also $\pivotatom\defas(\cost < \pivot)$ is learned as unit clause,
  although the latter is redundant from the logical perspective
  because $\minvalue<\pivot$.
  In fact, the unit clause $\pivotatom$ allows the SAT solver for
  reusing all the clauses in the form $\neg \pivotatom \vee C_i$ which
  have been learned when investigating the cost range \lpivotrange, by
  unit-resolving them into the corresponding clauses $C_i$.}.
(In Algorithm~\ref{algo:mixsearch} this is done implicitly, since \pivotatom 
is not popped from \vi before step \ref{algomix:poppivot}.)
\RSCHANGEONE{Notice that the above trick 
 is useful because the algorithm 
of \cite{demoura_cav06} is not ``\T-deduction-complete'', that is, it
is not guaranteed to  \T-deduce 
$\pivotatom$ from $\set{...,(\cost < \minvalue)}$.
}

In addition, the \larat-inconsistent assignment
$\mu\wedge(\cost < \minvalue)$ may be fed to \laratsolver and
the negation 
of the returned conflict $\neg\eta\vee\neg(\cost < \minvalue)$
s.t. 
$\eta\subseteq\mu$, 
can be
learned, preventing the SAT solver from generating 
any assignment containing $\eta$ in the future.

\marg{tightening}
\smallskip \noindent {\bf Tightening.}
In binary-search mode, if
 \laratsolver returns a conflict set 
$\eta\cup\{\pivotatom\}$, 
then it is further asked to find the maximum value $\maxvalue$
s.t. $\eta\cup\{(\cost<\maxvalue)\}$ is \larat-inconsistent. (This is done
with a simple modification of the algorithm in \cite{demoura_cav06}.)
\begin{itemize}
\item \RSCHANGEONE{If $\maxvalue\ge\currub$, then the clause
    $C^*\defas\neg\eta\vee\neg(\cost<\currub)$ is used do drive
    backjumping and learning instead of
    $C\defas\neg\eta\vee\neg\pivotatom$.  Since the unit clause
    $(\cost<\currub)$ is permanently assigned at level 0, this is
    equivalent to learning only $\neg\eta$, so that the dependency of
    the conflict from $\pivotatom$ is removed.  Eventually, instead of
    using $C$ to drive backjumping to level 0 and then to propagate
    $\neg\pivotatom$, the SMT solver may use $C^*$ (which is the same
    as using $\neg\eta$), then forcing the procedure to stop.}

\item \ignoreinshort{%
If $\currub>\maxvalue>\pivot$, then the  clauses 
$C_1\defas\neg\eta\vee\neg(\cost<\maxvalue)$ and 
$C_2\defas\neg\pivotatom\vee(\cost<\maxvalue)$ are 
used to drive backjumping and learning instead of 
$C\defas\neg\eta\vee\neg\pivotatom$.
\RSCHANGE{(Notice that $C$ can be inferred by resolving $C_1$ and $C_2$.)}
In particular, $C_2$ forces backjumping to level 1 
and unit-propagating the (possibly fresh) atom $(\cost<\maxvalue)$; 
eventually,  instead of using $C$ do drive
 backjumping to level 0 and then to 
propagate $\neg\pivotatom$, the SMT solver may use 
$C_1$ for backjumping to level  0 and
then to propagate $\neg(\cost<\maxvalue)$, restricting the range to 
$[\maxvalue,\currub[$ rather than to \rpivotrange. 
}

\end{itemize}

\RSCHANGEONE{%
Notice that tightening is useful because the algorithm 
 of \cite{demoura_cav06} is guaranteed neither to find the ``tightest'' 
theory conflict $\eta\cup\set{(\cost<\maxvalue)}$, nor to 
\T-deduce $(\cost<\maxvalue)$ from \set{...,\pivotatom}.
%
}

\ignoreinshort{%
\begin{example}
\marg{esempio tightening}
\label{ex:tightening}
Consider the formula $\vi\defas\psi\wedge(\cost \ge a+15)\wedge(a\ge
0)$ for some $\psi$ in the cost range $[0,16[$. With 
binary-search deciding $\pivotatom\defas(\cost<8)$, the \laratsolver
produces the lemma
$C\defas 
\neg(\cost \ge a+15)\vee
\neg(a\ge 0)\vee
\neg\pivotatom
$,
causing a backjumping step 
to level 0 on $C$ and the 
unit-propagation of $\neg\pivotatom$, restricting the range to $[8,16[$; it takes a sequence of
similar steps to progressively restrict the range to 
$[12,16[$, $[14,16[$, and $[15,16[$.
If instead the \laratsolver produces the lemmas
$C_1\defas
\neg(\cost \ge a+15)\vee
\neg(a\ge 0)\vee
\neg(\cost<15)
$
and
$
C_2\defas\neg\pivotatom \vee (\cost<15)
$,
then this first causes a backjumping step on $C_2$ to level 1 with
the unit-propagation of $(\cost<15)$, 
and then a backjumping step on $C_1$ to level zero with the
unit-propagation of $\neg(\cost<15)$, which directly restricts the range
to  $[15,16[$.
\end{example}
}


%

\smallskip \noindent{\bf Adaptive Mixed Linear/Binary Search Strategy.}
An adaptive version of the heuristic 
\dopivoting{} decides the next search mode according to
the ratio between the progress obtained in the latest 
binary- and linear-search steps and their respective costs.
If either  \ub or \lb is not present 
\RSCHANGE{---or if we are immediately after an \unsatres binary-search step, 
in compliance with
the strategy to avoid infinite ``Zeno'' sequences described in
\sref{sec:algorithms_offline}--- 
then the heuristic selects 
linear-search mode. 
Otherwise, 
it selects binary-search mode if and only if 
$$\left | \frac{\Deltaublin}{\Deltanconflin}\right |<\left |\frac{\Deltaubbin}{\Deltanconfbin}\right |,$$
where \Deltaublin and \Deltaubbin 
are respectively 
the variations of the upper bound \ub in the latest linear-search
and \satres binary-search steps performed, estimating the progress 
 achieved by such steps, 
whilst \Deltanconflin and \Deltanconfbin are respectively 
the number of conflicts produced in such steps, estimating their expense.
}

\smallskip
\RSCHANGETWO{%
  Overall, the inline version described in this section presents some potential
  computational advantages wrt. the offline version of
  Algorithm~\ref{algo:mixsearch}.  
  First, despite the incrementality of the calls to the SMT solver,
  suspending and resuming it may cause some overhead,
  because at every call the decision stack is popped to decision level
  0, so that some extra decisions, unit-propagations and early-pruning
  calls to the \Tsolver may be necessary to get back to the previous
  search status.
  Second, in Algorithm~\ref{algo:mixsearch} the procedure \minimize{}
  is invoked from scratch in a non-incremental way, whilst in the
  inline version it is embedded inside the \laratsolver, so that it starts
  the minimization process from an existing solution rather than from
  scratch.  
  \RSCHANGEFOUR{%
    Third, Algorithm~\ref{algo:mixsearch} requires computing the
    unsatisfiable core of \vi ---or at least checking if \pivotatom
    belongs to such unsat core--- which causes overhead.  Notice that
    the problem of computing efficiently minimal unsat-cores in SMT is
    still ongoing research (see \cite{cimattigs11_unsatcore}), so that
    in Algorithm~\ref{algo:mixsearch} there is a tradeoff between the
    cost of reducing the size of the cores and the probability of
    performing useless optimization steps.  }
%
}

\subsection{Extensions to \omlaratplus}  
\label{sec:algorithms_omlaratplus}  
We recall the terminology, assumptions, definitions and results of
\sref{sec:optsmt_results}.  
Theorems~\ref{teo:main}, \ref{teo:mued-extension} and
\ref{teo:muedi-extension} allow for extending to the \omlaratplus case
the procedures 
of \sref{sec:algorithms_offline} and
\sref{sec:algorithms_inline} 
as follows. 

As suggested by Theorem~\ref{teo:mued-extension}, straightforward \omlaratplus
extensions of the procedures for \omlarat{}  
of \sref{sec:algorithms_offline} and \sref{sec:algorithms_inline}
would be such that the SMT solver enumerates $ed$-extended 
satisfying truth assignments 
$\eta\defas\mu\wedge\mued$ as in Definition~\ref{def:mued-extension},
checking the \T-  and \larat-consistency  of its components
$\mut\wedge\mued$  and $\mu_{\larat}\wedge\mued$ respectively,
and then minimizing the $\mu_{\larat}\wedge\mued$  component. 
Termination is guaranteed by the fact that each \edmuext{} is a finite
set, whilst correctness and completeness is guaranteed by Theorems 
~\ref{teo:main} and \ref{teo:mued-extension}.

Nevertheless, as suggested in \sref{sec:optsmt_results}, 
minimizing $\mu_{\larat}\wedge\mued$ efficiently could be  
problematic due to the presence of negated interface equalities \neqij{}. 
Thus, alternative ``asymmetric'' procedures,
in compliance with the efficient usage of
\larat-solvers in SMT, 
should instead enumerate $edi$-extended satisfying
truth assignments $\rho\defas\mu\wedge\mueid$
as in Definition~\ref{def:muedi-extension},
checking the \T-  and \larat-consistency  of its components
$\mut\wedge\mued$  and $\mu_{\larat}\wedge\muei$ respectively,
and then minimizing the $\mu_{\larat}\wedge\muei$  component. 
This prevents from passing negated interface 
equalities to \minimize.  
As before, termination is guaranteed by the fact that each \edimuext{}
is a finite set, whilst correctness and completeness is guaranteed by Theorems 
~\ref{teo:main} and \ref{teo:muedi-extension}.

This motivates and explains the following \omlaratplus variants of the
offline and inline procedures of \sref{sec:algorithms_offline} and
\sref{sec:algorithms_inline} respectively.

Algorithm~\ref{algo:mixsearch} 
is modified as follows.
  First, \incrementalsmt{} in steps
\ref{algomix:smtsolverone} and \ref{algomix:smtsolver} is asked to
return also a $\larat\cup\T$-model \cali.
%
Then in step \ref{algomix:min} \minimize is invoked instead 
on \pair{\cost}{\mu_{\larat}\cup\muei}, s.t. 
$$
  \begin{array}{rll}
    \muei \defas & \{\eqij,\neg\dij,\neg\dji     &|\
    \eqij\in\IE{\mu},\     \cali\models\eqij\} \\
    \cup & \{\dij,\neg\dji &|\ 
    \eqij\in\IE{\mu},\  \cali\models\dij\} \\
    \cup & \{\dji,\neg\dij &|\ 
    \eqij\in\IE{\mu},\  \cali\models\dji\}.
  \end{array}
$$
\noindent
In practice, the negated strict inequalities $\neg\dij,\neg\dji$ are
omitted from \muei, because they are entailed by the corresponding
non-negated equalities/inequalities.

%

The implementation of an inline \omlaratplus{} procedures comes nearly
for free once the  SMT solver
 handles $\larat\cup\T$-solving by 
{\em Delayed Theory Combination} \cite{bozzanobcjrrs06}, 
with the strategy of case-splitting automatically 
disequalities \neqij into the two inequalities \dij and \dji, 
which is implemented in \mathsat:
%
the solver enumerates truth assignments in the form
$\rho\defas\mu_{\larat}\wedge\mueid\wedge\mu_{\T}$ as in
Definition~\ref{def:muedi-extension},
and passes 
$\mu_{\larat}\wedge\muei$ and 
$\mu_{\T}\wedge\mued$ 
to the \laratsolver and \Tsolver respectively.~%
(Notice that this strategy, although not explicitly described in
\cite{bozzanobcjrrs06}, implicitly
 implements points $(a)$ and  $(c)$ of
Theorem~\ref{teo:muedi-extension}.)
%
If so, then, in accordance with points $(b)$ and  $(d)$ of
Theorem~\ref{teo:muedi-extension},
 it suffices to apply \minimize{} to
$\mu_{\larat}\wedge\muei$, then learn $(\cost<\minvalue)$ and use it for
backjumping, as in \sref{sec:algorithms_inline}.
As with the offline version, in practice the negated strict inequalities are
omitted from \muei, because they are entailed by the corresponding
non-negated equalities/inequalities.

\section{Experimental evaluation}
\label{sec:expeval}

We have implemented both the offline and inline \omlarat procedures
and the inline \omlaratplus procedures of \sref{sec:algorithms} on top
of \mathsatfive~%
\footnote{\url{http://mathsat.fbk.eu/}.} 
 \cite{mathsat5_tacas13}; we refer to them
as \optmathsat.
\RSCHANGE{\mathsatfive is a state-of-the-art SMT solver 
which supports most of the quantifier-free SMT-LIB
theories and their combinations, and provides many other SMT 
functionalities (like, e.g., 
unsat-core extraction \cite{cimattigs11_unsatcore}, 
interpolation \cite{cimattigs10}, 
All-SMT \cite{cavada_fmcad07_predabs}).}
  
{We consider different configurations of \optmathsat, 
depending on the approach 
(offline vs. inline, denoted by ``-OF'' and ``-IN'') and on
the search schema (linear vs. binary vs. adaptive, 
denoted respectively by ``-LIN'', ``-BIN'' and ``-ADA'').%
~\footnote{Here ``-LIN'' means that \dopivoting always returns
\falseval, ``-BIN'' denotes the mixed linear-binary strategy
described in \sref{sec:algorithms_offline} to ensure termination,
{whilst ``-ADA'' refers to the adaptive strategy illustrated 
in \sref{sec:algorithms_inline}.}
}
\ignoreinshort{For example, the configuration \optlinin{}  
denotes the inline linear-search procedure.} 
We used only five configurations since 
the ``-ADA-OF'' were not implemented.}
%
%

Due to the absence of competitors on \omlaratplus, we evaluate the
performance of our five configurations of \optmathsat by comparing
them against {the commercial LGDP tool}
\gams~%
\footnote{\url{http://www.gams.com}.
} 
v23.7.1 \cite{gams} 
on \omlarat{} problems. 
{\gams is a tool for modeling and solving 
optimization problems, consisting of different language compilers,
which translate mathematical problems into representations 
required by specific solvers, like \cplex \cite{cplex}.}
\gams provides two reformulation tools, 
\logmip~
\footnote{\url{http://www.logmip.ceride.gov.ar/index.html}.}  
 v2.0 and \jams~%
\footnote{\url{http://www.gams.com/}.}
}
(a new version of the \emp~%
\footnote{\url{http://www.gams.com/dd/docs/solvers/emp.pdf}.}
 solver), s.t. both of
them allow for reformulating LGDP models by using either big-M (BM) or
convex-hull (CH) methods \cite{RamanGross94,SawayaGrossmann2012}.  We
use \cplex v12.2 \cite{cplex} (through an OSI/\cplex link) to solve
the reformulated MILP models. All the tools were executed using
default options, as suggested by the authors \cite{Vecchietti11}.
{We also compared \optmathsat against \mathsat 
augmented by Pseudo-Boolean (PB) optimization \cite{cimattifgss10} 
(we call it \pboptmathsat) on MaxSMT problems.}



\begin{remark}
\label{rem:infiniteprec}
Importantly, 
\mathsat and \optmathsat use {\em infinite-precision arithmetic}
whilst the \gams tools and \cplex implement
standard {\em floating-point arithmetic}.
Moreover
 the former handle strict inequalities natively  (see
 \sref{sec:background_smt}),  
whilst the \gams tools use
 an approximation with a very-small constant value ``{\tt eps}''
$\epsilon$
 (default $\epsilon\defas10^{-6}$), so that, e.g.,  
``{\em $(x>0)$ is internally rewritten into $(x \ge 10^{-6})$}''~%
\footnote{GAMS support team, email personal communication, 2012.}.
\end{remark}

The comparison is run on four distinct collections of benchmark problems:
\begin{itemize}
\item  (\sref{sec:expeval_lgdp}) LGDP problems, proposed by \logmip and
  \jams authors \cite{VecchGross04,SawayaG05,SawayaGrossmann2012};
\item (\sref{sec:expeval_smtlib}) \omlarat problems from SMT-LIB~%
\footnote{\url{http://www.smtlib.org/}.};
\item  (\sref{sec:expeval_sal}) \omlarat problems, coming from encoding 
parametric verification problems from the SAL~%
\footnote{\url{http://sal.csl.sri.com}.}
 model checker; 
\item (\sref{sec:expeval_pb}) the MaxSMT problems from \cite{cimattifgss10}.
\end{itemize}
\noindent
The encodings from LGDP to \omlarat and back are described in
\sref{sec:expeval_enc}.

%
\ignore{
\RSCHANGETHREE{We recently added to our comparison also a
  {\em parallel} version of \jamsBM, 
hereafter referred as \jamsBMmult{}, which runs \cplex in parallel
  mode on 4 cores.
The motivation, the description and the results of these tests will be
presented and discussed 
in \sref{sec:expeval_parallelgams}. 
(Hence, we invite the reader to ignore the \jamsBMmult{}
values in the tables until \sref{sec:expeval_parallelgams}.)
}
}

\smallskip
\RSCHANGETHREE{%
All tests were executed on two identical
2.66 GHz Xeon machines with 4GB RAM running Linux, 
using a timeout of 600 seconds for each run.  
In order to have a reliable and fair
measurement of CPU time, we have run only one process per PC at a time.
Overall, the evaluation consisted in $\approx40,000$ solver runs,
for a total CPU time of up to 276 CPU days.
}

The correctness of the minimum costs \minvalue found by \optmathsat have been 
cross-checked by another \smt solver, \yices~%
\footnote{\url{http://yices.csl.sri.com/}.}
by
checking the inconsistency within the bounds of 
$\varphi \wedge (\cost < \minvalue)$ and
the consistency of 
$\varphi \wedge (\cost = \minvalue)$ 
(if \minvalue is non-strict), or of 
$\varphi \wedge (\cost \le \minvalue)$ and
$\varphi \wedge (\cost = \minvalue+\epsilon)$  (if \minvalue is
strict), $\epsilon$ being some very small value.

All versions of \optimathsat passed the above checks.
On the LGDP problems (\sref{sec:expeval_lgdp}) 
all tools agreed on the final results, 
apart from tiny rounding errors by \gams tools;~%
\footnote{\gams+\cplex
often gives some errors $\le 10^{-5}$, which we believe are due to the printing
floating-point format: 
e.g., whilst \optmathsat reports the value 
{\tt 7728125177/2500000000} with infinite-precision arithmetic,
\gams+\cplex reports it as its floating-point approximation 
{\tt 3.091250e+00}.
}
on all the other problem collections (\sref{sec:expeval_smtlib},
\sref{sec:expeval_sal}, \sref{sec:expeval_pb}) instead,
the results of the \gams tools were affected by errors,
which we will discuss there. 


In order to make the experiments reproducible, 
{more detailed tables,}
the full-size
plots, a Linux binary of \optmathsat, the problems, and the results
are made available.~\footnote{\url{http://disi.unitn.it/~rseba/optimathsat2014.tgz}}
(We cannot distribute the \gams tools since
they are subject to licencing restrictions,  see \cite{gams}; 
however, they can be obtained at \gams url.
)

%
\subsection{Encodings.}
\label{sec:expeval_enc}
In order to translate LGDP models into 
\omlarat problems we use the encoding 
in \eqref{eq:ldgp_enc} of \sref{sec:optsmt_comparison},
namely \renc.~%
Notice that LGDP models are written in \gams language
which provides a large number of constructs.
Since our encoder supports only base constructs
(like equations and disjunctions),  
before generating the \renc encoding, 
we used the \gams Converter tool
for converting complex \gams specifications
(e.g. containing sets and indexed equations)
into simpler specifications.
\RSCHANGEONE{Notice also that in the \gams language the disjunction of
  constraints in \eqref{eq:ldgp_enc} must be described as nested
  if-then-elses on the Boolean propositions $Y^{jk}$, so that to avoid
  the need of including explicitly in $\phi$ the ``xor'' constraints
  discussed in the explanation of \eqref{eq:ldgp_enc}. Our encodings
  in both directions comply with this fact.  } 

In order to translate \omlarat problems into LGDP models we consider
two different encodings, namely \encone and \enctwo.
%

Since \gams tools do not handle negated equalities and strict
inequalities, with both encodings 
negated equalities $\neg (t_1 = t_2)$ or $(t_1 \neq t_2)$ in the input
\larat-formula   
$\varphi$ are first replaced by the disjunction of two
 inequalities $\neg (t_1 \le t_2) \vee \neg (t_1 \ge t_2)$) 
and strict inequalities $(t_1 < t_2)$ are rewritten as negated non-strict 
inequalities $\neg (t_1 \ge t_2)$.~%
\footnote{%
Here we implicitly assume
that the literals $\neg (t_1 = t_2)$, $(t_1 \neq t_2)$
 and $(t_1 < t_2)$ occur {\em positively} in
\vi; for negative occurrences the encoding is dual.%
}
Let $\varphi'$ be the \larat-formula obtained by $\varphi$ 
after these substitutions. 

%

{In \encone, which is inspired to the polarity-driven CNF
  conversion of  \cite{plaisted6},
we compute the Boolean abstraction ${\varphi'}^p$ of
$\varphi'$ (which plays the role of formula $\phi$ in  \eqref{eq:lgdp})
and then, for each \la-atom $\psi_i$ occurring positively
[resp. negatively] in $\varphi'$, we add the disjunction 
$\neg A_i \vee \psi_i$ [resp. $A_i \vee \neg \psi_i$], 
where $A_i$ is the Boolean atom of ${\varphi'}^p$ 
corresponding to the \la-atom $\psi_i$.%
}
\ignore{ 
\begin{eqnarray}
\textstyle
\label{eq:enc1}
\mbox{min } & \cost & \nonumber \\ 
\mbox{s.t. } & {\varphi'}^p & \nonumber\\
& \Bigr[
\begin{array}{c}
A_i \\
\psi_i  \\
\end{array} \Bigr] \vee \Bigr[
\begin{array}{c}
\mbox{Not } A_i \\
\mbox{Not } \psi_i  \\
\end{array} \Bigr] & \forall i\\
 & \cost \in \mathbb{R}, A_i \in \{True, False\}  \forall i & \nonumber
\end{eqnarray} 
} 
%
%

In \enctwo, 
first we compute the CNF-ization of $\varphi'$ 
using the \mathsatfive CNF-izer, 
and then we encode each non-unit clause 
$(l_{i1} \vee \ldots \vee l_{in}) \in \varphi'$ 
as a LGDP disjunction $[ Y^1_i \wedge l_{i1} ] \vee \ldots \vee [
Y^n_i \wedge l_{in}]$, where $Y^1_i, \ldots, Y^n_i$ are fresh Boolean
variables. 

%
\ignore{ 
\begin{eqnarray}
\textstyle
\label{eq:enc2}
\mbox{min } & \cost & \nonumber \\ 
\mbox{s.t. } 
& \Bigr[
\begin{array}{c}
Y^1_i \\
l_{i1}  \\
\end{array} \Bigr] \vee \ldots \vee \Bigr[
\begin{array}{c}
Y^n_i \\
l_{in}  \\
\end{array} \Bigr] & \forall i\\
 & \cost \in \mathbb{R}, Y_{ij} \in \{True, False\}  \forall i,j & \nonumber
\end{eqnarray} 
} 

\begin{remark}
 We decided to provide two different encodings for several reasons. 
\encone is a straightforward and very-natural encoding. 
However, we have verified empirically, and some discussion with \gams
support team confirmed it~%
\footnote{GAMS support team, email personal communication, 2012.},
that some \gams tools/options have
often problems in handling efficiently 
and even
correctly the
Boolean structure of the formulas $\phi$ in  \eqref{eq:lgdp}
 (see e.g. the number of problems terminated with error messages in
\sref{sec:expeval_smtlib}-\sref{sec:expeval_pb}). 
Thus, following also the suggestions of the \gams support team, 
we have introduced \enctwo, which eliminates any Boolean
structure, reducing the encoding substantially to a set of LGDP
disjunctions. 
Notice, however, that  \enctwo benefits from the CNF encoder
of \mathsatfive. 
\end{remark}


%
\subsection{Comparison on LGDP problems}
\label{sec:expeval_lgdp}

%
%
We have performed the first comparison over two distinct benchmarks, 
{\em strip-packing} and {\em zero-wait job-shop scheduling} problems,
which have been previously proposed as benchmarks for \logmip
and \jams by their authors
\cite{VecchGross04,SawayaG05,SawayaGrossmann2012}. 
%
We adopted the encoding of the problems into LGDP
given by the authors~\footnote{Examples are available at
\url{http://www.logmip.ceride.gov.ar/newer.html} and at
\url{http://www.gams.com/modlib/modlib.htm}.} 
and gave a corresponding \omlarat encoding. 
We refer to them as \dirgenerated benchmarks.
%

In order to make the results independent from the encoding used,
to investigate the correctness and effectiveness of the encodings
described in \sref{sec:expeval_enc}, 
and to check the robustness of the tools wrt. different encodings, 
we also generated formulas from \dirgenerated benchmarks
by applying the encodings \encone, \enctwo, and \renc;
we also applied the \encone/\enctwo and \renc encodings consecutively
to \smt formulas.
We refer to them as \encoded benchmarks.


 
\begin{figure}
\begin{tabular}{lll}
\scalebox{.5}{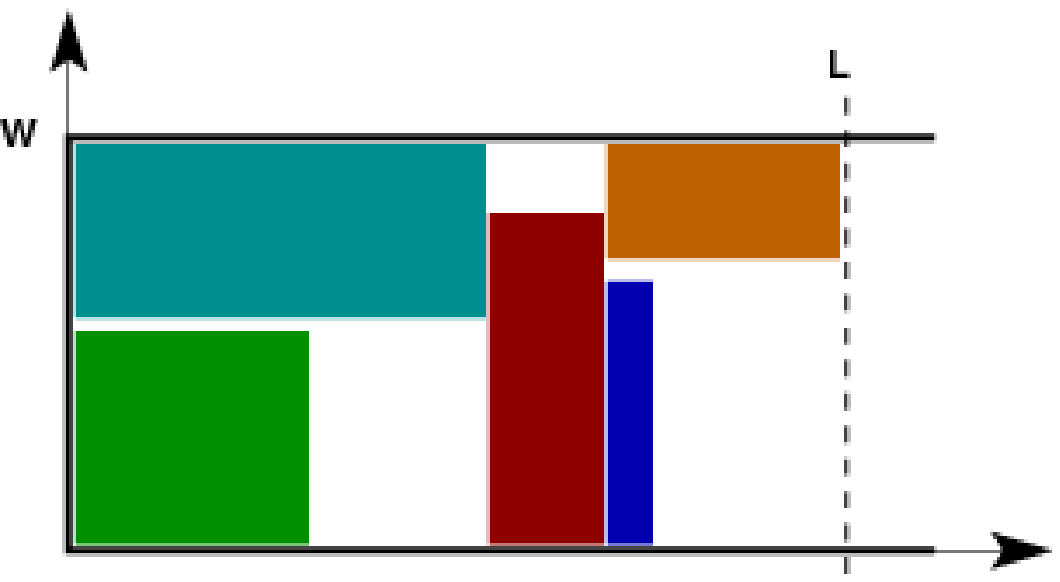} & &
\scalebox{.5}{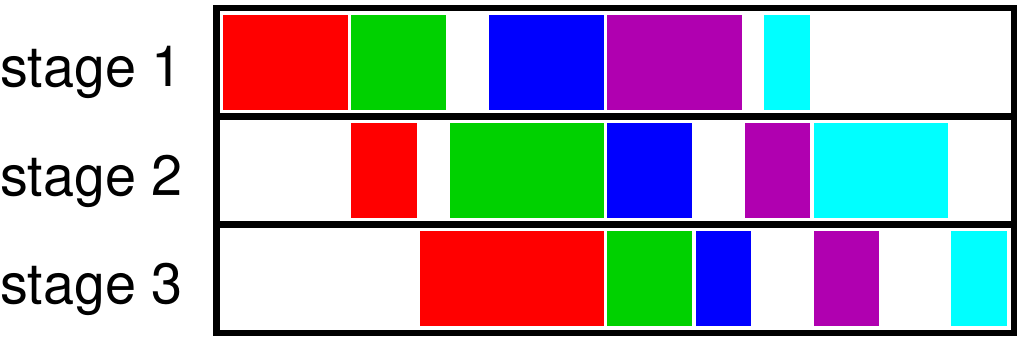} 
\end{tabular}
\caption{
\label{fig:lgdp}
Graphical representation of a strip-packing  (left) and 
of a zero-wait jobshop problem (right).
}
\end{figure}

\subsubsection{The strip-packing problem.}
\label{sec:expeval_sp}
\begin{figure}[t]
  \centering
{\setlength{\tabcolsep}{3.5pt}



{\scriptsize
\begin{tabular}{|l||r|r||r|r||r|r||r|r||r|r||r|r||r|r|}
\hline
\multirow{4}{*}{Procedure} & \multicolumn{14}{|c|}{Strip-packing} \\
\cline{2-15}
 & \multicolumn{6}{|c||}{$W=\sqrt{N}/2$} & \multicolumn{6}{|c||}{$W=1$} &
 \multicolumn{2}{|c|}{\multirow{2}{*}{Total}}\\
\cline{2-13}
  & \multicolumn{2}{|c||}{$N=9$} 
  & \multicolumn{2}{|c||}{$N=12$} 
  & \multicolumn{2}{|c||}{$N=15$} 
  & \multicolumn{2}{|c||}{$N=9$} 
  & \multicolumn{2}{|c||}{$N=12$} 
  & \multicolumn{2}{|c||}{$N=15$} 
  & \multicolumn{2}{|c|}{} \\

\cline{2-15}
  & \#s. & time 
  & \#s. & time 
  & \#s. & time 
  & \#s. & time 
  & \#s. & time 
  & \#s. & time 
  & \#s. & time \\
\hline \hline 

\multicolumn{15}{|c|}{Directly Generated Benchmarks} \\
\hline  
\shortoptlinof{} & 100 & 53 & 100 & 605 & 94 & 8160 & 100 & 749 & 89 & 3869 & 54 & 5547 & 537 & 18983\\
\hline 
\shortoptlinin{} & 100	& 12 & 100 & 144 & 100 & 3518 & 100 & 173 & 94 & 2127 & 74 & 6808 & \best{568} & \best{12782}\\
\hline 
\shortoptbinof{} & 100 & 50 & 100 & 625 & 89 & 8346 & 100 & 588 & 89 & 5253 & 45 & 5611 & 523 & 20473\\ 
\hline 
\shortoptbinin{} & 100 & 14 & 100 & 211 & 98 & 4880 & 100 & 202 & 94 & 2985 & 65 & 8101 & 557 & 16393\\
\hline
\shortoptadain{} & 100 & 13 & 100 & 192 & 99 & 5574 & 100 & 214 & 94 & 2675 & 63 & 7949 & 556 & 16617\\
\hline \hline
\shortjamsBM & 100 & 230 & 78 & 10177 & 12 & 1180 & 100 & 158 & 91 & 3878 & 51 & 6695 & \best{432} & \best{22318} \\
\hline 
\shortjamsCH & 100 & 2854 & 27 & 2393 & 1 & 417 & 100 & 1906 & 70 & 7471 & 17 & 4032 & 315 & 19073\\
\hline 
\shortlogmipBM & 100 & 229 & 78 & 10159 & 12 & 1192 & 100 & 157 & 91 & 3866 & 51 & 6720 & \best{432} & \best{22323} \\
\hline 
\shortlogmipCH & 100 & 2851 & 27 & 2414 & 1 & 424 & 100 & 1907 & 70 & 7440 & 17 & 4037 & 315 & 19073\\
\ignore{
\hline \hline 
 {\RSCHANGETHREE \shortjamsBMmult} & 100 & 99 & 92 & 7344 & 17 & 1730
 & 100 & 50 & 96 & 1992 & 67 & 6703 & \best{472} & \best{17918} \\
}
 \hline \hline 
\multicolumn{15}{|c|}{ \renc Encoded Benchmarks} \\
\hline  
\shortoptlinin{} & 100 & 12 & 100 & 144 & 100 & 3563 & 100 & 183 & 94 & 2169 & 73 & 6466 & 567 & 12537 \\
\hline \hline

\multicolumn{15}{|c|}{ \encone-\renc Encoded Benchmarks} \\
\hline 
\shortoptlinin{} & 100 & 13 & 100 & 166 & 100 & 5919 & 100 & 195 & 94 & 2156 & 74 & 7080 & 568 & 15529\\
\hline \hline

\multicolumn{15}{|c|}{ \enctwo-\renc Encoded Benchmarks} \\
\hline 
\shortoptlinin{} & 100 & 13 & 100 & 141 & 100 & 5574 & 100 & 172 & 94 & 2148 & 74 & 6650 & 568 & 12618\\
\hline \hline 

\multicolumn{15}{|c|}{ \encone Encoded Benchmarks} \\
\hline 
\shortjamsBM & 100 & 389 & 68 & 8733 & 12 & 1934 & 100 & 162 & 89 & 5565 & 47 & 7313 & \best{416} & \best{24096}\\
\hline 
\shortjamsCH & 99 & 980 & 46 & 6099 & 2 & 769 & 100 & 726 & 72 & 7454 & 17 & 3505 & 336 & 19533 \\
\hline 
\shortlogmipBM & 100 & 390 & 68 & 8723 & 12 & 1946 & 100 & 163 & 89 & 5547 & 47 & 7299 & \best{416} & \best{24068}\\
\hline 
\shortlogmipCH & 99 & 981 & 54 & 5480 & 12 & 735 & 100 & 725 & 74 & 7433 & 17 & 3542 & 346 & 18896 \\
\hline \hline

\multicolumn{15}{|c|}{ \enctwo Encoded Benchmarks} \\
\hline  
\shortjamsBM & 100 & 190 & 81 & 8460 & 11 & 2066 & 100 & 159 & 89 & 2960 & 56 & 8142 & 437 & 21977\\
\hline 
\shortjamsCH & 98 & 3799 & 24 & 2137 & 1 & 292 & 100 & 2402 & 68 & 7926 & 16 & 3429 & 307 & 19985\\
\hline 
\shortlogmipBM & 100 & 191 & 81 & 8462 & 11 & 2071 & 100 & 159 & 90 & 2964 & 56 & 8206 & \best{438} & \best{22053} \\
\hline 
\shortlogmipCH & 98 & 3807 & 24 & 2133 & 1 & 312 & 100 & 2388 & 68 & 7915 & 17 & 4027 & 308 & 20582\\
\hline 
\end{tabular}
}



} 
{\setlength{\tabcolsep}{2pt}
  \begin{tabular}{ccc}
%
%
\includegraphics[width=0.32\linewidth,bb=98 50 355 298]{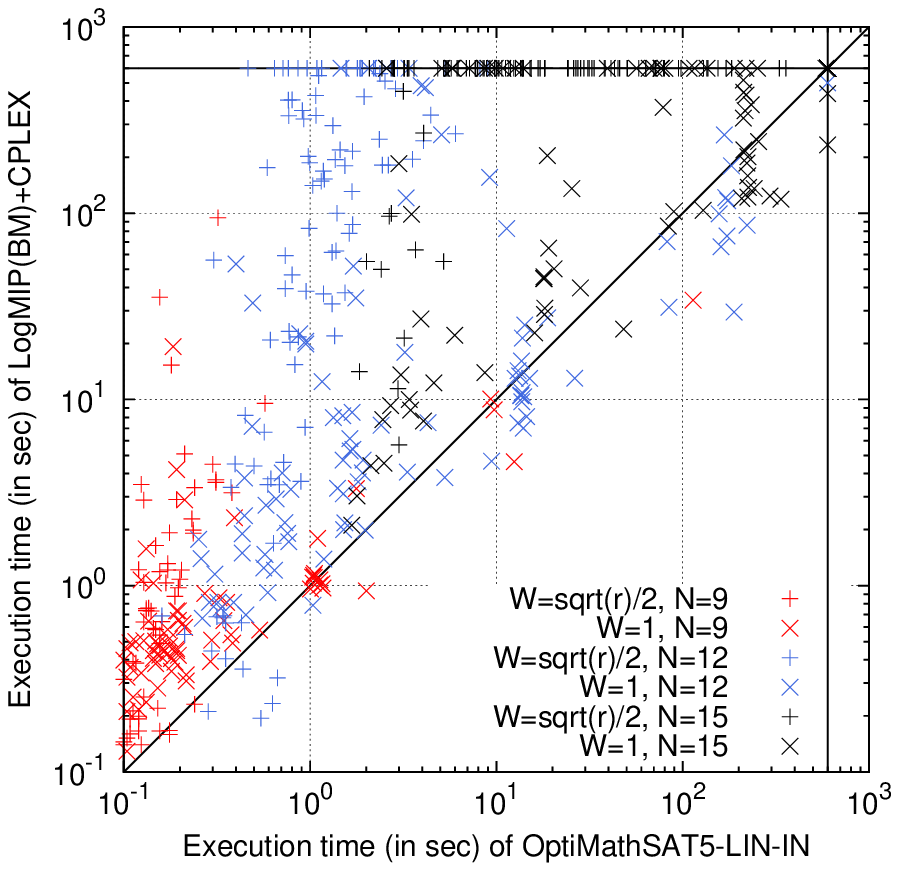} &
\includegraphics[width=0.32\linewidth,bb=98 50 355 298]{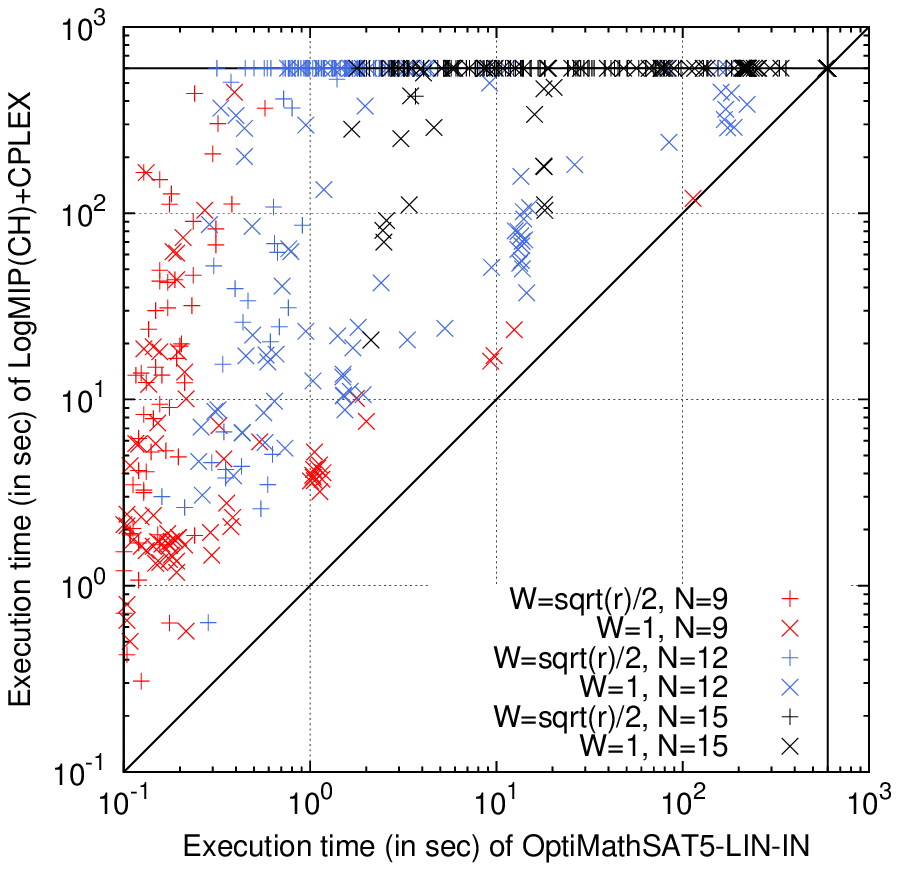} &
\includegraphics[width=0.32\linewidth,bb=98 50 355 298]{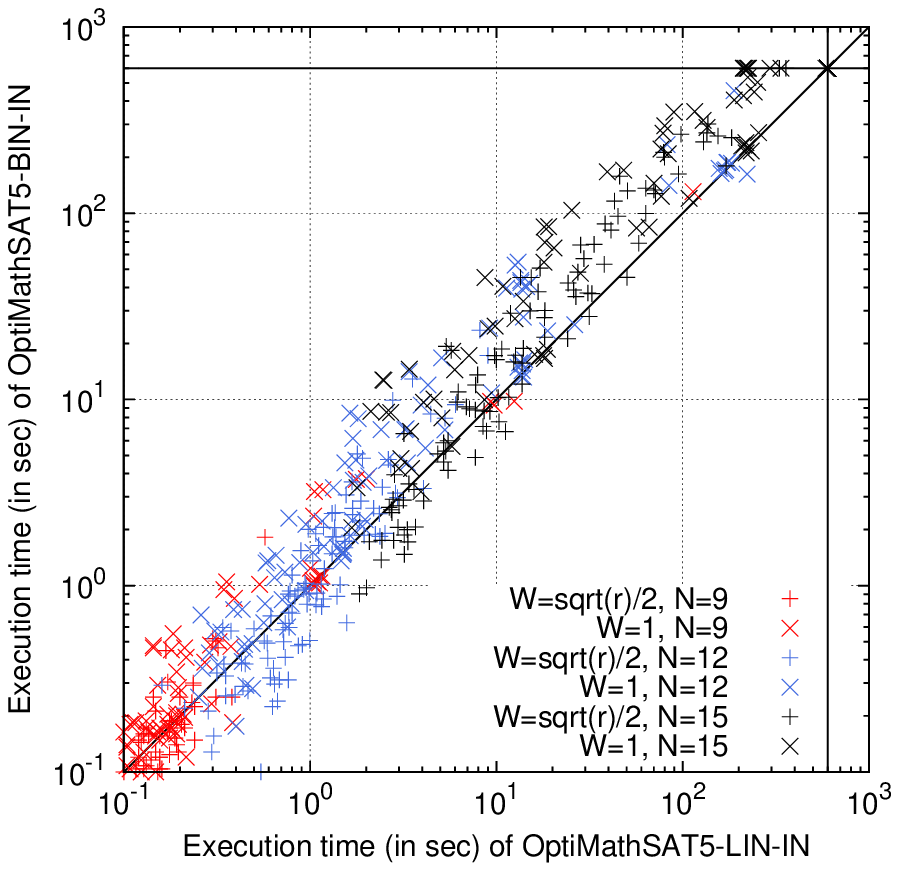}\\ 
 \end{tabular} 
 }
 \caption{Table: results (\# of solved instances, cumulative time in
   seconds for solved instances) for \optmathsat and \gams (using \logmip and \jams) on 100 random instances (including \dirgenerated and \encoded benchmarks) each of the strip-packing problem
   for $N$ rectangles, where $N=9,12,15$, and width
   $W=\sqrt{N}/2$,~$1$. 
(We use ``OM'' as shortcut for \optmathsat and omit ``+\cplex'' 
in the labels of \gams tools.)
%
Values highlighted in \best{bold} represent best performances.
Scatter-plots: comparison of the best configuration of \optmathsat (\optlinin{})
against \logmipBM (left), \logmipCH (center) and \optbinin{} (right)
on \dirgenerated benchmarks.
}
\label{fig:plots_sp}
\end{figure}


Given a set of $N$ rectangles 
of different length $L_i$ and height $H_i$, $i \in 1,..,N$,
and a strip of fixed width $W$ but unlimited length, 
the \emph{strip-packing} problem aims at
minimizing the  length $L$ of the filled part of the strip 
while filling the strip with
all rectangles, without any overlap and any rotation.
\ignoreinlong{
We considered the LGDP model provided by \cite{SawayaG05}.
}
{(See Figure~\ref{fig:lgdp} left.)}

\ignoreinshort{
The LGDP model provided by \cite{SawayaG05} is the following:
%
%
%
\begin{eqnarray}
\textstyle
\label{eq:lgdp_sp}
\mbox{min } & L  & \nonumber \\ 
\mbox{s.t. } & L  \ge x_i + L_i & \forall i \in N \nonumber\\
& \Bigr[
\begin{array}{c}
Y^1_{ij} \\
x_i+L_i \le x_j \\
\end{array} \Bigr] \vee
\Bigr[
\begin{array}{c}
Y^2_{ij} \\
x_j+L_j \le x_i \\
\end{array} \Bigr] & \\
& \vee \Bigr[
\begin{array}{c}
Y^3_{ij} \\
y_i-H_i \ge y_j \\
\end{array} \Bigr]  \vee \Bigr[
\begin{array}{c}
Y^4_{ij} \\
y_j-H_j \ge y_i \\
\end{array} \Bigr] & \forall i,j \in N, i< j \nonumber\\
& x_i \le \ub - L_i \mbox{  } & \forall i \in N \nonumber\\
& H_i \le y_i \le W \mbox{  } & \forall i \in N \nonumber\\
& L, x_i,y_i \in \mathbb{R}^1_+, Y_{ij}^1, Y_{ij}^2, Y_{ij}^3, Y_{ij}^4 \in \{True, False\} & \nonumber
\end{eqnarray}
where $L$ corresponds to the objective function to minimize
and every rectangle $j \in J$ is represented by the 
constants $L_j$ and $H_j$ (length and height respectively) and 
the variables  $x_j,y_j$ (the coordinates
of the upper left corner in the 2-dimensional space).
Every pair of rectangles $i,j \in N, i < j$
is constrained by a disjunction that
avoids their overlapping (each disjunct represents the 
position of rectangle $i$ in relation to rectangle $j$).
The size of the strip limits the position of each rectangle $j$:
the width of the strip $W$ and the upper bound \ub on the
optimal solution bound the $y_j$-coordinate and the height $H_j$
bounds the $x_j$-coordinate.
We express straightforwardly the LGDP model (\ref{eq:lgdp_sp}) 
into \omlarat as follows:
\begin{eqnarray}
\textstyle
\label{eq:sp_enc}
\begin{array}{lll}
\vi & \defas & (\cost = L) \wedge \bigwedge_{ i \in N} (L \ge x_i + L_i)\\ 
    & \wedge & \bigwedge_{ i,j \in N, i<j} \Bigl(
(x_i + L_i \le x_j) \vee 
(x_j + L_j \le x_i) \\
    &             & \vee (y_i - H_i \ge y_j) \vee  (y_j -H_j \ge y_i) \Bigr)\\
    & \wedge & \bigwedge_{ i \in N} (x_i \le \ub - L_i) \wedge \bigwedge_{ i \in N} (x_i \ge 0) \\
    & \wedge & \bigwedge_{ i \in N} (H_i \le y_i) \wedge \bigwedge_{ i \in N} (W \ge y_i) \wedge \bigwedge_{ i \in N} (y_i \ge 0) \\
\end{array}
\end{eqnarray}
} 

We randomly generated 
instances of the strip-packing problem
according to a fixed width $W$ 
of the strip and a fixed number of rectangles $N$. 
For each rectangle $j \in N$, length $L_j$ 
and height $H_j$ are selected in the interval $]0,1]$ uniformly at random.
The upper bound $\ub$ is computed with the same heuristic 
used by \cite{SawayaG05}, which sorts the 
rectangles in non-increasing order of width and fills the strip by
placing each rectangles in the bottom-left corner,
and the lower bound \lb is set to zero.
We generated 100 samples each for $9$, $10$ and $11$ rectangles 
and for two values 
of the width $\sqrt{N}/2$ and $1$ 
(Notice that with $W=\sqrt{N}/2$ the filled strip looks approximatively
like a square, whilst $W=1$ is half the average size of one rectangle.
)
%


\ignore{
The table of Figure \ref{fig:plots_sp} shows the number of solved
instances and their cumulative execution time
for different configurations of \optmathsat and \gams on 
\dirgenerated and \encoded benchmarks.
%
The scatter-plots of Figure \ref{fig:plots_sp} 
compare the best-performing version of \optmathsat, \optlinin{},
against \logmipG with BM and CH reformulation 
(left and center respectively)
and the two inline versions 
\optlinin{} and \optbinin{} (right) on \dirgenerated benchmarks.
}

\subsubsection{The zero-wait jobshop problem.}
\label{sec:expeval_js}\begin{figure}[t]
  \centering
{\setlength{\tabcolsep}{2.75pt}



{\scriptsize 
\begin{tabular}{|l||r|r||r|r||r|r||r|r||r|r||r|r||r|r|}
\hline
\multirow{3}{*}{Procedure} & \multicolumn{14}{|c|}{Job-shop} \\
\cline{2-15}
  & \multicolumn{2}{|c||}{$I=9,$} 
  & \multicolumn{2}{|c||}{$I=10,$}
  & \multicolumn{2}{|c||}{$I=11,$} 
  & \multicolumn{2}{|c||}{$I=12,$}
  & \multicolumn{2}{|c||}{$I=11,$}
  & \multicolumn{2}{|c||}{$I=11,$} 
  & \multicolumn{2}{|c|}{\multirow{2}{*}{Total}} \\
  & \multicolumn{2}{|c||}{$J=8$} 
  & \multicolumn{2}{|c||}{$J=8$}
  & \multicolumn{2}{|c||}{$J=8$} 
  & \multicolumn{2}{|c||}{$J=8$} 
  & \multicolumn{2}{|c||}{$J=9$} 
  & \multicolumn{2}{|c||}{$J=10$}
  & \multicolumn{2}{|c|}{}\\
\cline{2-15}
  & \#s. & time 
  & \#s. & time 
  & \#s. & time 
  & \#s. & time 
  & \#s. & time 
  & \#s. & time 
  & \#s. & time \\
\hline \hline

\multicolumn{15}{|c|}{Directly Generated Benchmarks} \\
\hline
\shortoptlinof{5} & 100 & 386 & 100 & 1854 & 97 & 9396 & 57 & 14051 & 100 & 9637 & 99 & 10670 & 553 & 45995\\
\hline 
\shortoptlinin{5} & 100  & 317 & 100 & 1584 & 100 & 8100 & 77 & 18046 & 100 &7738 & 100 & 7433 & \best{577} & \best{43228}\\ 
\hline
\shortoptbinof{5} & 100 & 726 & 100 & 3817 & 88 & 13222 & 38 & 12529 & 92 & 14183 & 90 & 13287 & 508 & 57764\\ 
\hline
\shortoptbinin{5} & 100 & 602 & 100 & 3270 & 97 & 12878 & 54 & 16234 &  96 & 13159 & 96 & 12350 & 543 & 58493\\
\hline
\shortoptadain{5} & 100 & 596 & 100 & 3230 & 97 & 12262 & 53 & 14810 & 96 & 12805 & 96 & 12125 & 542 & 55828\\
\hline \hline

\shortjamsBM & 100 & 268 & 100 & 1113 & 100 & 4734 & 87 & 17067 & 100 & 4941 & 100 & 6122 & \best{587} & \best{34245} \\ 
\hline
\shortjamsCH & 84 & 23830 & 4 & 1596 & 0 & 0 & 0 & 0 & 0 & 0 & 1 & 363 & 89 & 25789 \\
\hline
\shortlogmipBM & 100 & 267 & 100 & 1114 & 100 & 4718 & 87 & 17108 & 100 & 4962 & 100 & 6174 & \best{587} & \best{34343} \\
\hline 
\shortlogmipCH & 84 & 23871 & 4 & 1622 & 0 & 0 & 0 & 0 & 0 & 0 & 1 & 338 & 89 & 25831 \\
\hline \hline

\ignore{
 {\RSCHANGETHREE
 \shortjamsBMmult} & 100 & 128 & 100 & 343 & 100 & 1196 & 100 & 6973 & 100 & 1262 & 100 & 1478 & \best{600} & \best{9904} \\
 \hline \hline
}
\multicolumn{15}{|c|}{ \renc Encoded Benchmarks} \\
\hline  
\shortoptlinin{5} & 100 & 324 & 100 & 1571 & 100 & 7739 & 74 & 16494 & 100 & 7175 & 100 & 7504 & 574 & 40807 \\ 
\hline \hline

\multicolumn{15}{|c|}{ \encone-\renc Encoded Benchmarks} \\
\hline
\shortoptlinin{5} & 100 & 336 & 100 & 1578 & 100 & 7762 & 71 & 16589 & 100 & 7726 & 100 & 7706 & 571 & 41697\\ 
\hline \hline

\multicolumn{15}{|c|}{ \enctwo-\renc Encoded Benchmarks} \\
\hline 
\shortoptlinin{5} & 100 & 320 & 100 & 1533 & 100 & 7623 & 68 & 15120 & 100 & 7216 & 100 & 7598 & 568 & 39410 \\
\hline \hline

\multicolumn{15}{|c|}{ \encone Encoded Benchmarks} \\
\hline 
\shortjamsBM & 100 & 239 & 100 & 1128 & 100 & 5516 & 84 & 19949 & 100 & 6667 & 100 & 4176 & \best{584} & \best{37675} \\
\hline
\shortjamsCH & 100 & 14527 & 46 & 17887 & 0 & 0 & 0 & 0 & 1 & 497 & 0 & 0 & 147 & 32911 \\
\hline
\shortlogmipBM & 100 & 240 & 100 & 1122 & 100 & 5510 & 83 & 19489 &  100 & 6684 & 100 & 4180 & 583 & 37225 \\
\hline 
\shortlogmipCH & 100 & 14465 & 47 & 18206 & 0 & 0 & 0 & 0 & 1 & 495 & 0 & 0 & 148 & 33166 \\
\hline \hline

\multicolumn{15}{|c|}{ \enctwo Encoded Benchmarks} \\
\hline  
\shortjamsBM & 100 & 319 & 100 & 1865 & 100 & 12470 & 45 & 15704 & 97 & 13189 & 96 & 15773 & \best{538} & \best{59320} \\ 
\hline
\shortjamsCH & 95 & 22435 & 18 & 8030 & 2 & 671 & 0 & 0 & 1 & 526 & 3 & 1043 & 119 & 32723 \\
\hline
\shortlogmipBM & 100 & 319 & 100 & 1871 & 100 & 12440 & 45 & 15747 & 98 & 13661 & 95 & 15102  & \best{538} & \best{59140} \\
\hline 
\shortlogmipCH & 95 & 22401 & 18 & 7991 & 1 & 163 & 0 & 0 & 1 & 437 & 3 & 1020  & 118 & 32012 \\
\hline
\end{tabular}
}


}
{\setlength{\tabcolsep}{2pt}
  \begin{tabular}{ccc}
\includegraphics[width=.32\linewidth,bb=98 50 355 298]{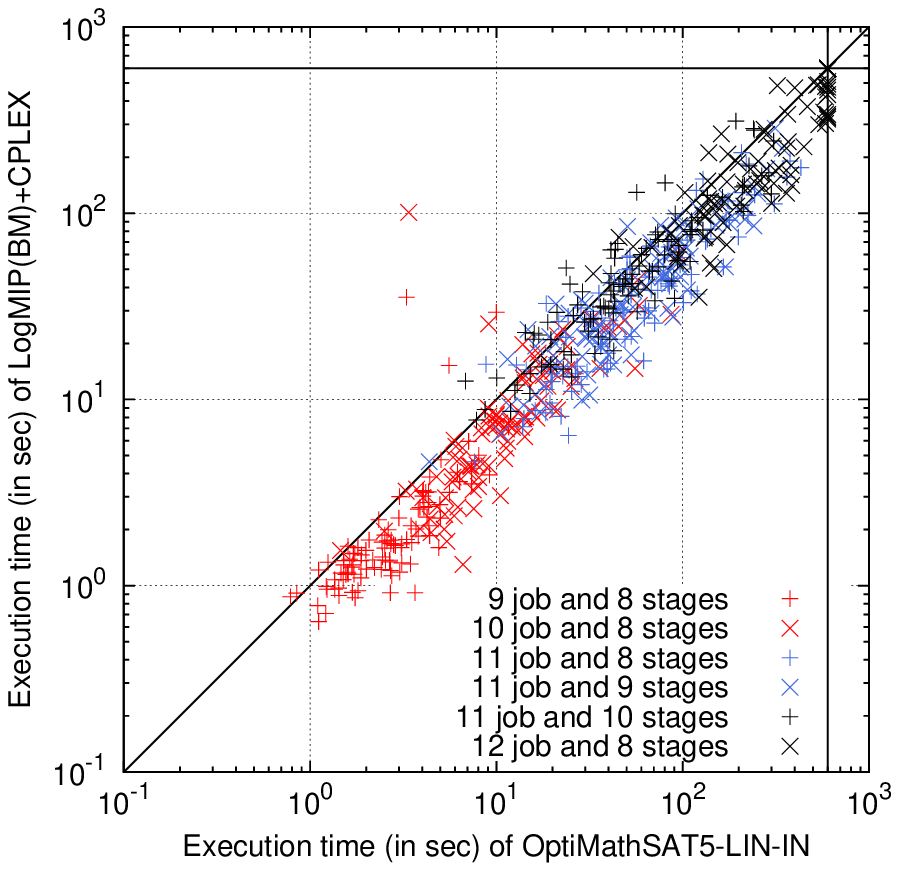} &
\includegraphics[width=.32\linewidth,bb=98 50 355 298]{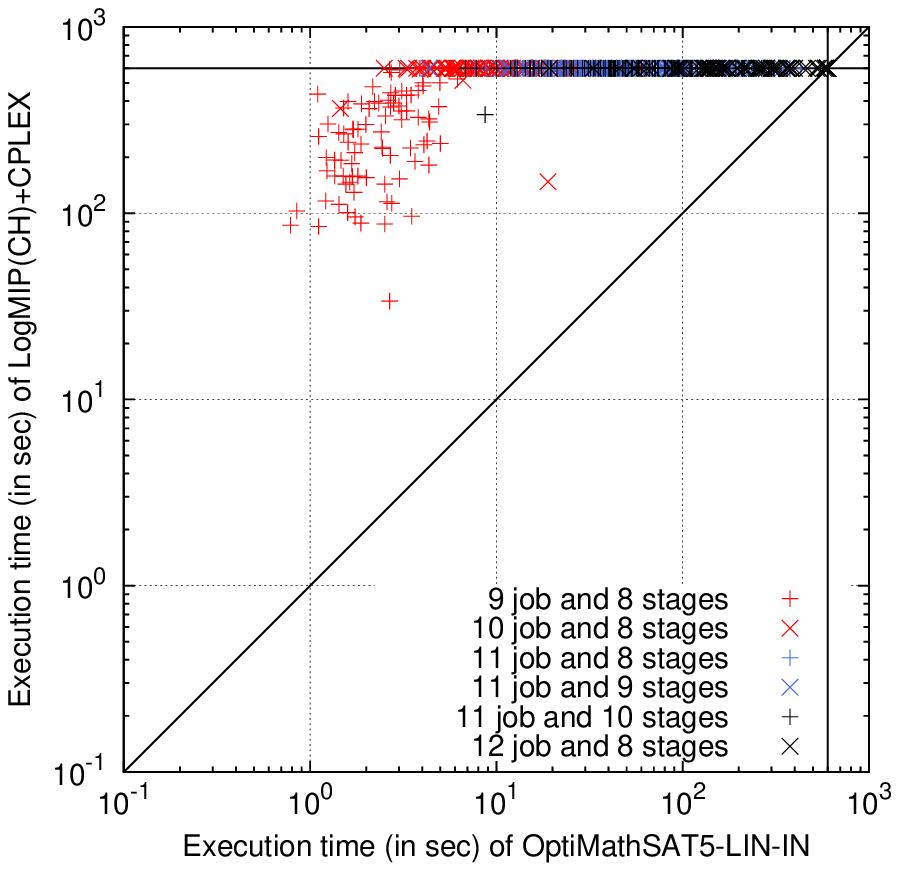} &
\includegraphics[width=.32\linewidth,bb=98 50 355 298]{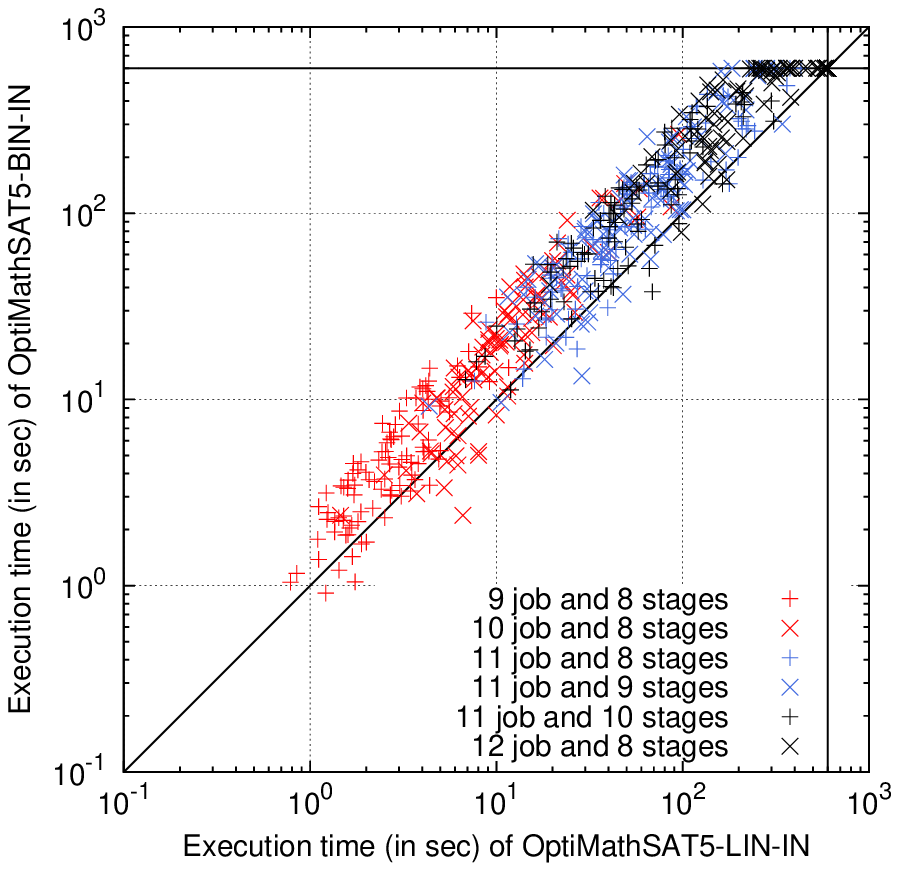} \\
\end{tabular}
}
\caption{Table: results (\# of solved instances, cumulative time in
   seconds for solved instances) for \optmathsat and \gams 
   on 100 random samples (including \dirgenerated and
   \encoded benchmarks) each of the job-shop problem 
   for $I=9,10,11,12$ jobs and $J=8$ stages and for 
   $I=11$ jobs and $J=9,10$ stage.
(We use ``OM'' as shortcut for OptiMathSAT and omit ``+\cplex'' 
in the labels of \gams tools.) 
Values highlighted in \best{bold} represent best performances.
Scatter-plots: comparison of the best configuration of \optmathsat (\optlinin{})
against \logmipBM (left), \logmipCH (center) and 
\optbinin{} (right) on \dirgenerated benchmarks.
\label{fig:plots_js}
}  
\end{figure}


Consider the scenario where there is a set $I$ of jobs
which must be scheduled sequentially on a set $J$ of consecutive stages 
with zero-wait transfer between them.
Each job $i \in I$  has a start time $s_i$ and 
a processing time $t_{ij}$ in the stage $j \in J_i$, $J_i$ being 
the set of stages of job $i$.
The goal of the \emph{zero-wait job-shop scheduling} problem 
is to minimize the makespan,
that is the total length of the schedule.
{(See Figure~\ref{fig:lgdp} right.)}

\ignoreinshort{
The LGDP model provided by \cite{SawayaG05} is:
\begin{eqnarray}
\textstyle
\label{eq:lgdp_js}
\mbox{min } & M & \nonumber \\ 
\mbox{s.t. } & M \ge \sum_{ j \in J_i} t_{ij} & \forall i \in I \nonumber \\
& \Bigr[
\begin{array}{c}
Y^1_{ik} \\
s_i +\sum_{ m \in J_i, m\le j} t_{im}  \le s_k +\sum_{ m \in J_k, m < j} t_{km}  \\
\end{array} \Bigr] \vee &  \\
& \Bigr[
\begin{array}{c}
Y^2_{ik} \\
s_k +\sum_{ m \in J_k, m \le j} t_{km}  \le s_i +\sum_{ m \in J_i, m < j} t_{im}  \\
\end{array} \Bigr] & \forall j \in C_{ik}, \forall i,k \in I, i<k \nonumber\\
 & M, s_i \in \mathbb{R}_+^1, Y^1_{ik}, Y^2_{ik} \in \{True, False\} & \forall i,k \in I, i<k \nonumber
\end{eqnarray}
where $M$ corresponds to the objective function to minimize
and every job $i \in I$ is represented by 
the variable $s_i$ (its start time) and 
the constant $t_{ij}$ (its processing time in stage $j \in J_i$).
For each pair of jobs $i,k \in I$ and for each stage $j$ 
with potential clashes (i.e $j \in  C_{ik} = \{J_i \cap J_k\}$), 
a disjunction ensures that no clash between jobs occur 
at any stage at the same time.
We encoded the corresponding LGDP model (\ref{eq:lgdp_js}) 
into \omlarat as follows: 
\begin{eqnarray}
\textstyle
\label{eq:js_enc}
\begin{array}{lll}
\vi & \defas & (\cost = M) \wedge \bigwedge_{i \in I} (M \ge s_i + \sum_{ j \in J_i} t_{ij})
\wedge \bigwedge_{i \in I} (s_i \ge 0) \\
 & \wedge & \bigwedge_{ j \in C_{ik}, \forall i,k \in I, i<k }
\left( (s_i + \sum_{ m \in J_i, m \le j} t_{im} \le s_k + \sum_{ m \in J_k, m < j} t_{km}) \right. \\
& & \hspace{3cm} \vee \left. (s_k + \sum_{ m \in J_k, m \le j} t_{km} \le s_i + \sum_{ m \in J_i, m < j} t_{im}) \right)\\
\end{array}
-\end{eqnarray}
} 

We generated randomly instances of the zero-wait
jobshop problem according to a fixed number of jobs $I$ 
and a fixed number of stages $J$. For each job $i \in I$, 
start time $s_i$ and processing time $t_{ij}$ of every job
are selected in the interval $]0,1]$ uniformly at random. 
We consider a set of 100 samples each 
for 9, 10, 11, 12 jobs and 8 stages and for 11 jobs and 9, 10 stages.
We set no value for \ub and $\lb=0$. 

\ignore{
The table of Figure \ref{fig:plots_js} shows the number of solved
instances and their cumulative execution time
for different configurations of \optmathsat and \gams on 
\dirgenerated and \encoded benchmarks.
%
%
The scatter-plots of Figure \ref{fig:plots_js} compare, 
on ``directly encoded'' benchmarks,
the best-performing version of \optmathsat, \optlinin{},
against \logmip with BM and CH reformulation (left and center respectively);
the figure also compares the two inline versions \optlinin{} and
\optbinin{} (right). 
}


\subsubsection{Discussion}
\label{sec:expeval_disc}
The table of Figure \ref{fig:plots_sp} shows the number of solved
instances and their cumulative execution time
for different configurations of \optmathsat and \gams on 
\dirgenerated and \encoded benchmarks.
%
The scatter-plots of Figure \ref{fig:plots_sp} 
compare the best-performing version of \optmathsat, \optlinin{},
against \logmipG with BM and CH reformulation 
(left and center respectively)
and the two inline versions 
\optlinin{} and \optbinin{} (right) on \dirgenerated benchmarks.

The table of Figure \ref{fig:plots_js} shows the number of solved
instances and their cumulative execution time
for different configurations of \optmathsat and \gams on 
\dirgenerated and \encoded benchmarks.
%
%
The scatter-plots of Figure \ref{fig:plots_js} compare, 
on ``directly encoded'' benchmarks,
the best-performing version of \optmathsat, \optlinin{},
against \logmip with BM and CH reformulation (left and center respectively);
the figure also compares the two inline versions \optlinin{} and
\optbinin{} (right). 

The results on the LGDP problems 
in Figures~\ref{fig:plots_sp}, \ref{fig:plots_js} suggest
some considerations.

Comparing the different versions of \optmathsat, 
we notice that:
\begin{itemize}
\item the inline versions
 (\inl) behave pairwise uniformly better than the
  corresponding offline versions (\of), which is not surprising;
\item overall  the \lin options  
seems to perform a little better than
  the corresponding \bin and \ada options
(although gaps are not dramatic).
\end{itemize}

\begin{remark}
%
We notice that with LGDP problems 
  binary search is not
  ``obviously faster'' than linear search,
in compliance with what stated in
point~\ref{item:binvslin}. in \sref{sec:algorithms_offline}. 
%
This is further enforced by the fact that 
in strip-packing \eqref{eq:sp_enc} [resp. job-shop \eqref{eq:js_enc}]
encodings, 
the cost variables $\cost\defas L$ [resp. $\cost\defas M$]
 occurs only in positive unit clauses in the form 
$(L \ge \tuple{term})$ [resp. $(M \ge \tuple{term})$]; 
thus, learning $\neg (\cost < \pivot)$ as a result of 
the binary-search steps with  \unsatres results
produces no constraining effect on the variables in $\tuple{term}$, 
and hence no substantial extra search-pruning effect 
due to the early-pruning technique of the SMT solver.
\label{remark:binvslin}
\end{remark}

Comparing the different versions of the \gams tools, we see that 
\logmip and \jams reformulations lead to substantially identical 
performance on both strip-packing and job-shop instances.
For both reformulation tools, the BM versions uniformly outperform
the CH ones, often dramatically.
%

Comparing the performances of the versions of \optmathsat against
these of the \gams tools, 
we notice that 
\begin{itemize}
\item on {\em strip-packing problems} all versions of \optmathsat
outperform all \gams versions, regardless of the encoding used.
E.g., the best \optmathsat version solved  $\approx 30\%$ 
more formulas than the best \gams version;

\item on {\em job-shop problems}  results are mixed.
 \optmathsat drastically outperforms the CH versions 
on all encodings and it slightly beats 
the BM ones on \twoencoded benchmarks,
whilst it is slightly beaten by the BM versions
on \dirgenerated and \oneencoded benchmarks.
E.g., the best \optmathsat version solved  $\approx 2\%$ 
less formulas than the best \gams version.

\end{itemize}
\noindent
Overall, we can conclude that \optmathsat performances on these
problems are comparable with, and most often significantly
better than, those of \gams tools.


\smallskip
We may wonder how these results are affected by the different
encodings used. 
(We recall from the beginning of \sref{sec:expeval} that 
all solvers agreed on the results, regardless of the encoding.)
In terms of performances, comparing the effects of the different encodings, 
we notice the following facts.

\begin{itemize}
\item On \optmathsat (\linin) the effects of the different encodings
  is substantially negligible, on both strip-packing and job-shop
  problems, since we have only very small 
  variations in the number of solved instances between \dirgenerated
  and \encoded instances, in the various encoding combinations.
  From this reason, we conclude that \optmathsat is robust wrt. the
  encodings of these problems.

\item On \gams tools the effects of the different encodings 
are more relevant, although very heterogeneous:
e.g., wrt. to \dirgenerated instances, \oneencoded 
solved formulas are slightly less 
with BM options, 
and up to much more 
with CH options;
\twoencoded solved formulas are slightly more  
on strip-packing and a little less on job-shop
with BM options, 
slightly less
on strip-packing and a much more on job-shop
with CH options. 
For this reason, in next sections we always report the results with
both encodings.
\end{itemize}

%
\begin{figure}[t]
  \centering
{\setlength{\tabcolsep}{2pt}
\begin{tabular}{ccc}
{\em solving/total time  ratio}& 
{\em minimization/total time ratio}& 
{\em certification/total time  ratio}\\
\includegraphics[width=.32\linewidth,bb=98 50 355 298]{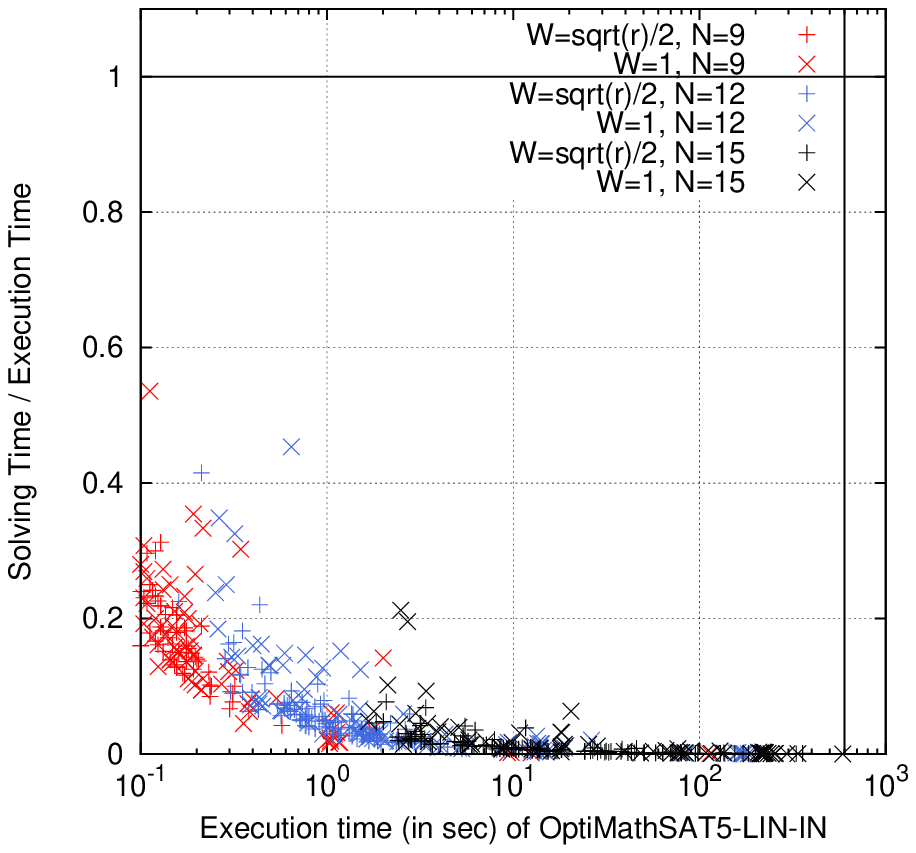} &
\includegraphics[width=.32\linewidth,bb=98 50 355 298]{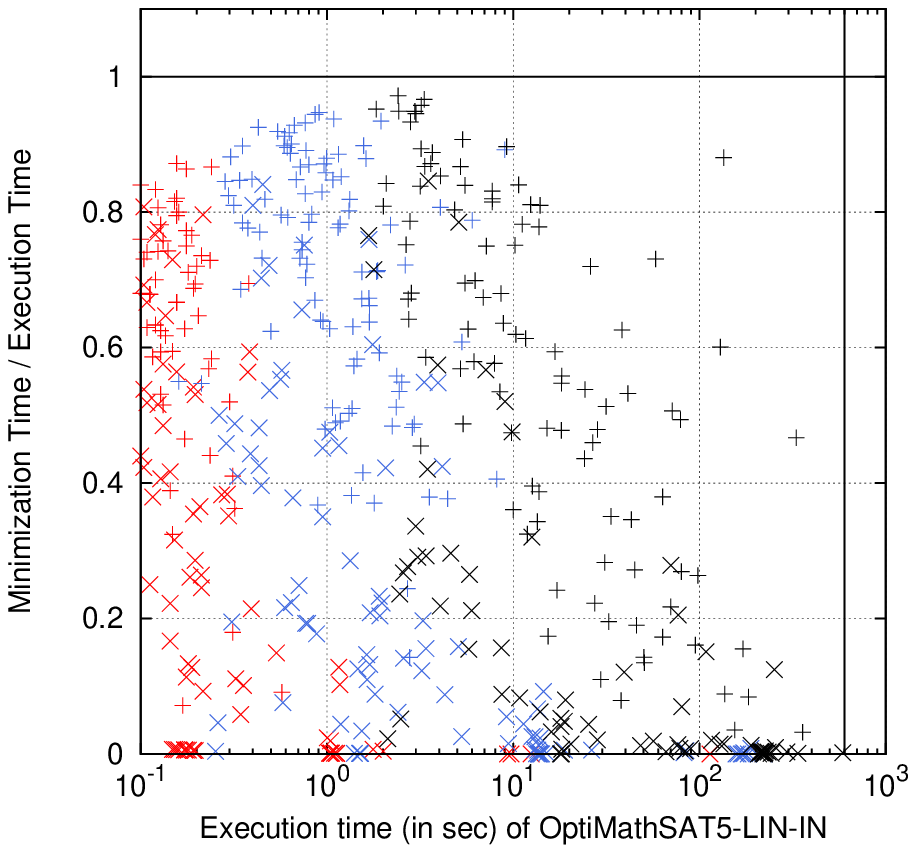} &
\includegraphics[width=.32\linewidth,bb=98 50 355 298]{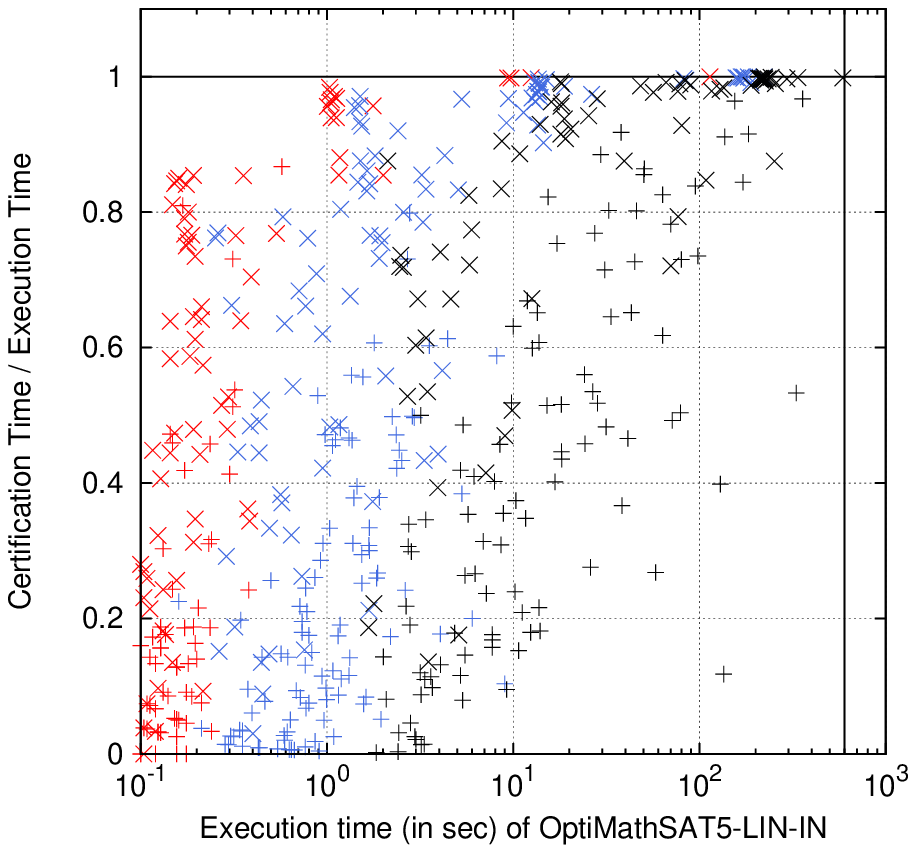}\\ 
\includegraphics[width=.32\linewidth,bb=98 50 355 298]{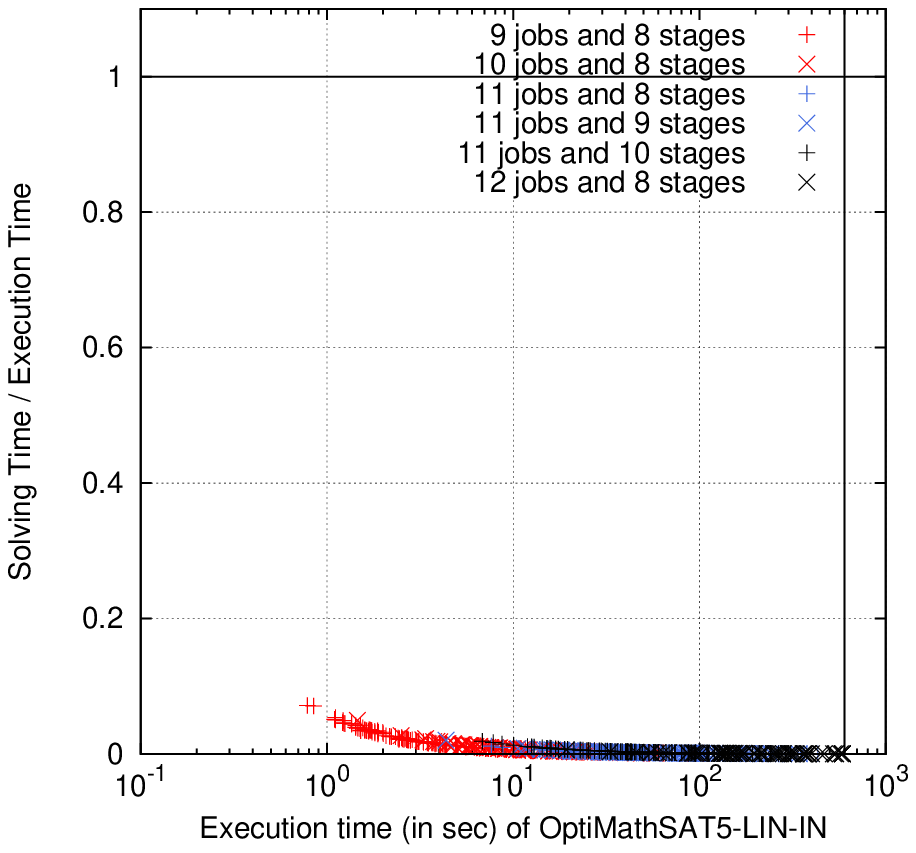} &
\includegraphics[width=.32\linewidth,bb=98 50 355 298]{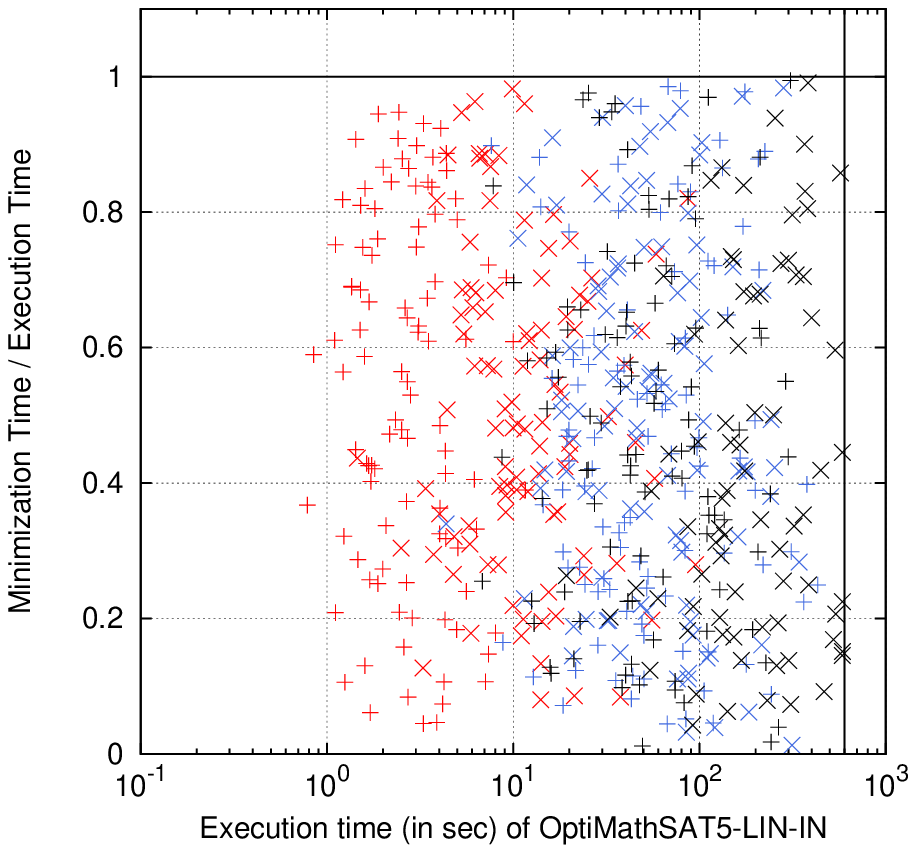} &
\includegraphics[width=.32\linewidth,bb=98 50 355 298]{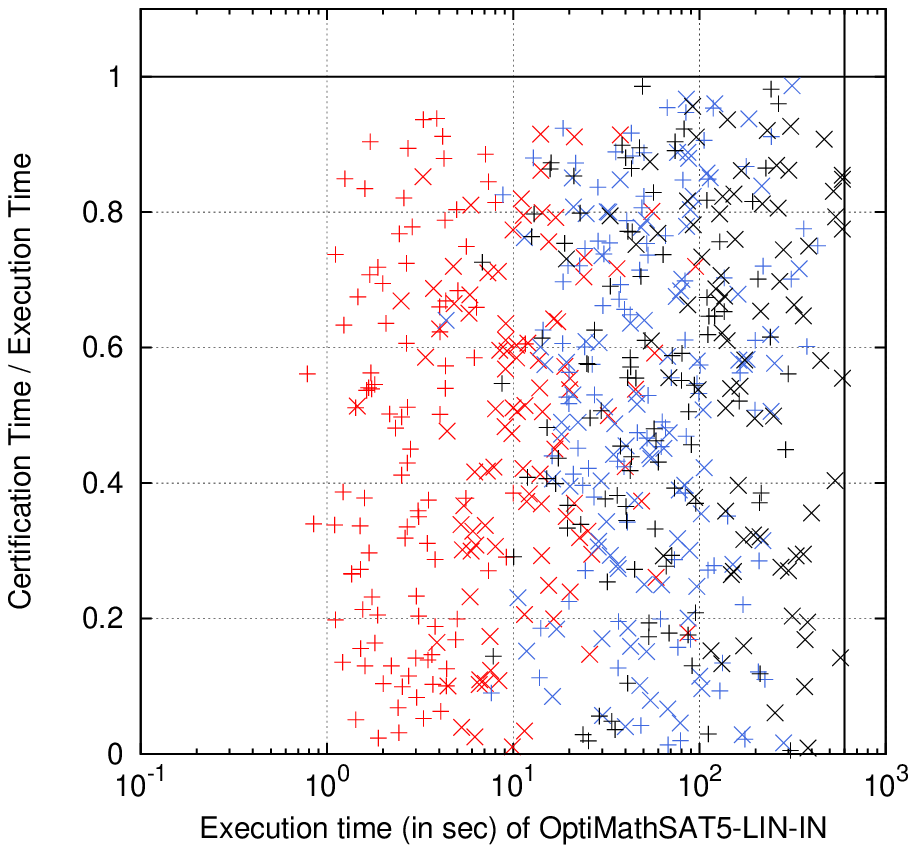}\\
 \end{tabular} 
 }
 \caption{Scatter-plots comparing solving, minimization 
 and certification time (left, center and right respectively) 
 with the execution time of  \optlinin{}  
 on \dirgenerated instances of strip-packing (top) 
 and job-shop (bottom). }
\label{fig:comp_time}
\end{figure}

\subsubsection{Analysis of \optmathsat performances.}
\label{sec:expeval_lgdp_analysis}
We want to perform a more fine-grained analysis of
the performances 
of the best version of \optmathsat, \optlinin{}. 
To this extent, we partition the total execution 
time taken on each problem into three consecutive
components:
\begin{itemize}
\item {\it solving time}, i.e.  the time spent on finding the first
  sub-optimal solution, 
\item {\it minimization time}, i.e.  the time required to search for the
  optimal solution, 
\item and {\it certification time}, i.e. 
 the time needed for checking there is no better solution.
\end{itemize}

Figure~\ref{fig:comp_time} reports, for all strip-packing (top) and
job-shop (bottom) instances, the ratios of the three components above 
over total execution time. 
%
%
(Notice the log scale of the x axis and the linear scale on the y axis.)
We notice a few facts:
\begin{itemize}
\item the solving time is nearly negligible, in particular on hardest
  problems. This tells us, among other facts, that \omlarat on these 
formulas is a much harder problem that plain \smtlarat on the same
formulas; 

\item the remaining time, on average, is either 
evenly shared between the minimization and the certification efforts (job-shop,
bottom)
or even it is mostly dominated by the latter, in particular on the
hardest problem (strip-packing, top).
\end{itemize}
\noindent
Overall, this suggests that on these instances \optlinin{} takes  on average
less than half of the total execution time to find the actual optimal 
 solution, and more than half to prove that there is no better
 one.

\ignore{
For all the formulas $\psi$ s.t. \optlinin{} reached the timeout, 
we also tried to detect the inconsistency of $\psi \wedge (\cost < \minvalue)$, 
where $\minvalue$ is the last suboptimal solution found,
for computing the number of formulas solved 
even though the timeout was reached. 
We used the solver \yices and a timeout of 1800 seconds 
but no run terminated within the timeout.
}

\subsection{Comparison on SMT-LIB problems}
\label{sec:expeval_smtlib}

\begin{figure}[t]
  \centering


{\scriptsize 
\begin{tabular}{|l||c|c|c|c|c|c|c|}
\hline
\multirow{2}{*}{Procedure} & \multicolumn{7}{|c|}{SMT-LIB/QF\_LRA formulas} \\
\cline{2-8}
& \#inst. & \#term. & \#correct.& \#err.msg. & \#wrong & \#unfeas. & time\\
\hline
\optlinin{5} & 194 & 194 & {194} & 0 & 0 & 0 & {1604}\\
\hline
\optbinin{5} & 194 & 194 & \best{194} & 0 & 0 & 0 & \best{1449}\\
\hline
\optadain{5} & 194 & 194 & 194 & 0 & 0 & 0 & 1618\\
\hline \hline
\multicolumn{8}{|c|}{ LDGP-SMT-Encoded Benchmarks (\encone-\renc) } \\
\hline
\optlinin{5} & 194 & 194 & 194 & 0 & 0 & 0 & 1820\\
\hline \hline
\multicolumn{8}{|c|}{ LDGP-SMT-Encoded Benchmarks (\enctwo-\renc) } \\
\hline 
\optlinin{5} & 194 & 194 & 194 & 0 & 0 & 0 & 1597\\
\hline \hline
\multicolumn{8}{|c|}{ LGDP-Encoded Benchmarks (\encone) } \\
\hline
\jamsBM & 194 & 171 & 116  & 52 & 0 & 3 & 1561\\
\hline
\jamsCH & 194 & 193 & 15  & 108 & 0 & 70 & 559\\
\hline
\logmipBM & 194 & 172 & \best{117}  & 52 & 0 & 3 & \best{2152}\\
\hline
\logmipCH & 194 & 193 & 15  & 108 & 0 & 70 & 576\\
\ignore{
 \hline \hline
 {\stchange \jamsBMmult} & 194 & 155 & 150 & 0 & 4 & 1 & 2990 \\
}
\hline \hline
\multicolumn{8}{|c|}{  LGDP-Encoded Benchmarks (\enctwo) } \\
\hline 
\jamsBM & 194 & 172 & \best{166}  & 0 & 4 & 2 & \best{6839}\\
\hline
\jamsCH & 194 & 105 & 104 & 0  & 0 & 1 & 9912\\
\hline
\logmipBM & 194 & 171 & 165  & 0 & 4 & 2 & 4103 \\
\hline
\logmipCH & 194 & 105 & 104 & 0 & 0 & 1 & 9649\\
\hline \hline
\ignore{  
  {\stchange \jamsBMmult} & 194 & 156 & 151 & 0 & 4 & 1 & 4647 \\
  \hline \hline
 {\RSCHANGETHREE \jamsBMmult} & 194 & 194 & 145 & 44 & 4 & 1 & 3131 \\
 \hline \hline
}
\end{tabular}
}

     
  \begin{tabular}{cc}
\includegraphics[width=0.42\linewidth,bb=97 50 355 298]{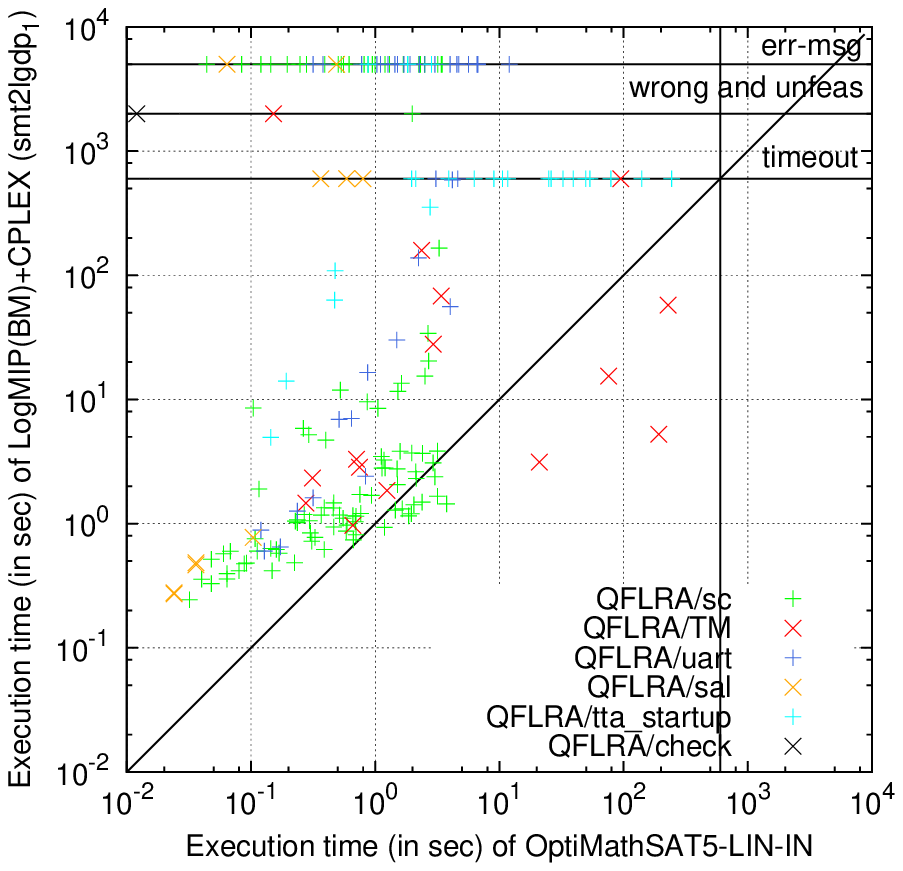} &
\includegraphics[width=0.42\linewidth,bb=97 50 355 298]{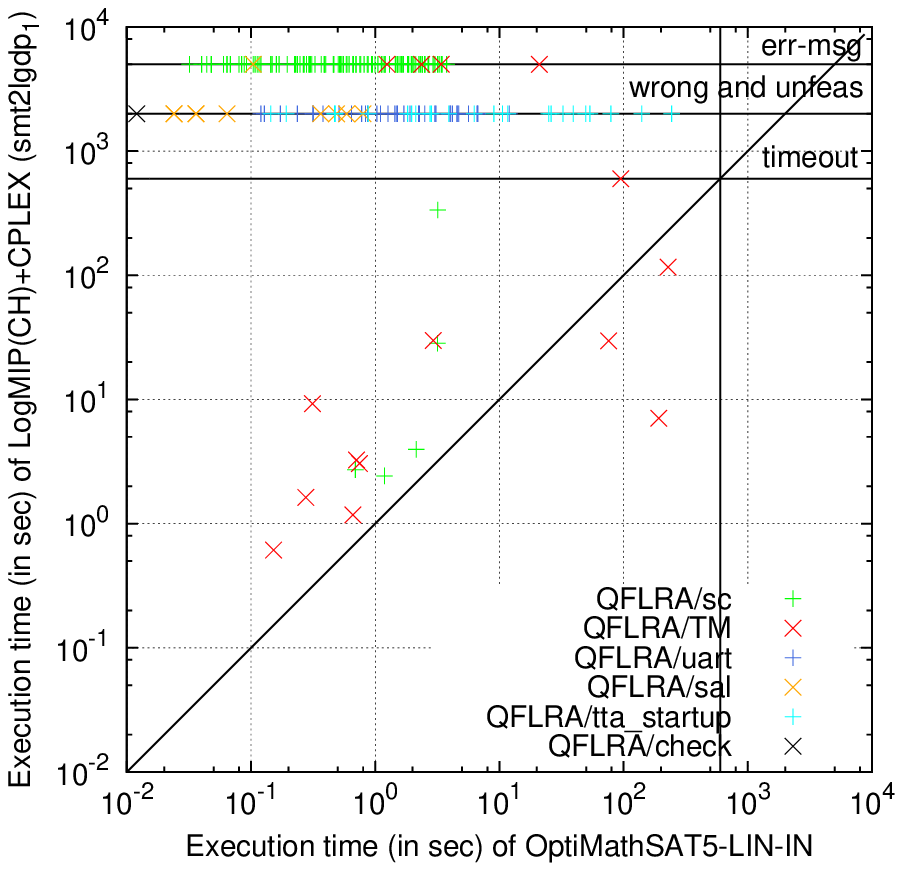} \\
\includegraphics[width=0.42\linewidth,bb=97 50 355 298]{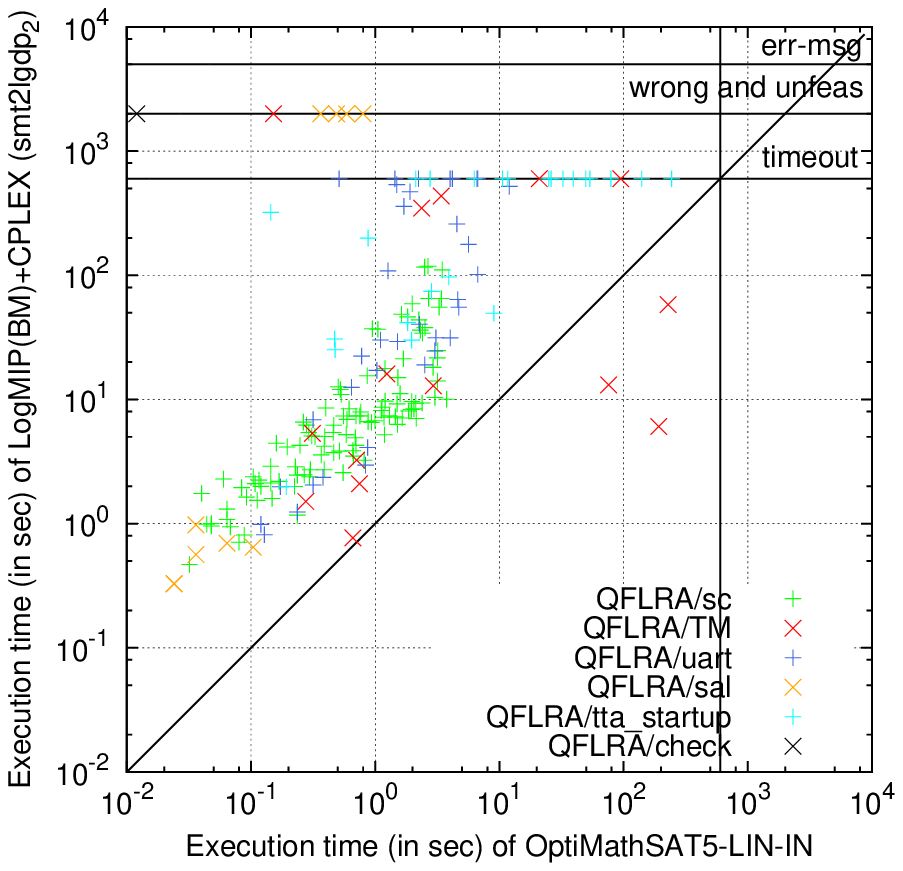} &
\includegraphics[width=0.42\linewidth,bb=97 50 355 298]{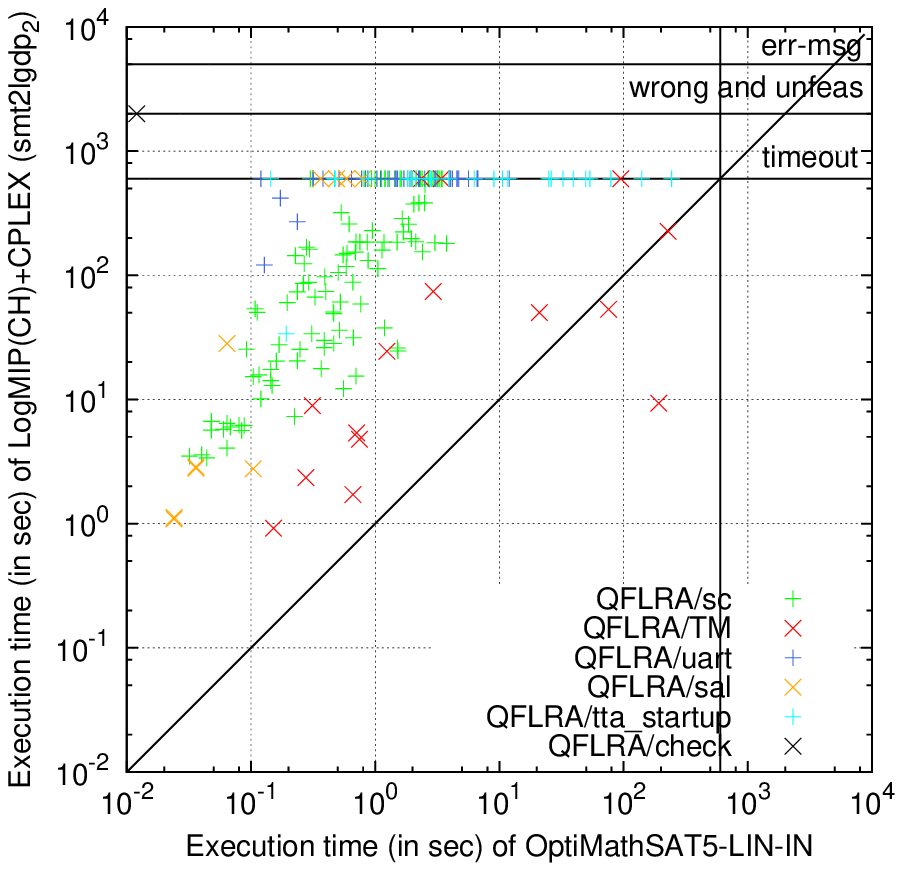} \\
 \end{tabular}
\caption{
Table:
Results 
for all the inline versions of \optmathsat and all the \gams tools, 
on a subset of SMT-LIB \larat satisfiable instances.
The columns report respectively:
\# of instances considered, 
\# of instances terminating within the timeout, 
\# of instances terminating with correct solution,
\# of instances terminating with error messages, 
\# of instances terminating returning a wrong minimum, 
\# of instances terminating wrongly returning ``unfeasible''. 
Scatter-plots: pairwise comparisons on the smt-lib
\larat satisfiable instances between \optlinin{}
and the two versions of \logmipG.
{LGDP models are generated using \encone (top) and \enctwo (bottom).}    
  \label{fig:generalresults}
  }
\end{figure}

 %

%
As a second comparison, 
in Figure~\ref{fig:generalresults} we compare \optmathsat 
against the \gams tools on the satisfiable \larat-formulas 
(QF\_LRA) in the SMT-LIB, augmented with
randomly-selected costs. 
\ignoreinshort{%
(Hereafter we do not consider the \of versions of \optmathsat.)
These instances are all classified as ``industrial'', 
because they come from the encoding of different real-world problems
in formal verification, planning and optimization.} 
They are divided into six categories, namely: 
 {\tt sc}, {\tt uart}, {\tt sal}, {\tt TM}, {\tt tta\_startup}, 
and {\tt miplib}.~%
{Notice that other SMT-LIB categories like {\tt spider\_benchmarks} 
and {\tt clock\_synchro} do not contain satisfiable instances and are thus
not reported here.} 

%
Since we have no control on the origin of each problem and
on the name and meaning of the variables, we selected iteratively one
variable at random as \cost variable, dropping it if the resulting
minimum was $-\infty$. This forced us to eliminate a few instances, in
particular all {\tt miplib} ones.
We used both 
\encone and \enctwo to encode these problems into  LGDP.
%

As before, to check for both correctness and effectiveness of the encodings, 
we also encoded the problems into LGDP by each encoding 
and encoded then back, to be fed to \optlinin{} (4th and 5th row). 
We notice that this caused substantial difference in neither
correctness  nor efficiency.

\smallskip 
We notice first that the results for \gams tools are affected by 
correctness problems, with both  encodings. 
Consider the encoding \encone.  Out of 194 samples, both GAMS tools
with the CH option returned ``unfeasible'' (i.e. inconsistent) on 70
samples and an error message (regarding some unsatisfied
disjunctions) on 108 samples. {The two versions with BM returned 3
unfeasible solutions and 52 solutions with error messages}.
%
Only 15 samples were solved correctly by GAMS tools 
with the CH option and {$117$ (with \logmip) or $116$ (with
  \jams)} samples with BM ones,  
whilst \optmathsat solved correctly all 194 samples.
(We recall that all \optmathsat results were cross-checked, 
and that the four GAMS tools were fed with the same files.) 
{With \enctwo encoding the number of  correctly-solved
  formulas increases,   
$104$ with CH option and $165$ (with \logmip) or $166$ (with \jams) with BM;} 
there are no error messages and
the number of unfeasible solutions of both GAMS
tools with the BM and CH options decreases 
 to 2 and 1 respectively,  
but the number of solutions with wrong minimum
 increases to 4  with the BM versions.

 Importantly, with both encodings, the results for \gams tools varied
 by modifying a couple of parameters from their default value, namely
 ``{\tt eps}'' and ``{\tt bigM Mvalue}''.  For example, on the
 above-mentioned {\tt sal} instance with \encone, with the default
 values the BM versions returned a wrong minimum value ``0'', the
 CH versions returned ``unfeasible'', whilst \optmathsat returned the
 correct minimum value ``2''; modifying {\tt eps} and {\tt bigM
   Mvalue}, the results become unfeasible also with BM options.
This highlights the fact that there are indeed some correctness and
robustness problems with the
\gams tools, regardless of the encodings used.  
~\footnote{We also
  isolated a subproblem, small enough to be solved by hand, in which
  the \gams tools returned evidently-wrong results, and notified it to
  the \gams support team, who reckoned the problem and promised to
  investigate it eventually.
}

\begin{figure}[t]
  \centering
{\setlength{\tabcolsep}{2pt}
\begin{tabular}{ccc}
{\em solving/total time  ratio}& 
{\em minimization/total time ratio}& 
{\em certification/total time  ratio}\\
\includegraphics[width=.32\linewidth,bb=98 50 355 298]{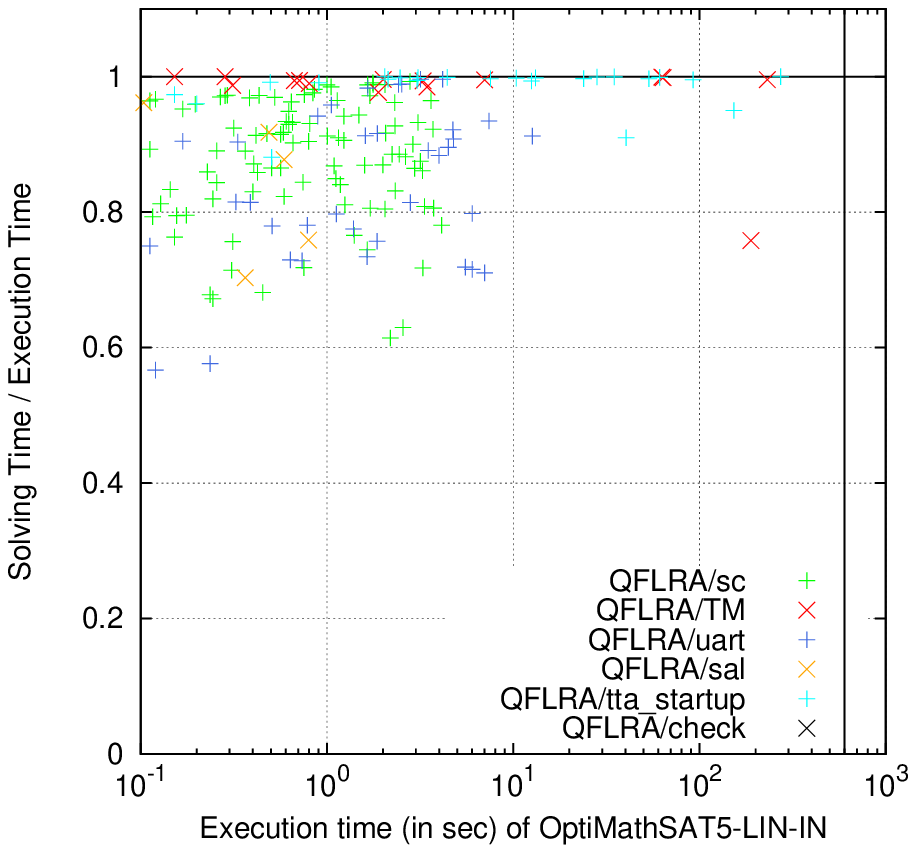} &
\includegraphics[width=.32\linewidth,bb=98 50 355 298]{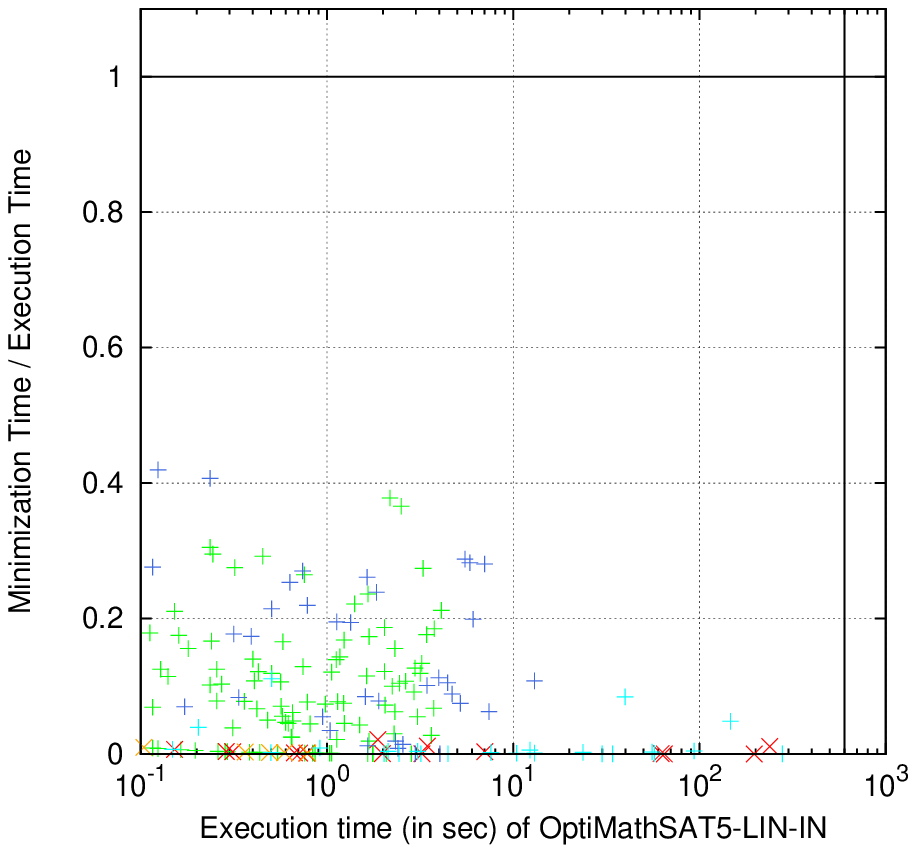} &
\includegraphics[width=.32\linewidth,bb=98 50 355 298]{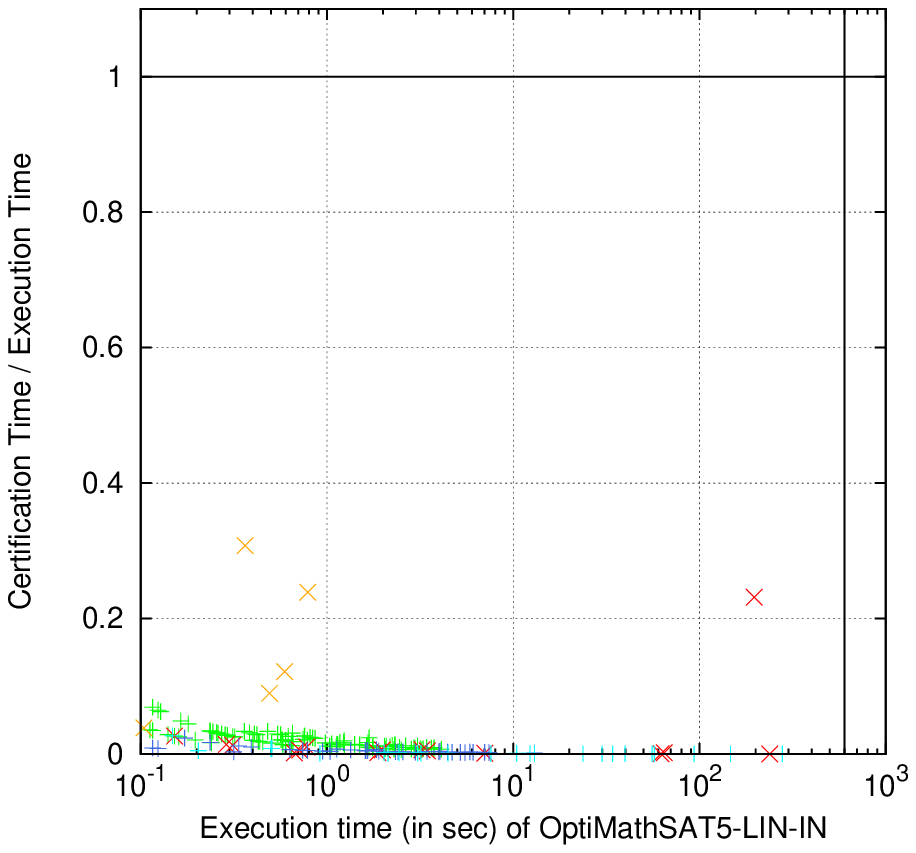}\\ 
 \end{tabular} 
 }
 \caption{\RSCHANGEONE{
Scatter-plots comparing solving, minimization 
 and certification time (left, center and right respectively) 
 with the execution time of  \optlinin{}  
 on  SMT-LIB instances. }}
\label{fig:comp_time_smtlib}
\end{figure}

\subsubsection{Discussion}
We conjecture that the problems with the \gams tools may be caused, at
least in part, by the fact that GAMS tools use floating-point rather
than infinite-precision arithmetic, and they introduce internally an
approximated representation of strict inequalities (see
Remark~\ref{rem:infiniteprec}).  Notice that, unlike with the LGDP problems in
\sref{sec:expeval_lgdp}, SMT-LIB problems do contain occurrences of
strict [resp. non-strict] inequalities with positive [resp. negative]
polarity.

From the perspective of the efficiency, all versions of 
\optmathsat solved correctly all problems within the timeout, 
the \binin version performing slightly better than the others;
\gams did not solve many samples
(because of timeout, wrong solutions and solutions with error messages). 
Looking at the scatter-plots, we notice that,
 with the exception of {a few samples}, 
\optmathsat always outperforms the \gams tools, 
often by more than one order magnitude. 
We notice that on these problems
 \enctwo is generally more effective than \encone
and less prone to errors.

\RSCHANGEONE{%
Finally, Figure~\ref{fig:comp_time_smtlib} reports for all SMT-LIB
instances the ratios of the solution, minimization and certification
times over the total execution time for  \optlinin{}. 
Unlike with Figure~\ref{fig:comp_time}, we notice that 
here the solution time is
dominating, 
the minimization time is significant, and the certification time is
nearly negligible. 
This means that on these instances \optlinin{} takes  on average
more than half of its execution time to find the first solution, 
less than half  to find the actual optimal 
 solution, and very little time to prove that there is no better
 one. 
 We conjecture that this is due to the fact that most satisfiable
 SMT-LIB instances come from the encoding of formal verification steps
 of bugged systems which, unlike with the LGDP problems of
 \sref{sec:expeval_lgdp}, have a limited number of solutions.
%
 }

\subsection{Comparison on SAL problems}
\label{sec:expeval_sal}

{\setlength{\tabcolsep}{3.7pt}
\begin{figure}[t]
  \centering


\footnotesize
\begin{tabular}{|l||c|c|c|c|c|c|c|}
\hline
\multirow{2}{*}{Procedure} & \multicolumn{7}{|c|}{SAL formulas} \\
\cline{2-8}
& \#inst. & \#term. & \#correct & \#err. msg. & \#wrong & \#unfeas. & time\\
\hline
\optlinin{5} & 392 & 385 & \best{385} & 0 & 0 & 0 & \best{44129}\\
\hline
\optbinin{5} & 392 & 382 & 382 & 0 & 0 & 0 & 45869\\
\hline
\optadain{5} & 392 & 381 & 381 & 0 & 0 & 0 & 44932\\
\hline \hline
\multicolumn{8}{|c|}{ LGDP-Encoded Benchmarks (\encone) } \\
\hline
\jamsBM & 392 & 24 & \best{4} & 19 & 1 & 0 & \best{1096}\\
\hline
\jamsCH & 392 & 46 & 0 & 0 & 0 & 46 & 0\\
\hline
\logmipBM & 392 & 24 & \best{4} & 19 & 1 & 0 & \best{1092}\\
\hline
\logmipCH & 392 & 46 & 0 & 0 & 0 & 46 & 0\\
\hline
\multicolumn{8}{|c|}{  LGDP-Encoded Benchmarks (\enctwo) } \\
\hline 
\jamsBM & 392 & 28 & \best{14} & 0 & 14 & 0 & \best{1456}\\
\hline
\jamsCH & 392 & 31 & 2 & 0 & 0 & 29 & 122\\
\hline
\logmipBM & 392 & 28 & \best{14} & 0 & 14 & 0 & \best{1428} \\
\hline
\logmipCH & 392 & 31 & 2 & 0 & 0 & 29 & 120\\
\ignore{
\hline \hline
{\RSCHANGETHREE \jamsBMmult} & 392 & 32 & \best{17} & 0 & 15& 0 &
\best{786}\\
}
\hline
\end{tabular}


%
\begin{tabular}{cc}
\includegraphics[width=0.45\linewidth,bb=98 50 355 298]{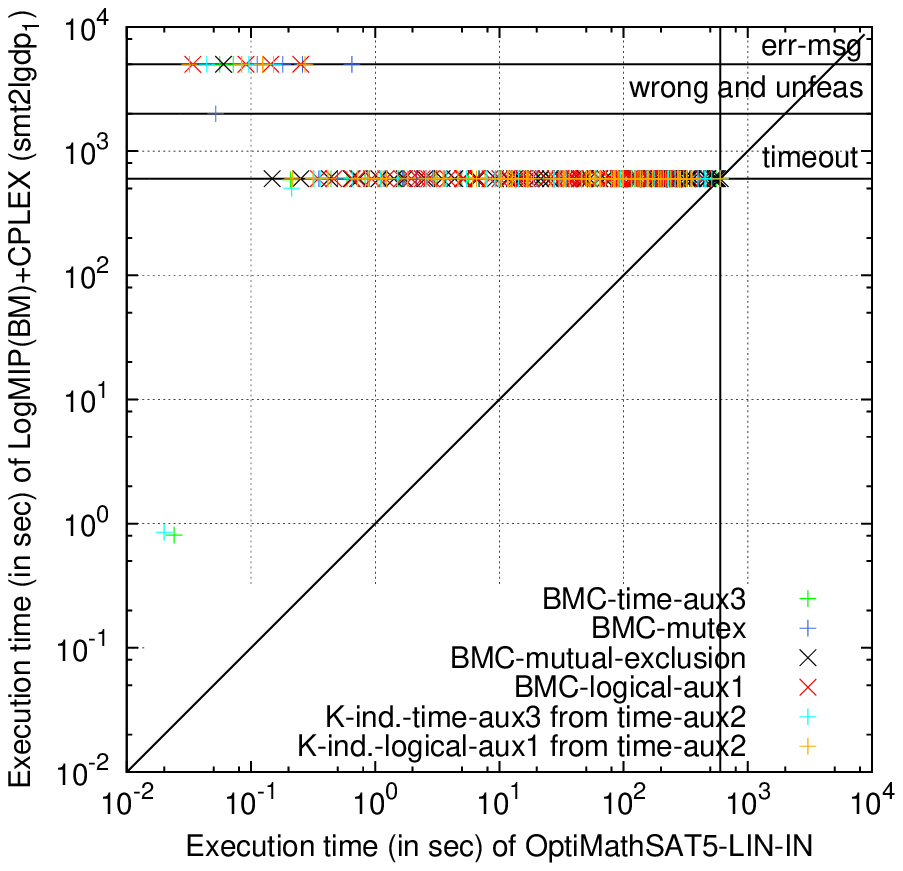} &
\includegraphics[width=0.45\linewidth,bb=98 50 355 298]{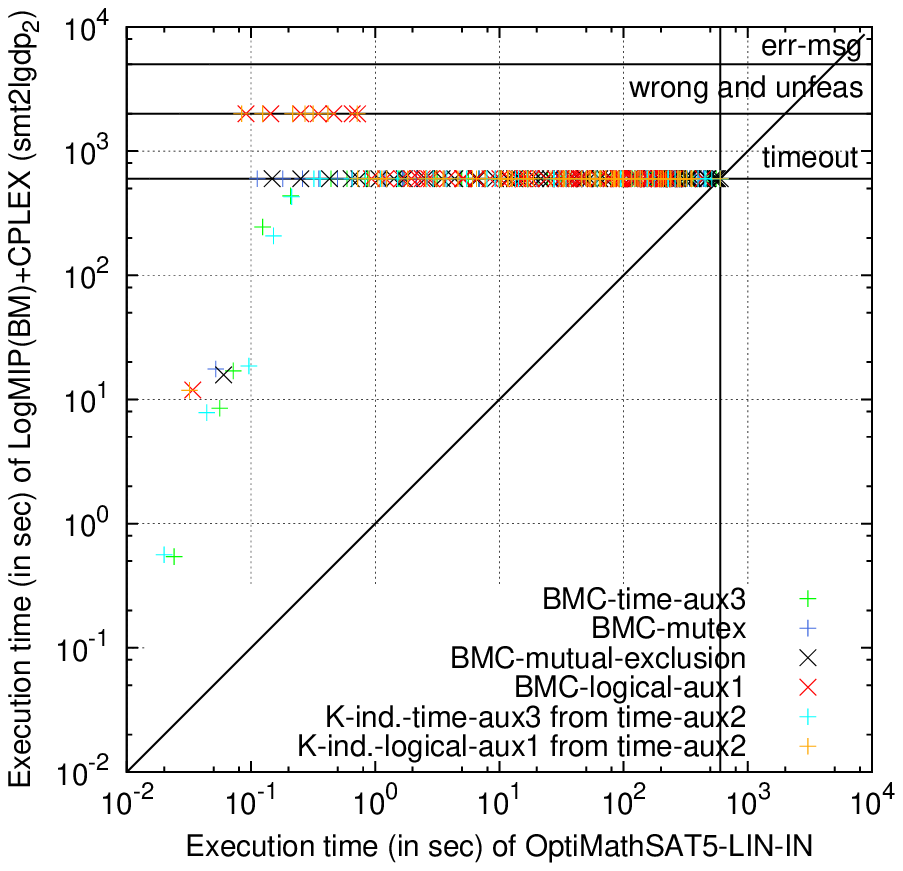}\\ 
 \end{tabular} 
 \caption{
Table:
Results 
for all the inline versions of \optmathsat and all the \gams tools, 
on formulas generated
from  SAL models of Fisher's protocol. 
The columns report respectively:
\# of instances considered, 
\# of instances terminating within the timeout, 
\# of instances terminating with correct solution,
\# of instances terminating with error messages (\gams tools only), 
\# of instances terminating returning a wrong minimum, 
\# of instances terminating wrongly returning ``unfeasible''.     
%
Scatter-plots: comparison of the best configuration
of \optmathsat (\optlinin{}) against \logmipBM on \encone 
and \enctwo encodings (left and right respectively).}
\label{fig:plots_sal}
\end{figure}
}

As a third comparison, 
in Figure~\ref{fig:plots_sal} we compare \optmathsat 
against the \gams tools on \larat-formulas obtained by using the SAL Model 
Checker 
on a set of 
bounded verification problems ---  Bounded Model Checking (\bmc) 
of invariants \cite{biere-cimatti-tacas} 
and K-Induction (\kind) \cite{SheeranSS00} --- of a well-known 
parametric timed system, Fisher's Protocol~%
\footnote{Problems available at
\url{http://sal.csl.sri.com/examples.shtml}}.

\marg{\bmc \& \kind}
\bmc [resp. \kind] takes a Finite-State Machine  $M$, an invariant
property $\Psi$ and  
an integer bound $k$, and produces a propositional  formula $\vi$ which is 
satisfiable [resp. unsatisfiable] if and only if there exists a $k$-step execution 
violating $\Psi$ [resp. a $k$-step induction proof that $\Psi$ is always
verified].
 The approach leverages to real-time systems by producing 
\smtlarat formulas rather than purely-propositional ones (see, e.g.,
\cite{acks_forte02}). 

\marg{Fisher protocol}
Fisher's Protocol ensures mutual exclusion among $N$ processes
using real-time clocks and a shared variable. 
The problem is parametric into two positive real values, 
$\delta_1$ and
$\delta_2$, describing the delays of some actions. 
It is known that mutual exclusion, and other properties included
in the SAL model, are verified if and only if $\delta_1 < \delta_2$.

We have produced our \omlarat problems as follows.
We fixed the value of $\delta_2$ (we chose $\delta_2=4$), and
then we generated six groups of formulas according to the
problem solved  (\bmc or \kind) and the  property addressed
(called {\tt mutex}, {\tt mutual-exclusion}, {\tt time-aux3} and 
{\tt logical-aux1}).
For each group, 
for increasing values of
$N \ge 2$ and for a set of sufficiently-big  values of $k\ge k^*$,%
~\footnote{For \bmc, $k^*$ 
is set to the smallest value of $k$ which makes the formula satisfiable,
imposing no upper bound on $\delta_1$; 
for \kind, $k^*$  is set to the smallest value of $k$ which
makes the formula encoding the inductive step  unsatisfiable,
imposing $\delta_2>\delta_1$). 
{In these experiments, $k^*$ ranges from $5$ to $10$, depending on the 
problem; also, for each problem, $k^*$ does not depend on $N$.}
}
we used SAL to produce the corresponding parametric 
\smtlarat formulas,%
\ignore{
~\footnote{The SAL tool {\tt sal-inf-bmc}
produced the formulas in the CVC3 language \cite{cvc3},
which were then converted into SMT-LIB compliant \larat-formulas
using the {\tt cvc2smt} encoder.
}
}
  and asked the tool under test to find 
 the minimum value of $\delta_1$
 which made the resulting formula \larat-satisfiable (we knew in
 advance from the  problem that, for $k$ big enough, this value is 
 $\delta_1=\delta_2=4.0$). 
As before, we used both \encone and \enctwo to encode 
the \omlarat benchmarks into LGDP.

\ignore{
\todo{SILVIA: se e' l'ok quanto detto sopra, taglierei i
  prossimi due paragrafi.\\}
Formulas were created by encoding the \bmc and \kind 
problems of the SAL model in CVC3 language \cite{cvc3} 
using the SAL tool {\tt sal-inf-bmc}, 
and then converting them into \larat-formulas
using the CVC3's encoder {\tt cvc2smt}.
\larat-formulas were then transformed into 
\omlarat benchmarks by fixing $\delta_2$ at the value 4 
and maximizing the variable $\delta_1$.
Overall, 392 \omlarat problems were generated. 
We used two LGDP encodings of \omlarat benchmarks, 
\encone and \enctwo.
 
We classified formulas in six groups according to the
solved problem (\bmc and \kind) and the verified property 
({\tt mutex}, {\tt mutual-exclusion}, {\tt time-aux3} and 
{\tt logical-aux1}).
For each group, we generated formulas using
a fixed bound $k$~\footnote{
For \bmc, $k$ 
is set to the first value which makes the formula satisfied. 
For \kind, $k$ is set to the first value which
makes the formula encoding the base case and the 
formula encoding the inductive step both satisfied.}
 and increasing values of $N$.
\todo{Silvia, non ho capito il senso della prossima sentence, in
  particolare il primo e il terzo punto:}
We stopped the evaluation of a solver when
no formula could be generated by {\tt sal-inf-bmc},  or
the timeout was reached three times consecutively
or unfeasible solution was found at least six times consecutively.
}

\subsubsection{Discussion}

The results are presented in Figure~\ref{fig:plots_sal}.
The three versions of \optmathsat  solved correctly 385, 382 and 381 
out of the 392 samples respectively, \optlinin{} being the best performer.
%

Considering the \gams tools with the encoding \encone,
the two  tools using BM solved on time and correctly only
4 samples over 392 and returned 19 solutions with
error messages  and 1 solution with wrong minimum, 
whilst the CH ones always returned ``unfeasible''.
(We recall that all \gams tools and options are fed the same inputs.)
Considering the encoding \enctwo, 
the \gams tools solved more problems correctly 
({14} with BM tools and 2 with CH),
but they returned wrong and unfeasible solutions
({14} wrong solutions for BM versions and 29
unfeasible for CH ones). No solution with error messages
was found.

The scatter-plots compare \optlinin{} 
with the best versions of \gams, \logmipBM, 
on both the encodings, 
showing that the former dramatically 
outperforms the latter, no matter the encoding used.


%
\subsection{Comparison on pseudo-Boolean SMT problems}
\label{sec:expeval_pb}
\begin{figure}[t]
  \centering
{\setlength{\tabcolsep}{2pt}


{\small
\begin{tabular}{|l||c|c|c|c|c|c|c|}
\hline
\multirow{2}{*}{Procedure} & \multicolumn{7}{|c|}{MaxSMT $/$ SMT+PB
  problems} \\
\cline{2-8}
& \#inst. & \#term. & \#correct. & \#err.msg. & \#wrong & \#unfeas. & time\\
\hline 
\optpblin{} & 675 & 636 & \best{636} & 0 & 0 & 0 & \best{19675} \\     
\hline
\optpbbin{} & 675 & 632 & 632 & 0 & 0 & 0 & 13024 \\     
\hline \hline
\multicolumn{8}{|c|}{\omlarat-Encoded Benchmarks} \\
\hline
\optlinin{} & 675 & 630 & 630 & 0 & 0 & 0 & 20744 \\     
\hline
\optbinin{} & 675 & 634 & 634 & 0 & 0 & 0 & 16502 \\     
\hline
\optadain{} & 675 & 637 & \best{637} & 0 & 0 & 0 & \best{18588} \\     
\hline \hline
\ignore{
\optlininm{}  & 675 & 633 & 633 & 0 & 0 & 0 & 21204 \\      
\hline
\optbininm{}  & 675 & 634 & 634 & 0 & 0 & 0 & 16039 \\     
\hline
\optadainm{}  & 675 & 636 & \best{636} & 0 & 0 & 0 &  \best{17505} \\     
\hline \hline
}
\multicolumn{8}{|c|}{ LGDP-Encoded Benchmarks (\encone) } \\
\hline
\jamsBM & 675 & 509 & {19} & 423 & 68 & 8 & {420} \\     
\hline
\jamsCH & 675 & 642 & 0 & 233 & 41 & 377 & 0 \\     
\hline
\logmipBM & 675 & 510 & \best{19} & 424 & 68 & 8 & \best{403} \\     
\hline
\logmipCH & 675 & 642 & 0 & 233 & 41 & 377 & 0\\     
\hline \hline

\multicolumn{8}{|c|}{  LGDP-Encoded Benchmarks (\enctwo)} \\
\hline 
\jamsBM & 675 & 449 & \best{92} & 9 & 351 & 6 & \best{1575} \\     
\hline
\jamsCH & 675 & 386 & 48 & 9 & 336 & 2 & 644 \\     
\hline
\logmipBM & 675 & 449 & {92} & 9 & 351 & 6 & {1650} \\     
\hline
\logmipCH & 675 & 383 & 48 & 9 & 333 & 2 & 674 \\     
\hline \hline
\ignore{
{\RSCHANGETHREE \jamsBMmult} & 675 & 479 & \best{95} & 9 & 367 & 6 & \best{4033} \\     
\hline \hline
}
\end{tabular}
}

}
{\setlength{\tabcolsep}{2pt}
\begin{tabular}{ccc}
\includegraphics[width=.32\linewidth,bb=97 50 355 298]{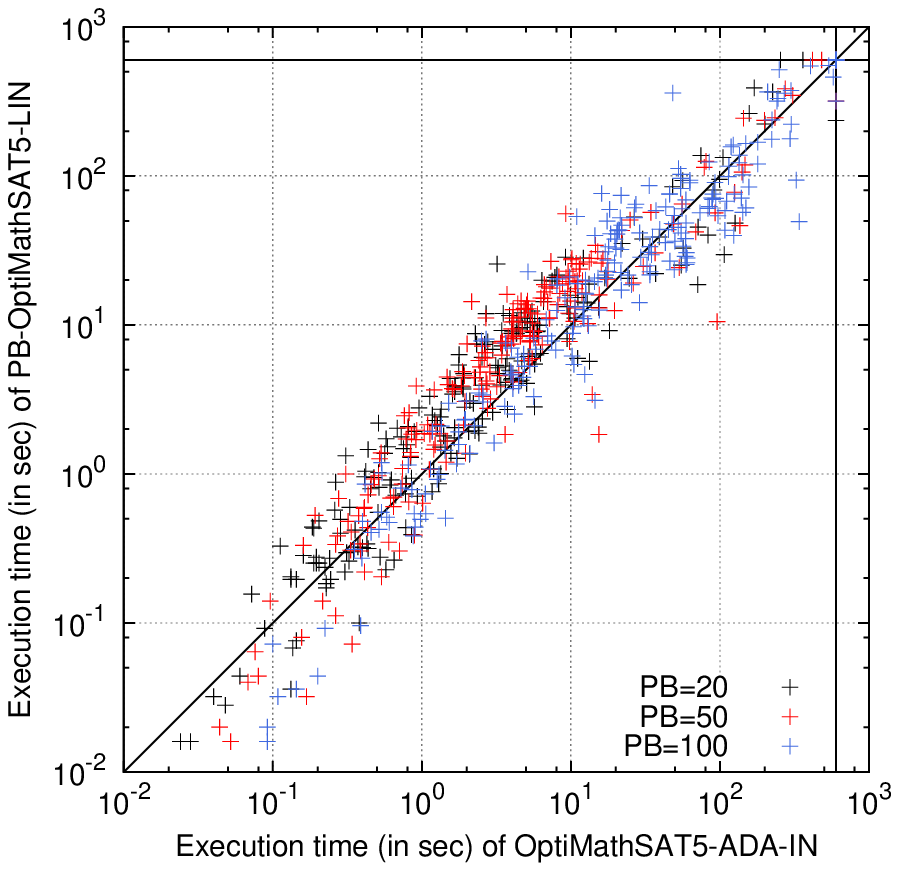} &
\includegraphics[width=.32\linewidth,bb=97 50 355
298]{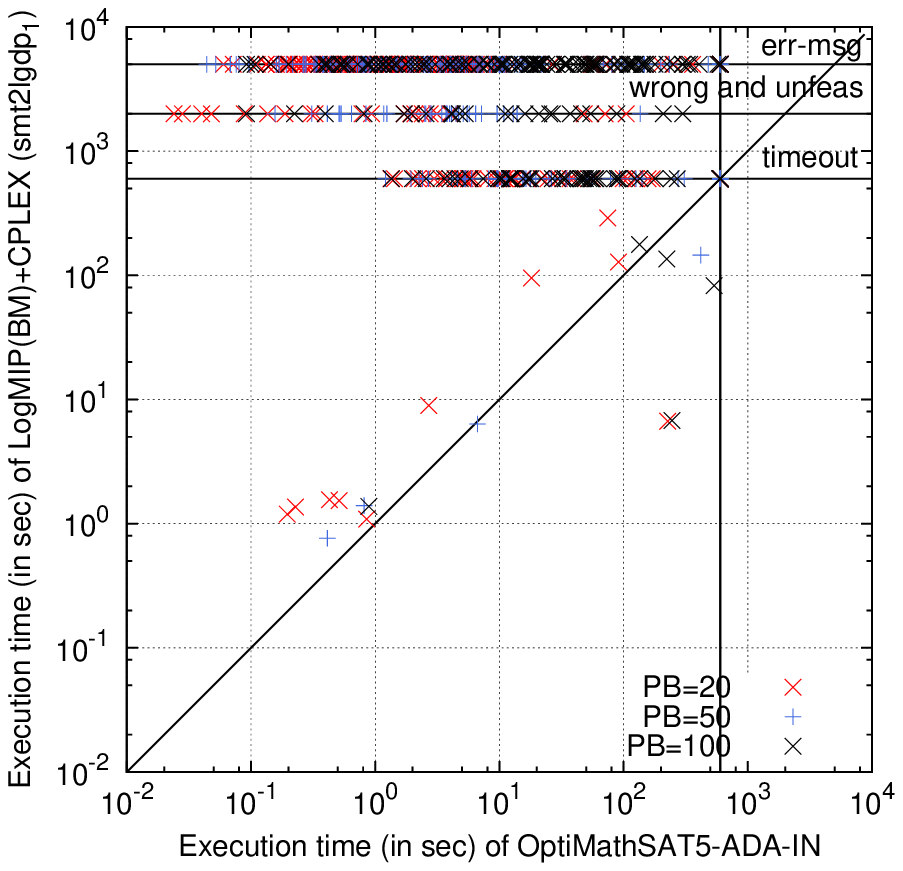} 
 &
 \includegraphics[width=.32\linewidth,bb=97 50 355
 298]{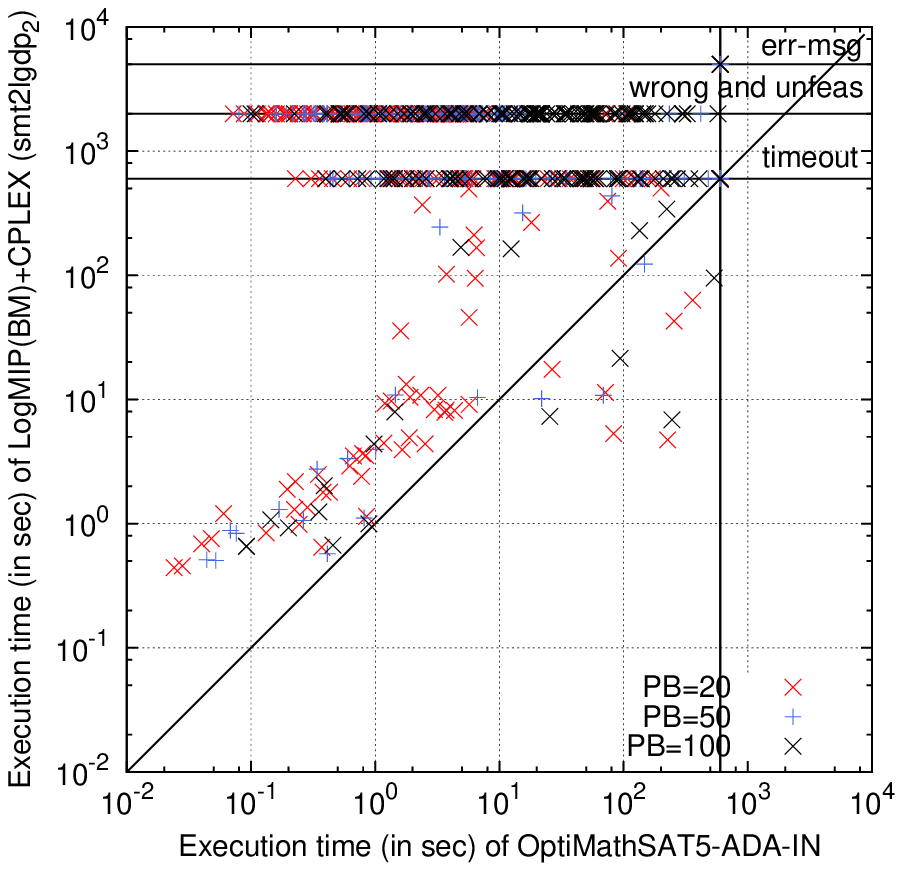}
\\ 
\end{tabular}}
\caption{
Table:
Results 
for \optmathsat, \pboptmathsat and  the \gams tools, on the
MaxSMT benchmarks from \protect\cite{cimattifgss10}.
The columns report respectively: 
\# of instances considered, 
\# of instances terminating within the timeout, 
\# of instances terminating with correct solution,
\# of instances terminating with error messages (\gams tools only), 
\# of instances terminating returning a wrong minimum, 
\# of instances terminating wrongly returning ``unfeasible''.     
%
Scatter-plots: comparison of the best configuration
of \optmathsat, \optadain{}, 
against
the best configuration of \pboptmathsat, \optpblin{} (left) 
and the best configuration of \gams tools, \logmipBM, on \encone 
and \enctwo encodings (center and right respectively).}
\label{fig:pb_results}
\end{figure}

As a fourth comparison, in Figure~\ref{fig:pb_results}
we evaluate \optmathsat  
on the problem sets used in \cite{cimattifgss10}
against the usual GAMS tools and against 
a recent reimplementation on \mathsatfive of the tool in
\cite{cimattifgss10}, namely  \pboptmathsat, for SMT with
Pseudo-Boolean constraints (see \sref{sec:optsmt}).
\footnote{A comparison against the tool in \cite{cimattifgss10} would
  not be fair, since the latter was based on the older and slower 
\mathsatfour. To witness  this fact, a comparison of these two
implementations is in \cite{cgss_sat13_maxsmt}.} 
\pboptmathsat is tested with both linear search and binary search
strategies (denoted with ``-LIN'' and ``-BIN'' respectively).

As described in \cite{cimattifgss10},
the problems consists of partial weighted MaxSMT problems which are 
generated randomly
starting from satisfiable \larat-formulas (QF\_LRA) in the
SMT-LIB, then converted into \smt problems with PB constraints, 
see \eqref{eq:maxsmt2pb-encoding} in  \sref{sec:optsmt}.
%
These problems are further encoded into \omlarat problems by means of the
encoding \eqref{eq:pb2smt-encoding} in \sref{sec:optsmt}, and hence
into LGDP problems by means of the usual two encodings. 
%

\subsubsection{Discussion}

The results are presented in Figure~\ref{fig:pb_results}.  The three
versions of \optimathsat solved respectively 630, 634 and 637 problems
out of 675 problems overall, whilst the two versions \pboptmathsat
solved respectively 636 and 632.  
Thus, despite they are both implemented on top of the same SMT solver
and \pboptmathsat is specialized for PB constraints, \optimathsat
performances are analogous to these of the more-specialized tool.
The various version of the \gams tool perform drastically worse: {with
  \encone they solve correctly only a very small number of samples (19
  with BM tools and even 0 with CH), returning error messages,
  unfeasible results or wrong minimum solutions on the remaining set
  of benchmarks;
with \enctwo  more samples are solved correctly and 
no error message is produced, but most problems produce a
wrong minimum solution. 
}
%

\begin{remark}
Notice that, unlike with LGDP problems (see
Remark~\ref{remark:binvslin}) and in part also with SMT-LIB and
SAL problems, with Pseudo-Boolean problems 
the cost variables 
 occurs in positive unit clauses in the form 
$(\cost = \tuple{term})$; 
thus, learning $\neg (\cost < \pivot)$ as a result of 
the binary-search steps with  \unsatres results
produces a constraining effect on the variables in $\tuple{term}$, 
and hence a pruning effect in
the search due to the early-pruning technique of the SMT solver. 
This might explain in part the fact that, unlike with
previous problems, here binary search  performs a little better than
linear search.
 \label{remark:binvslin2}
\end{remark}

The scatter-plots in Figure~\ref{fig:pb_results} compare the best
version of \optimathsat with these of \pboptmathsat and of the \gams
tools.
We see that \optadain{} performances are analogous to these of
\optpblin{}, and they are drastically superior to these of
\gams tools with both  encodings.

As a side note, in \cite{cgss_sat13_maxsmt} another empirical
evaluation is performed on MaxSMT problems ---although generated with a
slightly different random method from SMT-LIB benchmarks---
where \optmathsat performs equivalently better than \pboptmathsat and
the novel specialized MaxSMT tool presented there. 
We refer the reader to \cite{cgss_sat13_maxsmt} for details.





\RSCHANGETHREE{%
\subsection{Comparison against \gams with parallel \cplex on all problem sets}
\label{sec:expeval_parallelgams}
\begin{figure}[t]
  \centering
\RSCHANGETHREE{


\footnotesize
\begin{tabular}{|l||r|r|r|r|r|r|r||r|}
\hline
{Procedure} 
& \#inst. & \#term. & \#correct & \#err. msg. & \#wrong & \#unfeas. &
time & average\\
&&&&&&&& speedup \\
\hline
\multicolumn{9}{|c|}{Strip-packing LGDP problems (Directly Generated Benchmarks)  }  \\
\hline
  \optlinin{} & 600 & 568 & \best{568} & 0 & 0 & 0 & \best{12782} & --\\     
      \jamsBM & 600 & 432 &      {432} & 0 & 0 & 0 &      {22318} & --\\     
{\jamsBMmult} & 600 & 472 &      {472} & 0 & 0 & 0 &      {17918} & 5.98\\
\hline
\multicolumn{9}{|c|}{Job-shop  LGDP problems (Directly Generated Benchmarks)  }  \\
\hline 
  \optlinin{} & 600 & 577 &      {577} & 0 & 0 & 0 &      {43228} & --\\     
     \jamsBM & 600 & 587 &      {587} & 0 & 0 & 0 &      {34245} & --\\     
{\jamsBMmult} & 600 & 600 & \best{600} & 0 & 0 & 0 & \best{10465} & 9.54\\
\hline
\multicolumn{9}{|c|}{SMT-LIB problems (LGDP-Encoded Benchmarks (\enctwo)) }  \\
\hline 
\optlinin{5} & 194 & 194 & \best{194}  & 0 & 0 & 0 & \best{1604}& --\\
\jamsBM      & 194 & 172 &      {166}  & 0 & 4 & 2 &      {6839}& --\\
 {\jamsBMmult} & 194 & 155 & 150 & 0 & 4 & 1 & 2990 & 2.36\\
%
\hline 
\multicolumn{9}{|c|}{SAL problems (LGDP-Encoded Benchmarks (\enctwo))}  \\
\hline
\optlinin{5} & 392 & 385 & \best{385} & 0 & 0 & 0 & \best{44129}& --\\
\jamsBM & 392 & 28 &      {14} & 0 & 14 & 0 &      {1456}& --\\
{\jamsBMmult} & 392 & 32 &      {17} & 0 & 15& 0 &
     {\ 786}& 3.66\\
\hline
\multicolumn{9}{|c|}{MaxSMT $/$ SMT+PB problems (LGDP-Encoded Benchmarks (\enctwo))}  \\
\hline 
\optlinin{} & 675 & 630 & \best{630} & 0 & 0 & 0 & \best{20744} & --\\     
\jamsBM & 675 & 449 &      {92} & 9 & 351 & 6 &      {1575} & --\\     
{\jamsBMmult} & 675 & 479 &      {95} & 9 & 367 & 6 &      {4033} & 2.35\\     
\hline
\end{tabular}

     
}
\ \\
\RSCHANGETHREE{ 
  \begin{tabular}{ccc}
\ & & \\
Strip-packing LGDP & Job-shop LGDP & \\
\includegraphics[width=0.32\linewidth,bb=97 50 355 298]{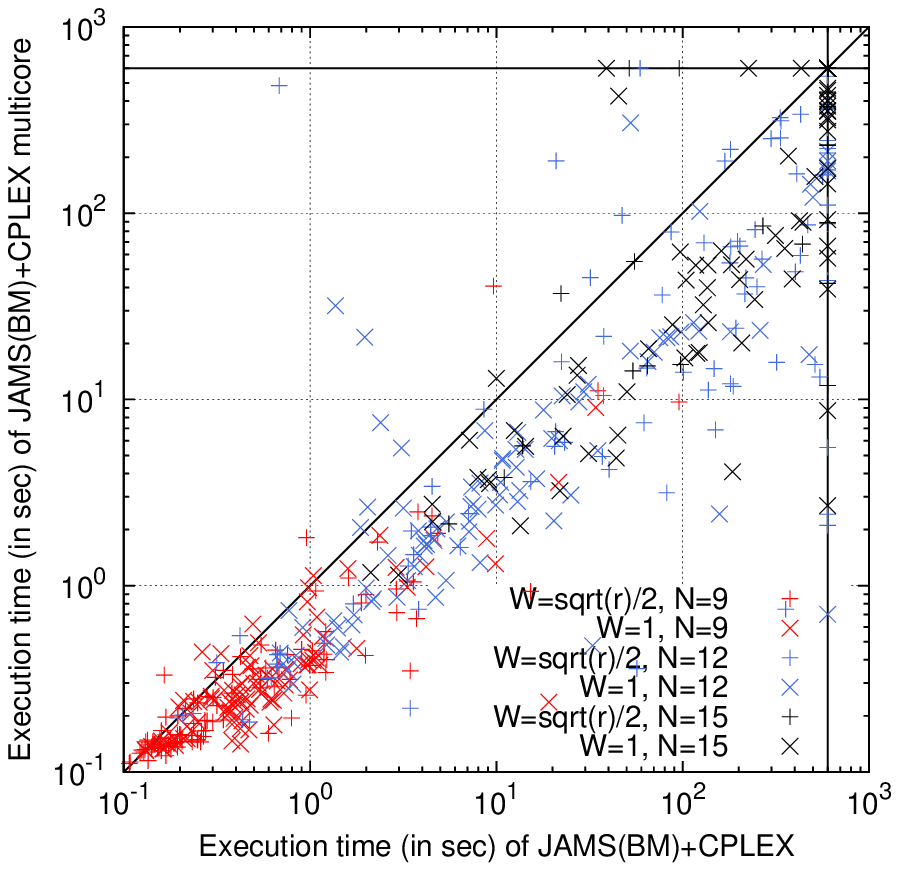} &
\includegraphics[width=0.32\linewidth,bb=97 50 355
298]{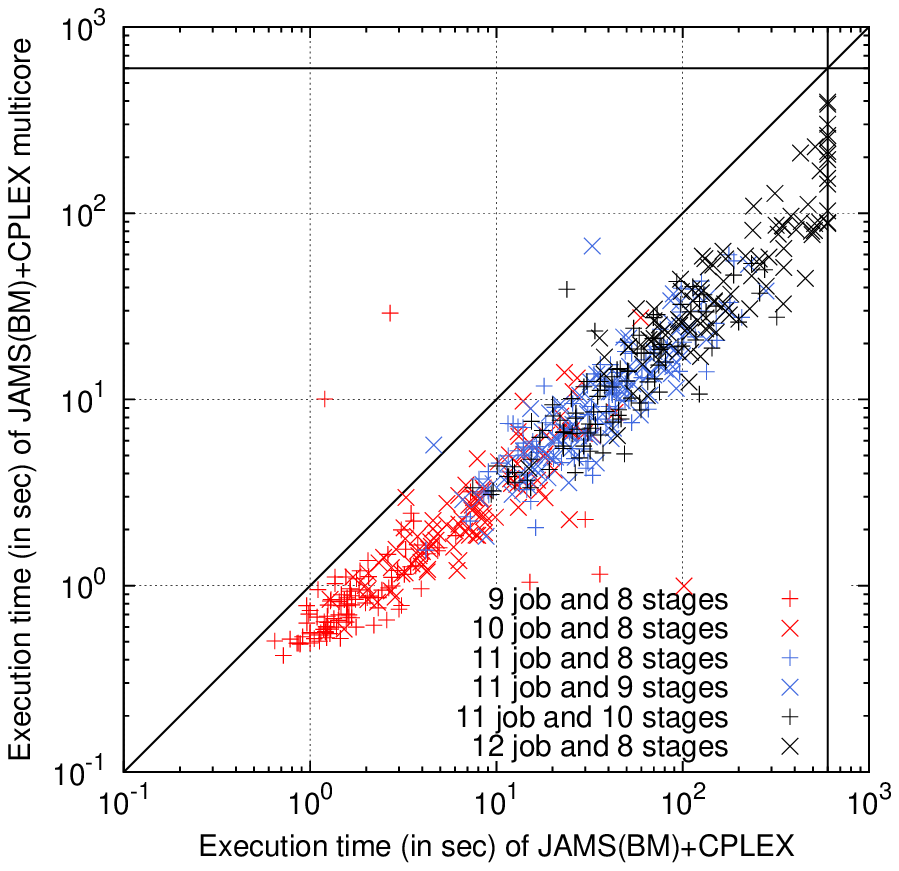} \\
\ & & \\
SMT-LIB & SAL & MaxSMT / SMT+PB\\
\includegraphics[width=0.32\linewidth,bb=97 50 355 298]{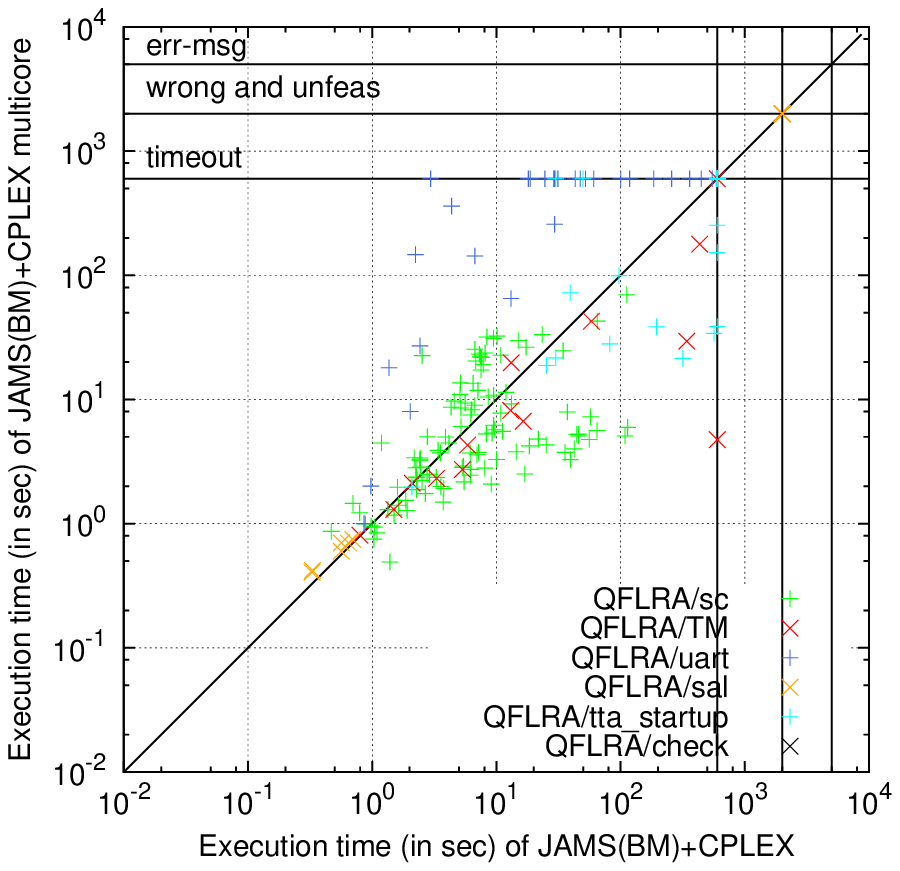} &
\includegraphics[width=0.32\linewidth,bb=97 50 355 298]{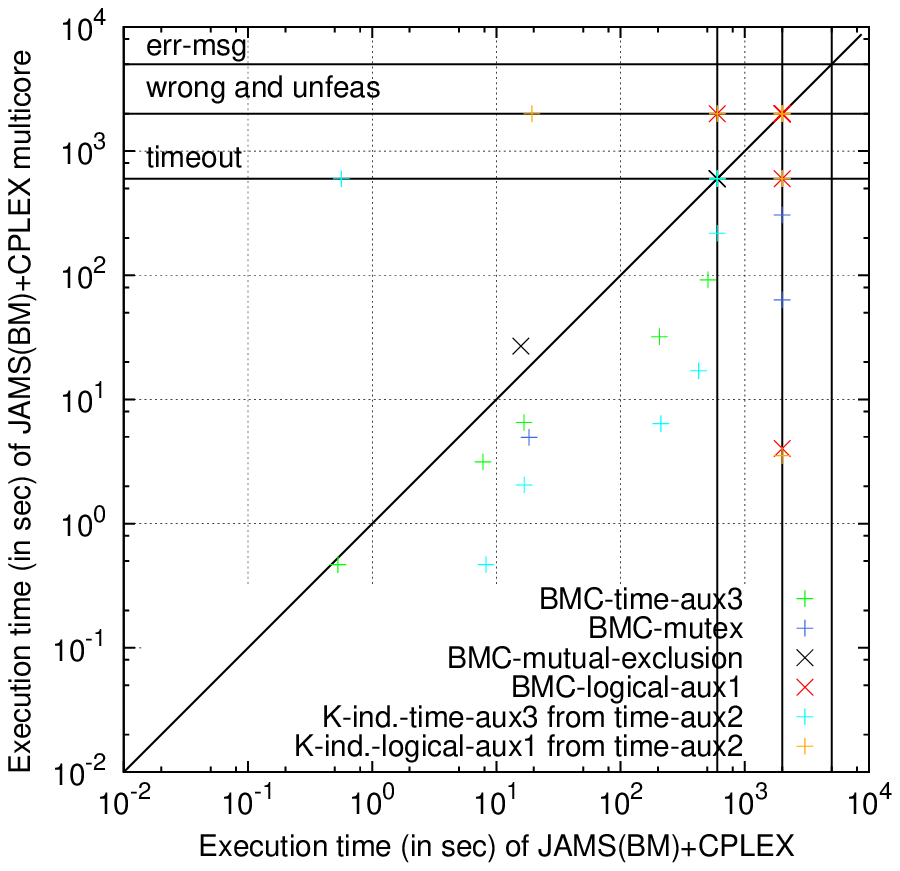} &
\includegraphics[width=0.32\linewidth,bb=97 50 355 298]{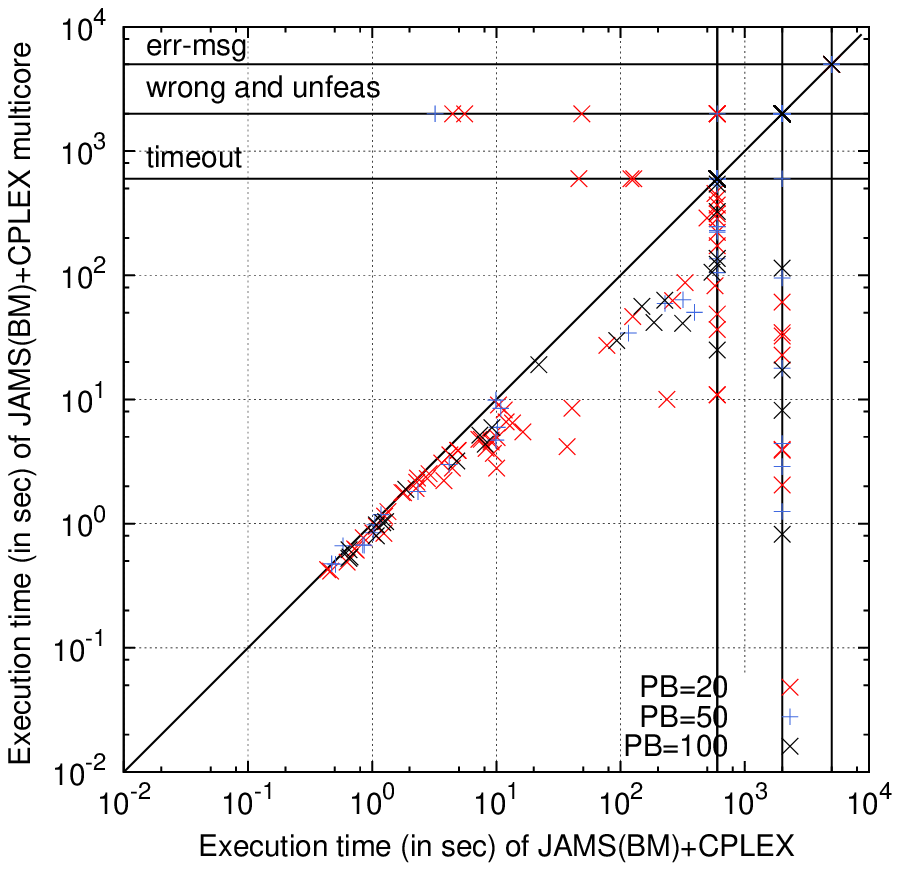}
\\
 \end{tabular}
}
\caption{
  \label{fig:multicoreresults}
\RSCHANGETHREE{ 
Table: comparison of \optlinin{}, \jamsBM and \jamsBMmult on the
five problem sets. Last column: average speedup for \jamsBMmult wrt. \jamsBM.
Scatterplots: pairwise comparison 
of \jamsBMmult vs. \jamsBM on the five problem sets.
}
  }
\end{figure}


As it is common practice in the SMT literature, 
in this paper we deal with {\em single-core} {\em sequential}
 procedures.
In fact, despite a couple of attempts \cite{wintersteigerhm09,kalinnikaswb10},
the parallelization of SMT-solving procedures
 is still an open research issue.
In particular \mathsatfive, and hence \optimathsat, provide no support 
for parallelization. 
Thus, although the issue of efficiently parallelizing OMT is
potentially a
very interesting research topic, it is definitely not in the
intended scope of this paper. 


Unlike with \mathsatfive and \optimathsat, however, \cplex provides
full support for multiple-core parallel solving.
This is an important benefit, since it allows for exploiting the
multiple-core CPUs of current PCs, reducing the elapsed time when
searching for a solution.
This gives  \gams a potential advantage wrt. \optimathsat which  the
previous tests could not reveal.~%

In order to investigate the actual relevance of this potential advantage, 
we have recently enriched our empirical investigation by
running all the tests in \sref{sec:expeval_lgdp}-\sref{sec:expeval_pb}
also on another \gams tool, namely \jamsBMmult{}:
this is  the  best-performing 
\gams \gams tool in the tests,  
\jamsBM, which uses instead the most recent version of \cplex, v12.6, {\em in parallel
  mode on four cores} (options {\tt opportunistic parallelmode}, {\tt 4 threads}).
Each of the four threads  is given a timeout of 600s.%
\footnote{%
We have a technical remark: in order  to use the same timeout
mechanism for all tools,
in all previous tests we have used the Linux command {\tt
  ulimit}  to handle the timeout for all \optimathsat, \gams and
\pboptmathsat versions. Unfortunately, {\tt
  ulimit} does not seem to work properly for multi-threaded processes, so
that for \jamsBMmult we had to use instead the \gams/\cplex internal
timeout mechanism, which we have assumed to be reliable. 
\ignore{
From some comparative tests performed in single-thread mode with the
two methods we have noticed no  difference.
Also, 
\jamsBMmult found no solution in more than $600s$ and some in a little
less than  $600s$.
}}
%
Therefore \jamsBMmult is given four times the CPU time resources
than its  competitors.

\ignore{
The results are displayed in the ``\jamsBMmult{}'' rows in the tables
of Figures~%
\ref{fig:plots_sp},
\ref{fig:plots_js},
\ref{fig:comp_time_smtlib},
\ref{fig:plots_sal}, and
\ref{fig:pb_results}
}

\subsubsection{Discussion}

The results are displayed in 
Figure~\ref{fig:multicoreresults}.
In the table, for each group of benchmarks
in \sref{sec:expeval_lgdp}-\sref{sec:expeval_pb}, we compare the
performances of the  best-performing \optimathsat tool, \optlinin{},
of the  best-performing \gams tool, \jamsBM,   and of its parallel
version, \jamsBMmult.
The results for the former two tools are taken from
Figures~\ref{fig:plots_sp}-\ref{fig:pb_results}.
In the last column we report the mean values of the speedup for \jamsBMmult wrt. \jamsBM
over the problems for which both tools terminated within the timeout. 
%
%
In the scatterplots we compare pairwise the performances 
of \jamsBMmult and \jamsBM on the five problem sets.

From Figure~\ref{fig:multicoreresults} we notice the following facts.
\begin{itemize}
\item The usage of \cplex in parallel mode on 
  the four cores pays off in terms of elapsed
  time: 
we notice a significant average speedup from
\jamsBM to \jamsBMmult,  ranging from $2.35$
to $9.54$ with the five problem sets.

\item The speedup is high and reasonably regular for the two LGDP problem sets,
  it is lower and quite irregular for the other three problem sets.

\item The speedup does not change the qualitative results of the
  evaluation in the previous sections: \optimathsat still performs
  better than all \gams tools, including \jamsBMmult,  on the Strip-packing,
  SMT-LIB, SAL and MaxSMT/SMT+PB problems sets, it performs worse on
  the Job-shop problem set.
\end{itemize}

Overall we can conclude that \optimathsat is very competitive with, and often
outperforms, \gams LGDP tools on the very-extensive set of problems we
have used to evaluate them, despite the possibility of \gams to
use \cplex in parallel mode on multiple-core CPUs.
This clearly demonstrates the potential of our novel OMT approach.

}
%

\section{Related work.} %
\label{sec:related}
%
\marg{Optimiz. in SMT}%
The idea of optimization in SMT was first introduced by Nieuwenhuis \&
Oliveras \cite{nieuwenhuis_sat06}, who presented a very-general logical 
framework of ``SMT with progressively stronger theories''
(e.g., where the theory is progressively strengthened by every
new approximation of the minimum cost),
and present implementations for MaxSMT based on this framework.
%

Cimatti et al. \cite{cimattifgss10} introduced the notion of
 ``Theory of Costs'' \calc to handle PB cost functions and constraints by
an ad-hoc and independent ``\C-solver'' in the standard lazy SMT
schema, and implemented a variant of MathSAT tool able to handle SMT
with PB constraints and to minimize PB cost functions. 

The SMT solvers \yices \cite{yices_sd} and \zthree \cite{z3_sd}
 also provide support for MaxSMT,
although there is no publicly-available document describing the
procedures used there.


Ans{\'o}tegui et al. \cite{AnsoteguiBPSV11} describe the evaluation of
an implementation of a MaxSMT procedure based on \yices, although
this implementation is not publicly available.

Cimatti et al. \cite{cgss_sat13_maxsmt}
presented a  ``modular'' approach for
MaxSMT, combining a lazy \smt{} solver with a
MaxSAT solver, which can be used as blackboxes.

We recall that  MaxSMT and SMT with PB functions can be encoded
into each other, and that both are strictly less general than 
the problem addressed in this paper (\sref{sec:optsmt}).




\smallskip
Two other forms of optimization in SMT, which are quite different from
the one presented in our work,  have been proposed in the
literature. 

Dillig et al. \cite{dilligdma12} addressed the problem of finding 
\emph{partial} models for quantified first-order
formulas modulo theories, which minimize the number of free
variables which are assigned a value from the domain.%
{Quoting an example from \cite{dilligdma12}, 
given the  formula {$\vi\defas(x+y+w>0)\vee(x+y+z+w<5)$},
the partial
  assignment $\{z=0\}$ satisfies
\vi because every total assignment extending it 
satisfies \vi and is minimum because there is no assignment satisfying \vi
which assigns less then one variable.}
They proposed a general procedure addressing the problem for every
theory \T admitting quantifier elimination, and implemented a version 
for \laint and \euf into the {\sc Mistral} tool.

Manolios and Papavasileiou \cite{manoliosp13} proposed the 
``ILP Modulo Theories'' framework as an alternative to SAT Modulo
Theories, which allows for combining 
Integer Linear Programming with decision procedures for 
signature-disjoint stably-infinite theories \T;~%
they presented a general algorithm by integrating the Branch\&Cut 
ILP method with \T-specific decision procedures, and implemented it
into the {\sc Inez} tool.
%
 Notice that the approach of \cite{manoliosp13}
  cannot combine ILP with
   \larat, since \laint and \larat are not signature-disjoint. 
 (See Definition 2 in  \cite{manoliosp13}.)
Also, the objective
function is defined on the Integer domain. 

We understand that 
neither of the above-mentioned works can handle the
problem addressed in this paper, and vice versa.~%

\onlyinsilviathesis{
\smallskip
\RSTODO{TAGLIARE QUANTO SEGUE?\\}
Closest in spirit to our work is a forthcoming paper from  Li et al. 
\cite{li_popl14}, to be presented at POPL 2014 conference.
%
It extends the \omlarat problem we introduced in
\cite{st-ijcar12} to ``multiple-objectives'',
by considering contemporarily 
a set of {\em independent} cost
variables for the input formula \vi, namely $\{\cost_1,...,\cost_k\}$,
so that the problem 
  consists in enumerating $k$ independent models for \vi, each minimizing one
  specific $\cost_i$.~%
\footnote{More precisely, in \cite{li_popl14} the objectives are {\em
    maximized}, but the problem is dual.}
(Intuitively, enumerating such models is in general more efficient than
solving one optimization problem at the time, because it allows for sharing the
SMT search steps among different cost objectives.)
Then \cite{li_popl14} proposes a multiple-objective generalization of the
linear-search algorithm  
of \cite{st-ijcar12}, and presents an
implementation called {\sc SYMBA} on top
of the \zthree SMT solver \cite{z3}.
The multiple-objective minimization is performed in two alternative
ways: an ``offline'' version, in which a sequence of blackbox calls to the SMT
solver allows for finding progressively-better solutions along one
objective direction, and a more efficient
``inline'' version, in which the simplex algorithm inside the
\larat-solver of Z3 is modified to find the optimum, as in our inline
version described in \cite{st-ijcar12} and 
in \sref{sec:algorithms_inline}. 

Looking at the empirical evaluation in
\cite{li_popl14}, where \optimathsat is invoked on k
distinct single-objective calls, we notice that 
it solves 1052 problems within the timeout, 
whilst the best ``offline'' {\sc SYMBA} version solves 1046 
and the ``inline'' version solves 1065; if invoked on single calls,
like \optimathsat, {\sc SYMBA}  solves 1045 problems. 
Notice that, if {\sc SYMBA} is restricted to work on single
objectives, we see no substantial algorithmic difference between the
two ``inline'' procedures, apart from the fact that they are built on
top of \mathsat and \zthree respectively.
}

\section{Conclusions and Future Work}
\label{sec:concl}


\ignoreinshort{In this paper we have introduced the problem of \omlaratplus,
an extension of \smtlaratplus with minimization of \larat terms,
and proposed two novel procedures addressing it.
We have described, implemented and experimentally evaluated this
approach, clearly demonstrating all its potentials. 
We believe that \omlaratplus{} and its solving procedures 
propose as a very-promising tools for a variety of optimization problems.} 

This research opens the possibility for several interesting future
directions. 
%
%
A short-term goal, which we are currently working at,
 is to extend the approach to \laint and to mixed 
$\larat$/$\laint$, by exploiting the solvers which are already present
in \mathsat{} {\cite{griggio-jsat11}}. 
\RSCHANGETHREE{As it is implicitly suggested in
\sref{sec:expeval_parallelgams}, a medium-term goal is to investigate 
the parallelization of OMT procedures, so that to exploit the power of
current multiple-core CPUs.}
A longer-term goal is to investigate the feasibility of extending
the technique to deal with non-linear constraints, possibly using 
MINLP tools as \Tsolver/\minimize{}. 

 %
\RSCHANGE{
\section*{Acknowledgements}
\label{sec:thanks}We want to thank the following people for their help and contributions:
{\bf Ignacio Grossmann},
{\bf Nicolas Sawaya},
{\bf Aldo Vecchietti},  
and the {\bf \gams support
  team} for providing help and useful suggestions vie email about LGDP and
the usage of the \gams tools;
our colleagues
{\bf Alessandro Cimatti},
{\bf Alberto Griggio} and
{\bf Patrick Trentin} for their feedback and suggestions;
%
%
the associate editor and the anonymous reviewers  for providing
many insightful comments and suggestions.

}

\bibliographystyle{acmtrans}
 \bibliography{sathanbook,rs_ownrefs,rs_refs,st_refs,mrg_a-l,mrg_m-z,rs_specific_refs}

\newpage
\appendix
\section{Appendix: Proof of the Theorems}
\label{sec:appendix}
\subsection{Proof of Theorem~\ref{teo:main}}

We first need proving the following lemmas.

\begin{lemma}
\label{lemma:main_totalass}
Let $\vi$ be a \laratplus-satisfiable \laratplus-formula and 
$\cale  \defas \{\eta_1,\ldots,\eta_n\}$
be the set of all 
total truth assignments propositionally satisfying $\vi$.
Then $\mincost(\vi)=min_{\eta_i\in\cale} \mincost(\eta_i)$.
\end{lemma}

\begin{proof}
  If $\vi$ is \laratplus-unsatisfiable, then
  $\mincost(\vi)=min_{\eta_i\in\cale} \mincost(\eta_i)=+\infty.$
  Otherwise, the thesis follows straightforwardly 
  from the fact that the set of the models of $\vi$ is the union of
  the sets of the models of the assignments in \cale.
  \qed
\end{proof}

\begin{lemma}
\label{lemma:partvstotalass}
  Let $\vi$ be a \laratplus-satisfiable \laratplus-formula and 
  $\mu$ be a \laratplus-satisfiable {\bf partial} assignment s.t. $\mu\pmodels\vi$.
Then there exists at least one \laratplus-satisfiable 
{\bf total} assignment $\eta$ s.t. 
$\mu\subseteq\eta$, $\eta\pmodels\vi$, and 
$\mincost(\mu)=\mincost(\eta)$. 
\end{lemma}
 
\begin{proof}
Let $\cali$ be a model for $\mu$, and hence for $\vi$. Then 
%
\begin{equation}
\label{eq:induced_assignment}
\eta \defas
\bigwedge_{\stackrel{\psi_i\in \atoms{\vi}}{\cali\models\psi_i}} \psi_i
\wedge
\bigwedge_{\stackrel{\psi_i\in \atoms{\vi}}{\cali\models\neg\psi_i}} \neg\psi_i
\end{equation}
By construction, $\eta$ is a total truth
assignment for \vi and it is \laratplus-satisfiable, 
$\mu\subseteq\eta$ and $\mincost(\eta)=\mincost(\mu)=\cali(\cost)$.
Since $\mu\subseteq\eta$, then $\eta\pmodels\vi$. 
\qed
\end{proof}

The proof of Theorem~\ref{teo:main} then follows.


\begin{reptheorem}{teo:main}
Let $\vi$ be a  \laratplus-formula and let 
$\calm  \defas \{\mu_1,\ldots,\mu_n\}$
be a complete collection of (possibly partial)
 truth assignments propositionally satisfying $\vi$.
Then $\mincost(\vi)=min_{\mu\in\calm} \mincost(\mu)$.
\end{reptheorem}

\begin{proof}
If $\vi$ is \laratplus-unsatisfiable, then 
$\mincost(\vi)=min_{\mu\in\calm} \mincost(\mu)=+\infty$ by
Definition~\ref{def:omlaratplus} and Theorem~\ref{teo:smt_foundations}. 
Otherwise, $\mincost(\vi)<+\infty$. Then:
\begin{itemize}
  \item[Proof of $\mincost(\vi)\le min_{\mu\in\calm} \mincost(\mu)$:]\ \\
    By absurd, suppose exists $\mu\in\calm$
    s.t. $\mincost(\mu)<\mincost(\vi)$.
    By Proposition~\ref{prop:cost_monotonicity}, $\mu$ is \laratplus
    satisfiable.
    By Lemma~\ref{lemma:partvstotalass}, there exists a
    \laratplus-satisfiable total assignment $\eta$ s.t.
    $\mu\subseteq\eta$, $\eta\pmodels\vi$, and
    $\mincost(\mu)=\mincost(\eta)$.
By lemma~\ref{lemma:main_totalass}, 
$\mincost(\eta)\ge\mincost(\vi)$, and hence $\mincost(\mu)\ge\mincost(\vi)$,
contradicting the hypothesis. 

  \item[Proof of $\mincost(\vi)\ge min_{\mu\in\calm} \mincost(\mu)$:]\ \\
    From Lemma~\ref{lemma:main_totalass} we have that
    $\mincost(\vi)=min_{\eta_i\in\cale} \mincost(\eta_i)$.
    Let $\eta\in\cale$ s.t. $\mincost(\vi)=\mincost(\eta)<+\infty$.
    Hence $\eta$ is \laratplus-satisfiable.
%
    Thus, there exists
    $\mu\in\calm$ s.t. $\mu\subseteq\eta$.
    $\mu$ is \laratplus-satisfiable since $\eta$ is
    \laratplus-satisfiable.
From Proposition~\ref{prop:cost_monotonicity},
$\mincost(\mu)\le\mincost(\eta)$, hence 
    $\mincost(\mu)\le\mincost(\vi)$.
    Thus 
    the thesis holds.
%
\end{itemize}
\end{proof}

\newpage
\subsection{Proof of Theorem~\ref{teo:mued-extension}}

\begin{reptheorem}{teo:mued-extension}
\label{teo:mued-extension2}
  Let $\mu$ be as in
  Definition~\ref{def:mued-extension}.   
Then 
\begin{aenumerate}
\item
$\mincost(\mu)=min_{\eta\in\edmuext} \mincost(\eta)$
\item forall $\eta \in\edmuext$, \\
$
\mincost(\eta) = 
\left \{ 
  \begin{array}{lll}
+\infty && \mbox{if $\mut\wedge\mued$ is
  \T-unsatisfiable or} \\  
 &&    \mbox{if $\mularat\wedge\mued$ is
  \larat-unsatisfiable} \\
\mincost(\mularat\wedge\mued) & \ \ \ \ &  \mbox{otherwise.}
  \end{array}
\right .
$

\end{aenumerate}
\end{reptheorem}

\begin{proof}
\ \\
\begin{aenumerate}
\item Let
$$
\mu' \defas \mu \wedge \bigwedge_{\eqij \in\IE{\mu}} (\eqij \vee \neqij).
$$
\noindent
$\mu$ and $\mu'$ are obviously \laratplus-equivalent, so that
$\mincost(\mu)=\mincost(\mu')$. By construction, \edmuext{} is the set
of all total truth assignments propositionally satisfying $\mu'$, so that 
$\mincost(\mu')=min_{\eta\in\edmuext} \mincost(\eta)$.
 
\item By Theorem~\ref{teo:NO},
$\eta$ is \laratplus-satisfiable if and only if
    $\mularat\wedge\mued$ is
    \larat-satisfiable
    and $\mut\wedge\mued$ is \T-satisfiable.
Thus, 
\begin{itemize}
\item if $\mut\wedge\mued$ is \T-unsatisfiable,
then $\eta$ is \laratplus-unsatisfiable, so that 
$\mincost(\eta) = +\infty$.
\item If $\mut\wedge\mued$ is \T-satisfiable
and $\mularat\wedge\mued$ is
    \larat-unsatisfiable, then $\eta$ is \laratplus-unsatisfiable, so
    that $\mincost(\eta) = \mincost(\mularat\wedge\mued)=+\infty$.

\item If $\mut\wedge\mued$ is \T-satisfiable
and $\mularat\wedge\mued$ is
    \larat-satisfiable, then $\eta$ is \laratplus-satisfiable. 
We split the proof into two parts.
  \begin{itemize}
    \item[{\bf $\le$ case:}]
Let $c\in\mathbb{Q}$ be the value of $\mincost(\mularat\wedge\mued)$.
Let $\mu'\defas \mu\wedge(\cost=c)$. Since $(\cost=c)$ is a
\larat-pure atom, then $\mu'=\mut'\wedge\mularat'$ s.t. $\mut'=\mut$ and
$\mularat'=\mularat\wedge(\cost=c)$, which are respectively 
\T- and \larat-pure and
\T- and \larat-satisfiable by construction. Let
$\eta'\defas\eta\wedge(\cost=c)$.
Since \IE{\mu}=\IE{\mu'}, then $\mu'$, $\mularat'$, $\mut'$ and
$\eta'$ match  
the hypothesis of Theorem~\ref{teo:NO}, from which we have that
$\eta'$ is \laratplus-satisfiable, so that $\eta$ has a model 
$\cali$ s.t. $\I(\cost)=c$. Thus, we have that
$\mincost(\eta)\le\mincost(\mularat\wedge\mued)$.

    \item[{\bf $\ge$ case:}]
%
%
Let $c\in\mathbb{Q}$ be the value of $\mincost(\eta)$.
Then $\eta\wedge(\cost=c)$ is \laratplus-satisfiable. 
We define $\mu'$, $\mularat'$, $\mut'$ and
$\eta'$  as in the ``$\le$'' case. As before, they match 
the hypothesis of Theorem~\ref{teo:NO}, from which we have that
$\mularat'$ is \larat-satisfiable. Hence,
\mularat has a model $\I$ s.t. $\I(\cost)=c$.
Thus, we have that
$\mincost(\eta)\ge\mincost(\mularat\wedge\mued)$. 

  \end{itemize}

\end{itemize}

\end{aenumerate}
\hfill\qed
\end{proof}

\newpage
\subsection{Proof of Theorem~\ref{teo:muedi-extension}}

\begin{reptheorem}{teo:muedi-extension}
\label{teo:muedi-extension2}
  Let $\mu$ be as in
  Definition~\ref{def:mued-extension}.  
Then 
\begin{aenumerate}
\item $\mu$ is \laratplus-satisfiable iff some $\rho\in\edimuext$ 
is \laratplus-satisfiable. 
\item
$\mincost(\mu)=min_{\rho\in\edimuext} \mincost(\rho)$.
\item forall $\rho\in\edimuext$,
 $\rho$ is \laratplus-satisfiable iff $\mut\wedge\mued$ is
  \T-satisfiable and $\mularat\wedge\mue\wedge\mui$ is
  \larat-satisfiable.  
\item forall $\rho \in\edimuext$, \\
$
\mincost(\rho) = 
\left \{ 
  \begin{array}{lll}
+\infty && \mbox{if $\mut\wedge\mued$ is
  \T-unsatisfiable or} \\  
 &&    \mbox{if $\mularat\wedge\mue\wedge\mui$ is
  \larat-unsatisfiable} \\
\mincost(\mularat\wedge\mue\wedge\mui) & \  \ &  \mbox{otherwise.}
  \end{array}
\right .
$
\end{aenumerate}
\end{reptheorem}

\begin{proof}
Let
\begin{eqnarray}
\label{eq:augmentedass}
  \mu^{*}&\defas &
\mu 
\wedge 
\bigwedge_{\eqij\in \cali\cale(\mu)}
\left ( 
\begin{array}{lll}
  &(\eqij\vee\dij\vee\dji) \wedge \\
  &(\neg\eqij\vee\neg\dij) \wedge \\
  & (\neg\eqij\vee\neg\dji) \wedge\\
  & (\neg\dij\vee\neg\dji) 
 \end{array}
 \right ) 
\end{eqnarray}
All clauses in the right conjuncts in \eqref{eq:augmentedass}
are \larat-valid, hence
$\mu$ and $\mu^{*}$ are \laratplus-equivalent, so that
$\mincost(\mu)=\mincost(\mu^{*})$. 
By construction, \edimuext{} is the set
of all total truth assignments propositionally satisfying $\mu^{*}$.

\begin{aenumerate}
\item By Theorem~\ref{teo:smt_foundations}, $\mu^{*}$ is 
\laratplus-satisfiable iff some $\rho\in\edimuext$ 
is \laratplus-satisfiable, from which the thesis.

\item $\mincost(\mu)=\mincost(\mu^{*})=min_{\rho\in\edimuext} \mincost(\rho)$.

\item 
We consider one
$\rho\in\edimuext$.
 $\rho=\mut\wedge\mularat\wedge\mue\wedge\mud\wedge\mui$. 
We notice that all literals in \mui are \larat-pure,
s.t. it is  the \larat-pure part of $\rho$ (namely, $\rho_{\larat}$).
Thus, by Theorem~\ref{teo:NO},
$\rho$ is \laratplus-satisfiable iff 
    $\overbrace{\mularat\wedge\mui}^{\rho_{\larat}}\wedge\overbrace{\mue\wedge\mud}^{\mued}$ is
    \larat-satisfiable
    and $\mut\wedge\overbrace{\mue\wedge\mud}^{\mued}$ is \T-satisfiable.
By construction, $\mui\models_{\larat}\mud$. 
Thus, $\mularat\wedge\mui\wedge\mue\wedge\mud$ is
    \larat-satisfiable iff $\mularat\wedge\mui\wedge\mue$ is
    \larat-satisfiable. Thus the thesis holds.

\item We consider one
$\rho\in\edimuext$ and partition it as in point (c).
From point (c), if $\mut\wedge\mued$ is \T-unsatisfiable or
$\mularat\wedge\mui\wedge\mue$ is
    \larat-unsatisfiable, then $\rho$ is \laratplus-unsatisfiable, so
    that $\mincost(\rho)=+\infty$.
Otherwise, $\rho$ is \laratplus-satisfiable. 


  \begin{itemize}
    \item[{\bf $\le$ case:}]
Let $c\in\mathbb{Q}$ be the value of
$\mincost(\mularat\wedge\mue\wedge\mui)$. 
Let $\mu'\defas \mu\wedge(\cost=c)$.
Since $(\cost=c)$ is a
\larat-pure atom, then $\mu'=\mut'\wedge\mularat'$ s.t. $\mut'=\mut$ and
$\mularat'=\mularat\wedge(\cost=c)$, which are respectively 
\T- and \larat-pure. 
Also, $\mut'\wedge\mued$ is \T-satisfiable 
and $\mularat'\wedge\mue\wedge\mui$ is \larat-satisfiable 
by construction. 
Let
$\rho'\defas\rho\wedge(\cost=c)$.
Since \IE{\mu}=\IE{\mu'}, then also $\mu'$, $\mularat'$, $\mut'$ and
$\rho'$ match  the hypothesis of this theorem. 
Thus,  by point (c), 
$\rho'$ is \laratplus-satisfiable, so that $\rho$ has a model 
$\cali$ s.t. $\I(\cost)=c$.
Therefore we have that $\mincost(\rho)\le\mincost(\mularat\wedge\mue\wedge\mui)$. 

    \item[{\bf $\ge$ case:}]
Let $c\in\mathbb{Q}$ be the value of $\mincost(\rho)$.
Then $\rho\wedge(\cost=c)$ is \laratplus-satisfiable. 
We define $\mu'$, $\mularat'$, $\mut'$ and
$\rho'$  as in the ``$\le$'' case. 
As before, they also match 
the hypothesis of this theorem, so that by point (c)
$\mularat'\wedge\mue\wedge\mui$ is \larat-satisfiable. 
Thus, $\mularat'\wedge\mue\wedge\mui$ has a model $\I$
s.t. $\I(\cost)=c$. 
Therefore we have that
$\mincost(\rho)\ge\mincost(\mularat\wedge\mue\wedge\mui)$. 


 \end{itemize}

\end{aenumerate}
\ \hfill\qed

\end{proof}

\end{document}